\documentclass[structabstract]{aa}  
\usepackage{subfig,graphicx}
\usepackage{txfonts}
\usepackage{natbib}
\begin{document}
   \title{GOODS-{\it Herschel}: evidence for a UV extinction bump in galaxies at $z > 1$}


   \author{V. Buat\inst{1}
   \and  E. Giovannoli \inst{1}
          \and S. Heinis \inst{1}
         \and V. Charmandaris\inst{2,3,4}
          \and D. Coia\inst{5}
            \and E. Daddi\inst{6}
            \and M. Dickinson\inst{7}
             \and D. Elbaz \inst{6}
            \and H.S. Hwang \inst{6}
            \and G. Morrison\inst{8,9}
            \and K.~Dasyra\inst{6,4}
            \and H.~Aussel\inst{6}
            \and B.~Altieri\inst{5}
\and H. Dannerbauer\inst{6}
\and J.~Kartaltepe\inst{7}
\and R.~Leiton\inst{6}
\and G.~Magdis\inst{6}
\and B.~Magnelli\inst{10}
\and P.~Popesso\inst{10}
          }

   \offprints{V. Buat}
   \institute{Laboratoire d'Astrophysique de Marseille, OAMP, Universit\'e Aix-marseille, CNRS, 38 rue Fr\'ed\'eric Joliot-Curie, 13388 Marseille cedex 13, France\\
              \email{veronique.buat@oamp.fr}
         \and University of Crete, Department of Physics and Institute of Theoretical \& Computational Physics, GR-71003 Heraklion, Greece
\and IESL/Foundation for Research \& Technology-Hellas, GR-71110 Heraklion, Greece
\and Chercheur Associ\'e, Observatoire de Paris, LERMA (CNRS:UMR8112), 61 Av. de l\'\ Observatoire,
F-75014, Paris, France
         \and Herschel Science Operation Centre, ESAC, ESA, PO Box 78, 28691 Villanueva de la Ca{\~n}ada, Madrid, Spain
        \and Laboratoire AIM-Paris-Saclay, CEA/DSM/Irfu - CNRS - Universit\'e Paris Diderot, CE-Saclay, F-91191 Gif-sur-Yvette, France
         \and National Optical Astronomy Observatory, 950 North Cherry Avenue, Tucson, AZ 85719, USA
         \and Institute for Astronomy, University of Hawaii, Honolulu, HI, 96822, USA
         \and Canada-France-Hawaii telescope, Kamuela, HI, 96743, USA
         \and Max-Planck-Institut f\"ur Extraterrestrische Physik (MPE), Postfach 1312, 85741, Garching, Germany
           }

   \date{}

 
  \abstract
   {Dust attenuation curves in external galaxies are  useful to study their dust  properties as well as to interpret their intrinsic spectral energy distributions. These functions are not very well known in the UV range either at low or high redshift. In particular the presence or absence of a UV bump at 2175 $\AA$ remains an open issue which has consequences on the interpretation of broad band colours of galaxies involving the UV range.}
 {We want to study  the dust attenuation curve in the UV range at $z >1$  where the UV is redshifted into the visible, and with $Herschel$ data to constrain dust emission and a global dust attenuation. In particular we will search for the presence of a UV bump  and the related implications on dust attenuation determinations.}
{We use deep photometric data of the Chandra Deep Field South obtained  with  intermediate and broad band  filters  by the MUSYC project  to sample the UV rest-frame of galaxies with  $1<z <2$. 
{\it Herschel}/PACS  and  {\it Spitzer}/MIPS data are used to measure the dust emission. 30 galaxies were  selected  with  high S/N in all  bands. Their SEDs from the UV to the far-IR are fitted using the  CIGALE code  and the characteristics of the  dust attenuation curve are obtained as Bayesian outputs of the SED fitting process. }
{ The  mean dust attenuation curve we derive exhibits  a significant UV bump  at 2175 $\AA$ whose amplitude  corresponds  to 35$\%$ (76 $\%$)  that of the Milky Way (Large Magellanic Cloud: LMC2 supershell) extinction curve.  An analytical expression of the average attenuation curve ($A(\lambda)/A_{\rm V}$)  is given, it is found slightly steeper than  the  Calzetti et al.  one, although at a 1$\sigma$ level.  Our  galaxy sample is used to study  the derivation of the  slopes of the UV continuum  from broad band colours, including the rest-frame GALEX $FUV-NUV$ colour. Systematic errors induced by the presence of the bump are quantified. We compare dust attenuation factors measured with  CIGALE   to the slope of the UV continuum and find that there is a large scatter   around the relation  valid for local starbursts ($\sim 0.7$ mag). The uncertainties on the determination of the UV slope   lead to an extra systematic  error of the order of 0.3 to 0.7 mag on dust attenuation when a filter overlaps the UV bump.}


       
   \maketitle
%

\section{Introduction}
Stellar light in galaxies is absorbed and scattered by the interstellar medium, because of the presence of dust grains. On a galaxy scale, the process is usually quantified by the introduction of an average attenuation function which reflects the combined effects of  absorption and scattering in a complex geometry. \\
The most important feature in the extinction curves of the Milky Way (MW) and Large Magellanic Cloud (LMC) is the so called UV bump, a strong absorption feature at 2175 $\AA$ which is not observed  in the Small Magellanic Cloud (SMC) bar, although a line of sight through the SMC wing exhibits an extinction curve with a strong UV bump \citep{li02,cartledge05}. The exact origin of this feature is still under discussion \citep[e.g.][and references therein]{draine89,xiang11}. When dealing with average attenuation curves, the search for a bump is  complicated by the effects of scattering or  geometry \citep{inoue06,noll07,panuzzo07} which may affect the amplitude of the resulting bump in an average attenuation curve even when assuming  a local Milky Way like extinction curve. \\
The best way to study UV bumps  in external galaxies is to perform  UV rest-frame spectroscopy: \citet{calzetti94} deduced the net attenuation curves of local starburst galaxies  by comparing the UV spectra of  dusty and dust-free spectra of  local starbursts: they did not find any  bump.  Conversely,  \citet{noll05,noll09a} observed a significant 2175 $\AA$ absorption feature in  the spectra of  star forming galaxies at $z\sim 2$ consistent with an extinction curve intermediate between the SMC and LMC ones.  \\
The 2175 $\AA$ feature has also been detected in the spectral energy distribution of some distant galaxies  hosting    gamma-ray bursts \citep{liang09,liang10}, the extinction curves are found to display a large diversity of shapes  \citep[][and references therein]{liang09,liang10,zafar11}. We  also refer to \citet{xiang11} for an extensive review on the detection of the UV bump in Galactic and extragalactic interstellar medium.\\
Broad band colours are also used to search for evidence of a UV bump. From an analysis of the observed B-R colours of galaxies at $z\sim 1$  \citet{conroy10a}  disfavours  the presence of UV bumps as strong as that in the Milky Way whereas \citet{burgarella05} analysed the broad bands SEDs of nearby galaxies observed by GALEX and IRAS and conclude to the presence of a bump in the attenuation curve of these galaxies. In all cases, the use of broad band colours which overlap the bump makes difficult to disentangle the effects of a dust extinction curve with or without a bump, of  dust/stars geometry and of various star formation histories \citep{inoue06,panuzzo07,johnson07a,johnson07b,buat11} leading to contradicting conclusions on the presence of a UV bump in  nearby galaxies observed by GALEX and SDSS \citep{conroy10b,johnson07a,johnson07b}. \\
The presence and characteristics of the UV bump, and far UV rise in the extinction and attenuation curves can provide information on the interstellar medium properties \citep{noll09a}. Beyond the studies of the interstellar medium, an  reliable attenuation curve  in the UV range is very useful to derive star formation rates in galaxies since the UV emission is commonly used as a star formation tracer.  For example, \citet{wije10} compared star formation rates deduced from luminosities measured in  the GALEX bands and the H$\alpha$ and [OII] emission lines and concluded that they must remove the 2175 $\AA$ feature from the attenuation curve to obtain the best agreement between the different SFR estimators. The   attenuation curve of \citet{calzetti00} is commonly used, especially in high-z studies and does not include a bump: if  such a bump is present, omitting it leads to incorrect dust attenuation corrections \citep{ilbert09}. \\
The global amount of dust attenuation is robustly estimated by comparing dust and stellar emission, through the $L_{\rm IR}/L_{\rm UV}$ ratio. When IR data are not available 
 the shape of the UV continuum ($< 3000$  $\AA$) has been proposed  as a proxy to measure the amount of attenuation.
 \citet{calzetti94} showed that  the UV continuum of local starburst galaxies can be  fit by  a power law  ($ f_\lambda ({\rm erg ~cm^{-2} s^{-1}  \AA^{-1}}) \propto \lambda^\beta$)  for $\lambda > 1200 $  $\AA$. They calculated $\beta$ by defining 10 windows in the IUE spectra of their galaxies  avoiding the 2175 $\AA$ feature.  \citet{Meurer99} found a relation between  $\beta$ and  the amount of dust attenuation measured by $L_{\rm IR}/L_{\rm UV}$.  This  local starburst relation is widely used to estimate dust attenuation in high redshift galaxies using a broad band  colour  in the UV rest-frame  of the galaxies as a proxy of  $\beta$ \citep[e.g.][]{burgarella07,elbaz07,daddi07,reddy08}. If a bump is present in the dust attenuation curve, one must be cautious by choosing broad band filters which do not overlap the bump in order to estimate reliable slopes outside the wavelength range affected by the bump, as was done to derive the \citet{Meurer99} relation. This has been  illustrated by studies of the local universe based on GALEX FUV and NUV data: the NUV band largely overlaps the UV bump and the FUV-NUV colour is found to be sensitive to the presence or absence of this feature as well as to the recent star formation history \citep{panuzzo07,inoue06}.\\
In this paper we take advantage of the availability of high quality deep optical data obtained through intermediate-band filters in the Chandra Deep Field South (CDFS) to   sample the rest-frame UV of galaxies at high redshift. These data are combined with the  $Herschel$ observations of the CDFS  field as part of the open time key program GOODS-$Herschel$  project (P.I. D. Elbaz).  The galaxies detected  in the  far-IR   are expected to experience a substantial extinction, which    helps to detect the imprint of the bump in their SED. The combination of UV and IR data will allow us to compare the properties of the UV continuum to the dust attenuation at work in these objects. In a similar vein, \citet{ilbert09} used broad and intermediate band filters to derive photometric redshifts in the COSMOS field and found that a UV bump is necessary to explain the UV rest-frame fluxes of some starburst galaxies but they did not discuss the amplitude of the bump and  dust attenuation in  these galaxies.\\
The construction of the sample is explained in section 2. We want to use only  sources detected with high signal to noise ratio (SNR) in a large number of filters which sample well the rest-frame UV wavelength region affected by the bump.  This criterium leads to a redshift selection between 0.95 and 2.2. 
The dust attenuation curve as well as the amount of dust obscuration is  constrained by fitting the whole SEDs of the galaxies with the CIGALE code, which is described in section 3. Section 4 is devoted to the results about the presence of the UV bump and section 5 to the consequences on the determination of the slope of the UV continuum and dust attenuation in galaxies. The conclusions are presented in section 6. 
   

\section{Selection of the sample}
   \begin{figure}
   \centering
  \includegraphics[width=8cm]{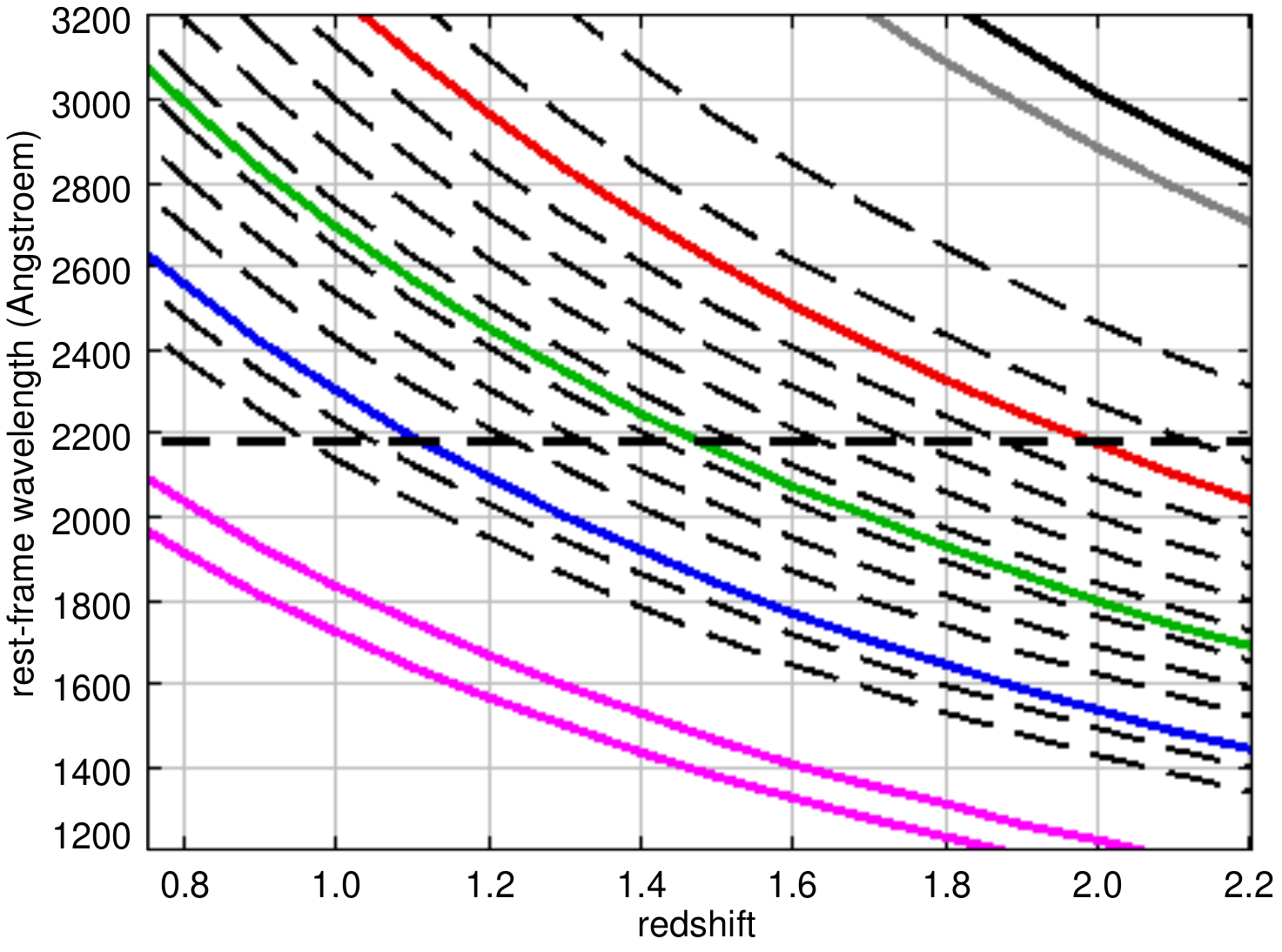}
  
   \caption{Rest-frame ultraviolet wavelengths sampled with our dataset as a function of redshift. The solid curves correspond to broad band filters (from bottom left corner  to top right corner: U,U38,B,V,R,I and z) and the dashed curves to intermediate-band filters (from bottom left corner to top right corner: IA-427,445,484,505,527,550,574,598,624,651 (coincident with the red solid curve for the R-band), 679 and 738. The FWHM of the intermediate filters varies from 20 to 34 nm. The horizontal dotted line corresponds to 2175 $\AA$,  the central wavelength of the expected bump.}
              \label{lambda}%
 \end{figure}

%

\begin{table*}
\caption{Galaxy sample}             
\label{sample}      
\begin{tabular}{|l|l|l|l|l|l|l|l|l|l|r|l|l|}
\hline
  \multicolumn{1}{|c|}{N} &
  \multicolumn{1}{c|}{RA (2000)} &
  \multicolumn{1}{c|}{DEC (2000)} &
   \multicolumn{1}{c|}{z} &
  \multicolumn{1}{c|}{$\log(L_{\rm IR})$} &
  \multicolumn{1}{c|}{$A_{\rm UV}$} &
  \multicolumn{1}{c|}{$E_{\rm b}$} &
  \multicolumn{1}{c|}{$err_{E_{b}}$} &

  \multicolumn{1}{c|}{$\gamma$} &
  \multicolumn{1}{c|}{$err_{\gamma}$} &
  \multicolumn{1}{c|}{$\delta$} &
  \multicolumn{1}{c|}{$err_{\delta}$} &
  \multicolumn{1}{c|}{$\chi^2_{min}$} \\
 \multicolumn{1}{|c|}{} &
  \multicolumn{1}{c|}{deg} &
  \multicolumn{1}{c|}{deg} &
  \multicolumn{1}{c|} {}&
  \multicolumn{1}{c|}{$L_{\odot}$} &
  \multicolumn{1}{c|}{mag} &
  \multicolumn{1}{c|}{} &
  \multicolumn{1}{c|}{} &
  \multicolumn{1}{c|}{$\AA$} &
  \multicolumn{1}{c|}{$\AA$} &
  \multicolumn{1}{c|}{} &
  \multicolumn{1}{c|}{} &
  \multicolumn{1}{c|}{} \\
\hline
  1 & 03:32:36.19 & -27:42:58.2 & 0.967 & 11.45 & 1.14 & 0.98 & 0.67 & 339.32 & 107.00 & 0.09 & 0.03 & 3.16\\
  2 & 03:32:20.72 & -27:44:53.7 & 0.969 & 11.50 & 2.80 & 1.74 & 0.31 & 361.66 & 102.58 & -0.30 & 0.00 & 2.69\\
  3 & 03:32:16.45 & -27:44:48.7 & 0.976 & 11.49 & 2.98 & 1.15 & 0.61 & 348.02 & 107.33 & 0.03 & 0.07 & 2.63\\
  4 & 03:32:31.80 & -27:49:58.4 & 0.98\tablefootmark{2}  & 11.49 & 2.89 & 1.18 & 0.54 & 360.33 & 107.93 & -0.16 & 0.09 & 1.43\\
  5 & 03:32:42.28 & -27:47:46.1 & 0.998 & 11.43 & 1.98 & 1.68 & 0.34 & 376.98 & 101.85 & -0.20 & 0.12 & 2.31\\
  6 & 03:32:37.39 & -27:44:07.0 & 1.018 & 11.28 & 3.90 & 1.51 & 0.46 & 350.17 & 106.16 & -0.24 & 0.06 & 1.76\\
  7 & 03:32:06.43 & -27:47:28.7 & 1.021 & 11.69 & 3.75 & 1.25 & 0.54 & 370.61 & 106.19 & -0.29 & 0.03 & 2.30\\
  8\tablefootmark{1}  & 03:32:15.78 & -27:46:30.2 & 1.023 & 11.35 & 1.68 & 1.05 & 0.65 & 347.67 & 107.83 & -0.21 & 0.05 & 3.72\\
  9 & 03:32:36.02 & -27:44:23.8 & 1.038 & 11.34 & 1.98 & 1.55 & 0.45 & 351.24 & 106.58 & -0.02 & 0.05 & 1.65\\
  10 & 03:32:10.58 & -27:47:06.1 & 1.045 & 11.42 & 2.83 & 0.95 & 0.62 & 352.28 & 108.69 & -0.10 & 0.09 & 1.24\\
  11 & 03:32:28.98 & -27:49:08.2 & 1.094 & 11.59 & 2.01 & 1.63 & 0.43 & 375.24 & 102.58 & -0.29 & 0.02 & 5.58\\
  12 & 03:32:39.87 & -27:47:15.0 & 1.095 & 11.43 & 1.93 & 1.79 & 0.29 & 400.42 & 92.73 & -0.27 & 0.07 & 2.25\\
  13 & 03:32:25.86 & -27:50:19.6 & 1.095 & 11.52 & 2.40 & 1.42 & 0.52 & 355.97 & 106.19 & -0.16 & 0.06 & 2.30\\
  14\tablefootmark{1} & 03:32:17.18 & -27:52:20.5 & 1.097 & 12.09 & 5.61 & 1.03 & 0.39 & 368.57 & 104.51 & -0.30 & 0.02 & 4.56\\
  15 & 03:32:34.85 & -27:46:40.5 & 1.099 & 11.46 & 4.19 & 1.61 & 0.43 & 373.69 & 102.00 & -0.18 & 0.08 & 1.08\\
  16 & 03:32:34.63 & -27:53:24.5 & 1.107 & 11.61 & 2.47 & 1.04 & 0.64 & 344.75 & 107.87 & -0.03 & 0.10 & 1.12\\
  17\tablefootmark{1} & 03:32:29.97 & -27:45:29.8 & 1.209 & 12.07 & 1.13 & 0.87 & 0.67 & 328.56 & 106.76 & 0.08 & 0.04 & 15.92\\
  18 & 03:32:43.48 & -27:45:56.5 & 1.220 & 11.45 & 2.54 & 1.35 & 0.55 & 336.36 & 105.68 & -0.21 & 0.06 & 2.70\\
  19 & 03:32:47.17 & -27:51:06.1 & 1.224 & 11.82 & 4.66 & 1.10 & 0.50 & 337.70 & 107.00 & -0.29 & 0.03 & 1.22\\
  20\tablefootmark{1}  & 03:32:35.97 & -27:48:50.4 & 1.309 & 11.54 & 1.24 & 0.87 & 0.67 & 337.88 & 107.68 & 0.02 & 0.07 & 2.70\\
  21\tablefootmark{1} & 03:32:29.93 & -27:43:01.0 & 1.37\tablefootmark{2}  & 12.12 & 4.29 & 1.62 & 0.32 & 408.06 & 89.38 & 0.01 & 0.06 & 1.71\\
  22\tablefootmark{1} & 03:32:28.79 & -27:47:55.6 & 1.383 & 11.69 & 4.02 & 0.66 & 0.47 & 331.44 & 107.91 & -0.21 & 0.07 & 3.36\\
  23 & 03:32:34.03 & -27:50:28.8 & 1.384 & 11.82 & 3.32 & 1.45 & 0.38 & 369.56 & 102.64 & 0.08 & 0.04 & 2.50\\
  24 & 03:32:41.33 & -27:45:38.2 & 1.52\tablefootmark{2}  & 11.91 & 3.69 & 1.30 & 0.51 & 332.05 & 107.31 & -0.20 & 0.08 & 1.53\\
  25\tablefootmark{1}  & 03:32:37.76 & -27:52:12.3 & 1.609 & 12.33 & 4.63 & 1.03 & 0.51 & 333.88 & 111.60 & -0.04 & 0.05 & 6.39\\
  26 & 03:32:37.71 & -27:50:00.4 & 1.619 & 12.43 & 5.21 & 0.94 & 0.36 & 352.65 & 105.73 & -0.09 & 0.07 & 2.88\\
  27 & 03:32:17.82 & -27:46:39.7 & 1.956 & 11.93 & 2.85 & 0.83 & 0.60 & 368.08 & 107.77 & 0.01 & 0.08 & 1.69\\
  28 & 03:32:41.77 & -27:46:56.4 & 1.994 & 12.00 & 3.32 & 1.03 & 0.50 & 351.00 & 107.88 & 0.02 & 0.09 & 2.85\\
  29 & 03:32:40.05 & -27:47:55.4 & 1.997 & 12.40 & 3.79 & 1.39 & 0.40 & 328.89 & 103.03 & -0.13 & 0.10 & 2.30\\
  30 & 03:32:40.75 & -27:49:26.1 & 2.130 & 12.12 & 3.02 & 1.33 & 0.53 & 350.01 & 107.06 & -0.08 & 0.11 & 1.41\\
\hline\end{tabular}

\tablefoot{IR luminosities($\log(L_{\rm IR})$), dust attenuation ($A_{\rm UV}$) and parameters relative to the attenuation curve (the amplitude and width of the UV bump,$E_b$ and  $\gamma$, and the steepness of the curve, $\delta$) and their corresponding errors are obtained  from the  Bayesian analysis performed with CIGALE.\\
\tablefoottext{1}{X-ray emitter }\\
\tablefoottext{2}{No spectroscopic redshift, the photometric redshift  is used.}
}

\end{table*}
As part of the GOODS-$Herschel$ key program \citep{elbaz11}, the $Herschel$ Space Observatory \citep{pilbratt10} surveyed a small area  of the Great Observatories Origins Deep Survey Southern field  (GOODS-S):  $10'\times 10'$ centered on the CDFS  were observed at 100 and 160 $\mu$m over 264 hours  by the PACS instrument \citep{poglitsch10}.   Source extraction on the PACS images was performed at the prior position of ${\it Spitzer}$ 24 $\mu$m sources as described in \citet{elbaz11}.\\
Cardamone et al. (2010, the MUSYC project) compiled a uniform catalog of optical and infrared photometry for sources in GOODS-S, incorporating the GOODS {\it Spitzer} IRAC and MIPS data \citep{dickinson03}.   In particular, they used deep intermediate-band imaging from the Subaru telescope to provide photometry with finer wavelength sampling than is possible from standard broad band data.  This was done in part to enable more accurate photometric redshifts, but for our purposes here it provides a valuable means of tracing the detailed shape of the UV rest-frame spectrum and to look for the presence of a UV absorption bump.\\
 We start with the GOODS-$Herschel$ catalogue of 533 sources detected  at 100 $\mu$m, and identify 325 sources with a SNR (flux over error) at  100 $\mu$m larger than 5.  We further restrict the sample to the 235 sources also detected at 160 $\mu$m with a SNR $ > 3$ and we cross-correlate them  with the MUSYC catalogue of 84402 sources with photometric redshifts. The tolerance radius between IRAC   and optical coordinates is taken to be 2 arcsec, it results in a sample of  219 sources cross-matched. Total fluxes in optical and {\it Spitzer} bands are computed from aperture fluxes as described in  \citet{cardamone09}.  \\
Before selecting sources according to their redshift we must choose the bands we will use: the redshift range will be selected to ensure a good sampling of the UV range.
All the optical  broad bands of the MUSYC survey are considered (U, U38, B, V, R,I,z) and   J,H,K bands are added when available.   We also add data from intermediate band filters whose 5$\sigma$ depth is fainter than 25 ABmag (table 2 of \citet{cardamone09}). 
Practically we use IA-427,445,484,505,527,550,574,598,624,651,679 and 738. Figure 1 illustrates the  rest-frame wavelengths corresponding to these filters as a function of z. \\
In order to ensure a good  photometric sampling around the UV bump (2175 $\AA$) we select galaxies with  redshifts   between 0.95 and 2.2. In this redshift range we have more than 10 photometric bands available in the UV (1300-3000 $\AA$) and a good sampling around 2175 $\AA$ (Fig.\ref{lambda}).  76  galaxies fulfill the redshift condition.  To ensure that sources at $z < 1.6$  have data below rest-frame 1800 $\AA$, they must have been detected in the U and U38 filters above 5$\sigma$  (see Table 3 of \citet{cardamone09}). This leads us to  39 galaxies all of which are also strongly detected in the other broad bands as well as above 5$\sigma$ in the IA-427 filter. When $z > 1.6$, measurements beyond the U filter are sufficient, so we select sources   detected  with SNR $ > 5 $ in the B band, adding an additional   constrain of a  5$\sigma$ detection an  intermediate filter (IA-427 for $z<1.8$ and IA-484 for $z>1.8$), 13 galaxies are thus added to the selection. At the end we have selected 52 galaxies. As a further check, we retrieve the HST images at the coordinates of these 52 sources using the GOODS cutout service V2.0 provided by the Multimission Archive at STScI   (http://archive.stsci.edu).  We  discard 22 sources  with bright   neighbours  closer than  4 arcsec  to the $Herschel$ detection and  whose IR measurements are uncertain (the FWHM of the PSF at 100 $\mu$m is 6.7 arcsec).   Half of the remaining sample of 30 galaxies has a disk-like morphology, the other half is dominated by compact sources with few interacting-peculiar galaxies while 28   have spectroscopic redshifts. 
 The sample of galaxies is presented in Table 1. The HST images for each target are available as online material.
7 galaxies are identified as X-ray emitters by \citet{cardamone08} they are flagged in Table 1. \\
As described earlier in this section, the present study is based on galaxies strongly detected from the UV to the far-IR, whose rest-frame UV spectrum is sampled by several intermediate-band filters. These highly selective criteria lead to a rather small sample of objects which we will study in detail in this paper in order to derive the parameters of a dust attenuation curve. Our analysis is complementary to other studies at intermediate and high redshifts which are based on larger samples but they lack far-IR data. These studies have concluded that a UV bump must be introduced in the attenuation curve but they were not able to quantify neither the amplitude of the bump nor the modification of the attenuation curve \citep{ilbert09,kriek11}

       \begin{figure*}
\centering
\includegraphics[width=15cm,angle=-90]{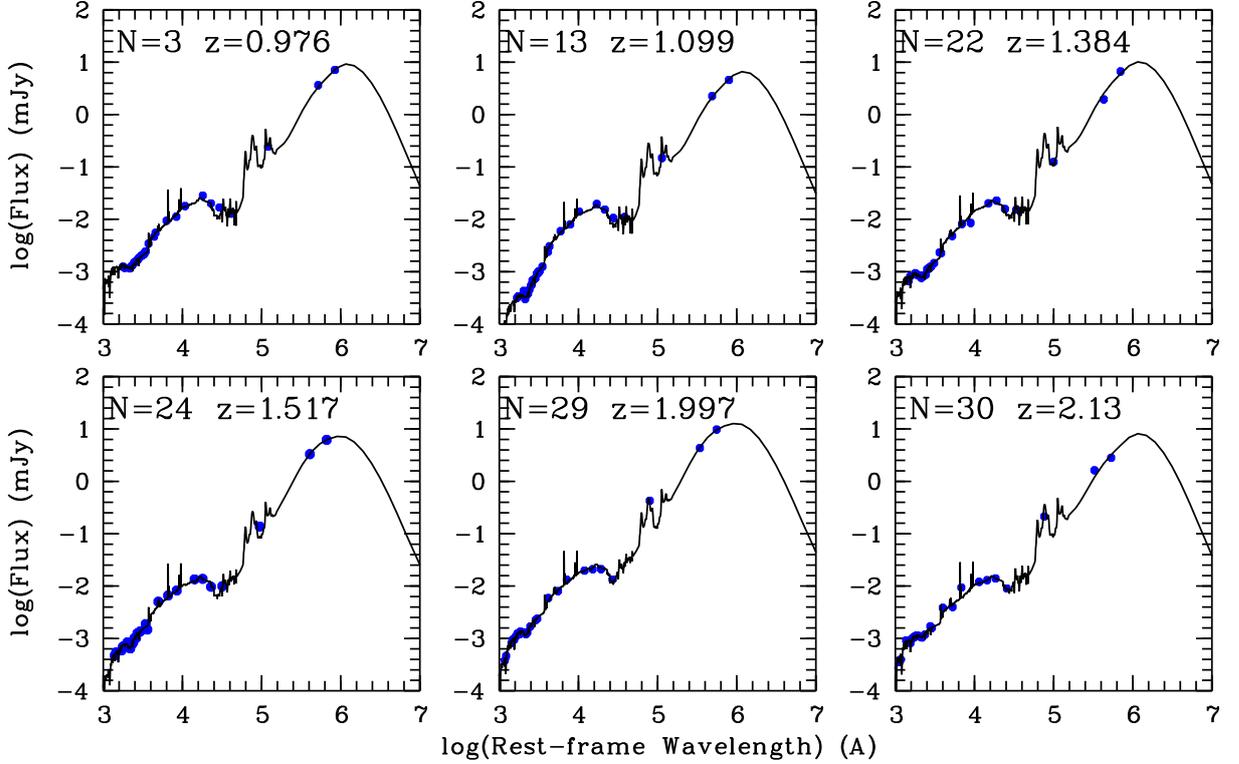}
\caption{Examples of  SEDs fitted with CIGALE. The best model ($\chi^2$ minimization) is plotted together with the observed fluxes.}
\label{fullSED}
\end{figure*}
  \begin{figure*}
 \centering
  \includegraphics[width=15cm,angle=-90]{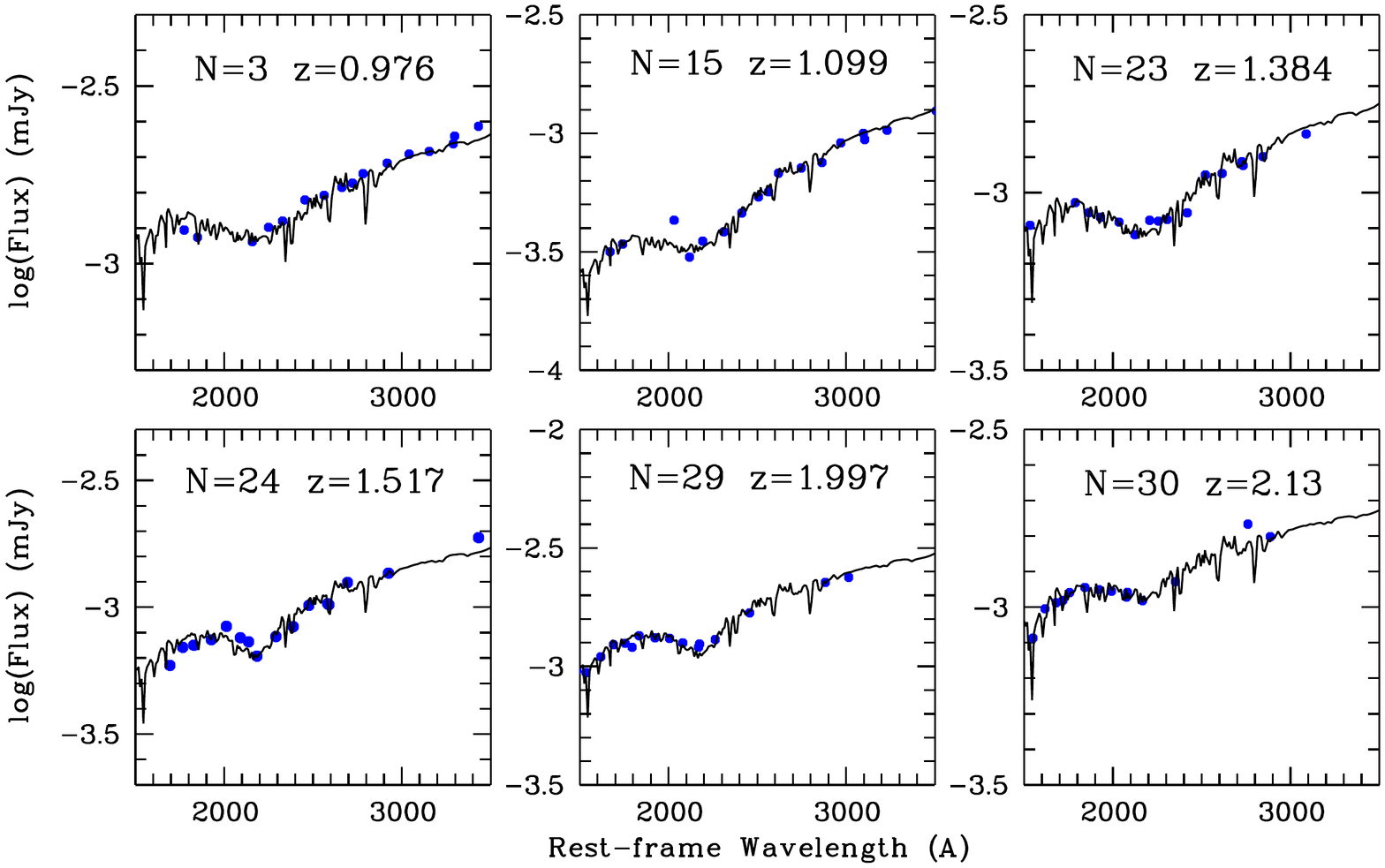}
   \caption{Spectral energy distributions for $\lambda<3500$  $\AA$  for the same objects as in Fig.\ref{fullSED}, the units, symbols and lines are the same as in Fig.\ref{fullSED} except for  the scale of the x-axis which is now linear.}
             \label{SED-bump}%
    \end{figure*}

       \section{SED fitting with CIGALE}
           SED fitting is  performed with the CIGALE code (Code Investigating GALaxy Emission) developped by  \citet{noll09b}.  This is a physically-motivated code which derives  properties of galaxies by fitting their UV-to-far-IR SEDs. CIGALE combines a UV-optical stellar SED and a dust, IR-emitting component and fully conserves the energetic balance between the dust absorbed stellar emission and its e-emission in IR. We refer to \citet{giovannoli10} for details on the use of CIGALE  to fit SEDs of  actively star forming galaxies observed both in UV and far-IR (rest-frame).  In the present work, the sampling of the UV range is very good so we expect to constrain the dust attenuation curve.  To model the attenuation by dust, the code  uses the attenuation law of \citet{calzetti00} as a baseline, and offers the possibility of varying the steepness of this law and  adding a bump centered at 2175 $\AA$. We refer  to \citet{noll09b} for a complete description of the dust attenuation prescription. The dust attenuation is described as:
    \begin{equation}
     A(\lambda) = {A_{\rm V} \over 4.05} ~(k'(\lambda)+D_{\lambda_0,\gamma,E_ {\rm b}}(\lambda) ) \left(\lambda \over {\lambda_{\rm V}} \right)^{\delta}
    \end{equation}
     where  $\lambda_V = 5500$  $\AA$, $k'(\lambda)$ comes from \citet{calzetti00} (Eq.4) and $D_{\lambda_0,\gamma,E_ {\rm b}}(\lambda)$,  the Lorentzian-like Drude profile  commonly used to describe the UV bump \citep{fitz90,noll09a} is defined as:
    
     \begin{equation}
     D_{\lambda_0,\gamma,E_{\rm b}} = {{E_{\rm b} \lambda^2 \gamma^2} \over { (\lambda^2-\lambda_0^2)^2+\lambda^2 \gamma^2}}
     \end{equation}
   The  peak amplitude above the continuum, full width half maximum (FWHM) and central wavelength of  the bump ($E_ {\rm b} $, $\gamma$ and $\lambda_0$) are free parameters of the code\footnote{$E_{\rm b}$ and $\gamma$ are related to the original parameters $c_3$ and $\gamma_{\rm FM}$ introduced by \citet{fitz90}: $E_b =  c_3/\gamma_{\rm FM}^2$ and $\gamma = \gamma_{\rm FM}\times \lambda_0^2$ }.
   There is no indication for variations of the central wavelength of the bump either in the Milky Way or in external galaxies \citep[][and references therein]{noll09a}, therefore we fix the central wavelength of the bump $\lambda_0$  to 2175 $\AA$. The parameter $\gamma$, indicating the FWHM, can vary between 200  and 500   $\AA$ in order to cover the range of values  found in previous studies: the FWHM found in the Milky Way is 437 $\AA$ \citep{fitz07} while  \citet{noll09a} found a FWHM of 274 $\AA$ for their composite spectrum of galaxies with $1.5<z<2.5$. 
After several trade-offs, the amplitude of the bump $E_{\rm b}$ is chosen between 0 and 2. For comparison the  corresponding value for the extinction curve of the Milky Way is 3.52 \citep{fitz07}, 1.63 for the  LMC2 supershell \citep{gordon03}, and between 0.5 and 1 for   $ z \sim 2$ galaxies \citep{noll09a}.\\
 Finally we choose $\delta$ between -0.3 and 0.1, this range of values is also chosen after several trade-offs, excluding values never chosen by the code. \\
 The stellar population created by CIGALE is the combination of an old, exponentially decreasing,  component (starting at the age of the universe at the redshift of the galaxy) and  a young burst of constant star formation  beginning  between $0.3 ~10^8$ and $3 10^8$ years ago. The e-folding rate of the old stellar population is found to be $2\pm 0.4$ Gyr. Since the UV emission reaches a stationary state after  $\sim 10^8$ years of constant star formation,   the general shape of the  UV SEDs analyzed in this work   is  predominantly modified  by the effect of dust attenuation  \citep{calzetti94}.  \\ The attenuation correction is applied to each component. To take into account the different distributions of stars of different ages, we adopt a visual attenuation $A_{\rm V}$ for the old stellar component equal to half that used for the young component. The exact value of this reduction factor (called $f_{att}$ in the code)  is very difficult to constrain even if a value lower than 1 is  always found \citep{noll09b,buat11}. In a first step we ran CIGALE with  $f_{att}$ as a free parameter with values comprised between 0 and 1, we obtained a mean value of 0.5 for $f_{att}$. To reduce the computation time we decided to freeze the value of $f_{att}$, after checking that it did not modify the values obtained for the other output parameters of the code.\\
 Dust luminosities, $\log(L_{\rm IR})$ between 8 and 1000 $\mu$m, are computed by fitting \citet{dale02} templates and are linked to the attenuated stellar population models (the stellar luminosity absorbed by the dust is re-emitted in IR). The validity of \citet{dale02} templates to measure total IR luminosities of sources detected by $Herschel$ is confirmed by the studies of \citet{elbaz10,elbaz11}.  Non-thermal sources,  not linked to the absorbed stellar population can also contribute to the dust emission.  CIGALE  includes dust-enshrouded AGNs by adding two PAH-free  templates from \citet{siebenmorgen04} as described in  \citet{buat11}. These two templates differ by the amount of silicate absorption and their spectral distribution beyond 20 microns is similar to the mean template constructed  by  \citet{mullaney11}. The spectrum   below  20 microns is representative of type 2 AGNs.\\
 The AGN contribution to $L_{\rm IR}$ is found to be lower than 10$\%$  except for N-17,  the only X-ray source of our sample  with broad emission lines, and N-25 which has  the highest X-ray luminosity ($\log(L_{\rm X}/L_{\odot}) = 44.4 $) and exhibits a power-law spectral energy distribution. For these two sources the AGN fraction is found to be  larger than 50$\%$. The current version of CIGALE is unable to fit accurately these SEDs leading to minimum $\chi^2$ values equal to 16 and 6.4 for N-17 and N-25 respectively. These two sources will be discarded when the output parameters of the code will be discussed. The case of the other X-ray sources is discussed in section 4.2.
Examples of full SED fitting are shown in Fig.~\ref{fullSED}. The figures for the full sample are available as online material. \\
All the  output parameters are estimated through the Bayesian analysis. Results  for the total IR luminosity, $L_{\rm IR}$, the dust attenuation at 1530 $\AA$, $A_{\rm UV}$, and the peak amplitude of the bump, $E_ {\rm b} $, are listed in Table 1 together with their errors, also evaluated by a Bayesian calculation \citep{giovannoli10}. The range of $\log(L_{\rm IR}/L_\odot)$ is 11.25-12.4 : all the  sources are luminous infrared galaxies ($\log(L_{\rm IR}/L_\odot) > 11$, LIRG)  or ultra luminous infrared galaxies ($\log(L_{\rm IR}/L_\odot) > 12$, ULIRG). The mean  dust attenuation is found to be  quite large with $<A_{\rm V} >= 0.92 \pm 0.34$ mag and  $<A_{\rm UV} >= 3.09 \pm 1.11$ mag. Such a large dust attenuation is expected for LIRGs and ULIRGs and is favourable to search for the imprint of UV bumps in the spectrum of these galaxies.

    \section {Evidence for a bump at 2175 $\AA$}

      \subsection{Amplitude of the bump from SED fitting with CIGALE}

\begin{figure}
   \centering
  \includegraphics[width=9cm]{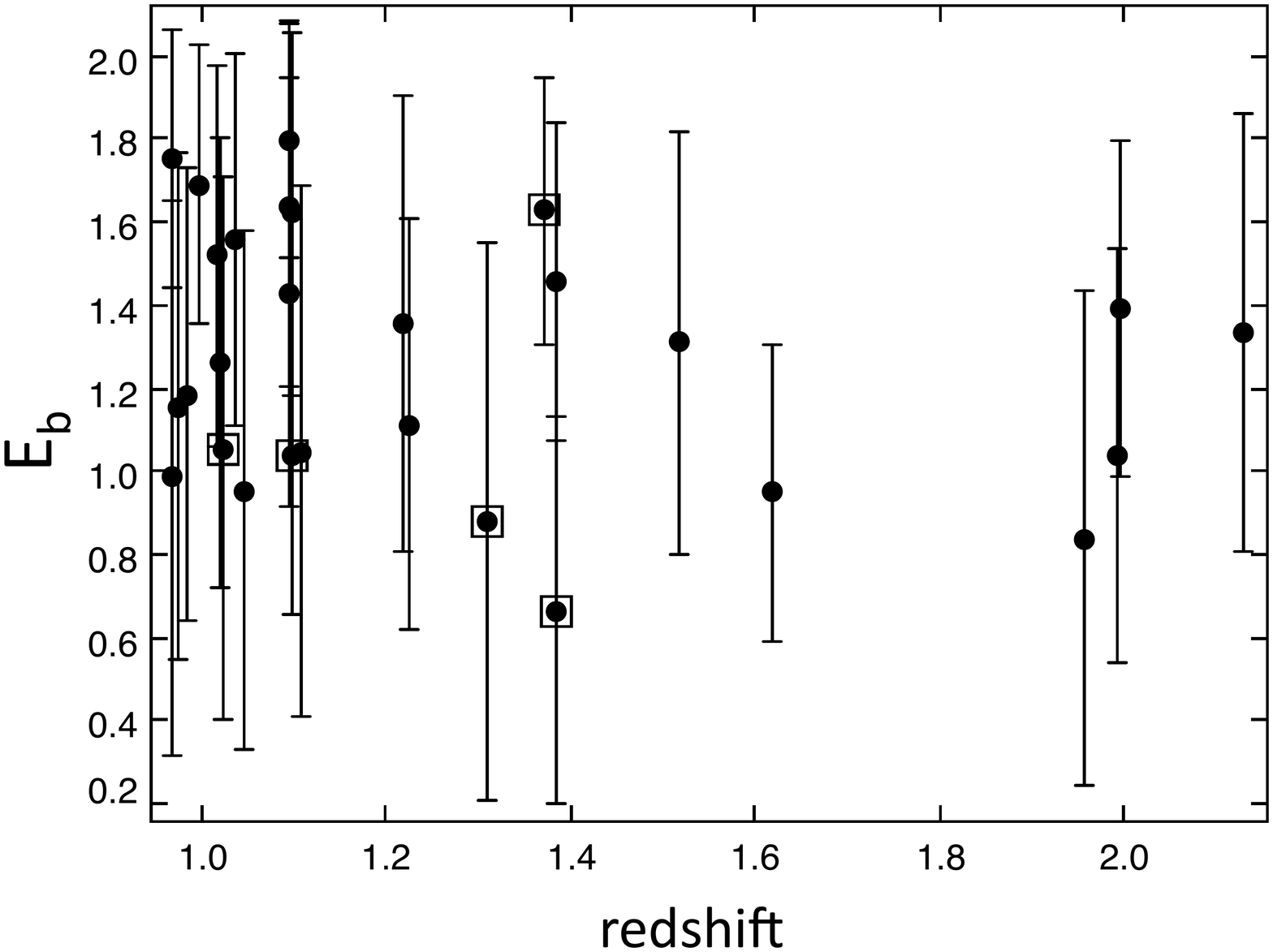}
   \caption{Amplitude estimated by our Bayesian analysis plotted as a function of redshift of the sources. Galaxies emitting in X-rays are plotted with empty squares. Standard deviations are reported as error bars}
              \label{bump}%
    \end{figure}

\begin{figure}
    \centering
      \includegraphics[width=9cm]{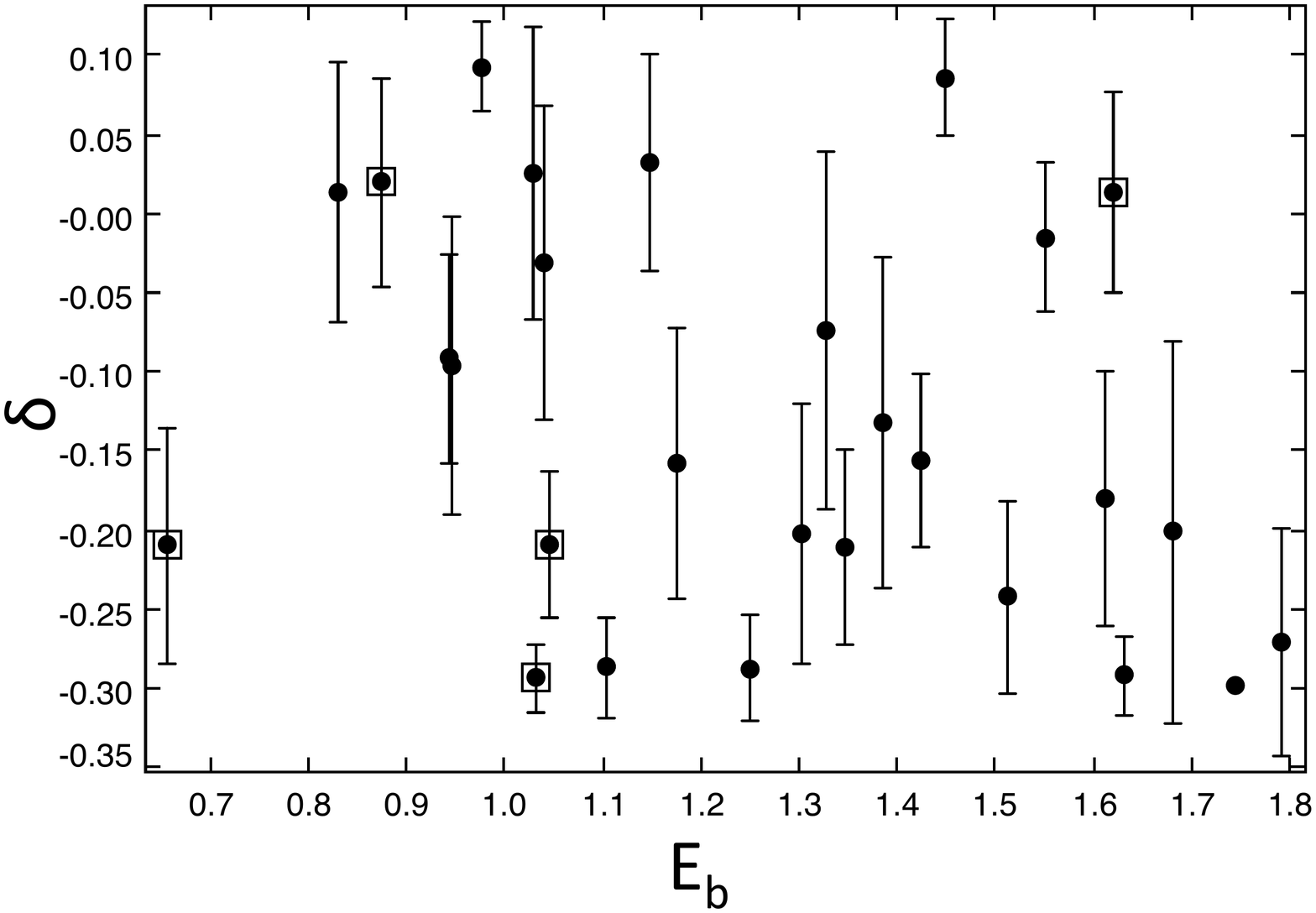}
   \caption{$\delta$ versus the amplitude of the bump $E_b$. X-ray emitters are encircled with squares }
              \label{delta}%
\end{figure}

 Some examples of SEDs in the UV range are shown in Fig.~\ref{SED-bump}. The figures for the whole sample are  available as online material. These few examples show the excellent sampling of the UV SED achieved thanks to the use of medium filters.  A dip at 2175 $\AA$ is  clearly seen in  most of the observed SEDs. The best models obtained by the  $\chi^2$ minimization  are found to reproduce the data very well. They  all correspond to a bump amplitude larger than 0. The median value of $E_b$ for the best models is 1.5.  In Fig.~\ref{bump} the Bayesian estimate of  $E_ {\rm b} $ is plotted against the redshift of the sources. All the sources are found to exhibit a bump. The uncertainty given by the Bayesian analysis is quite large but a null amplitude  is always excluded at the $1\sigma$ level. No trend is found  either with the redshift or dust luminosity. The mean value $<E_ {\rm b} >  = 1.26\pm 0.30$ corresponds to  $35 \%$ of the amplitude of the bump in   the average extinction curve of the MW and to $76\%$ of that  in  the LMC2 supershell. (Note  that we are dealing with attenuation curve and not extinction ones). \citet{noll09a} measured the amplitude of the UV bump in 68 spectra of galaxies with $1.5<z<2.5$ as part of the GMASS survey with a formalism similar to that used in the present work. They detected robustly a UV bump for 24 $\%$ of their sample with  $E_b = 0.9\pm 0.1$ while  $E_b = 0.5\pm 0.3$ for the remaining 76$\%$ of their sample.  Our result is marginally consistent with the mean amplitude  they measured for their  robust detections of a UV bump.  \\
   The  FWHM  of the bump is found to be 356 $\pm$ $20 \AA$, intermediate between the values  found in the MW (and the LMC) and  in galaxies at $z \sim 2$ \citep{noll09a}.\\
    The mean slope of the attenuation curve, $\delta$, is found to be -0.13$\pm$ 0.12, no trend is found for $\delta$ as a function of $E_b$ as shown in Fig.\ref{delta}. Our result is marginally consistent with the Calzetti et al. law which corresponds to $\delta=0$.

     \subsection{ X-ray emitters}
  
\begin{figure}
   \centering
  \includegraphics[width=9cm]{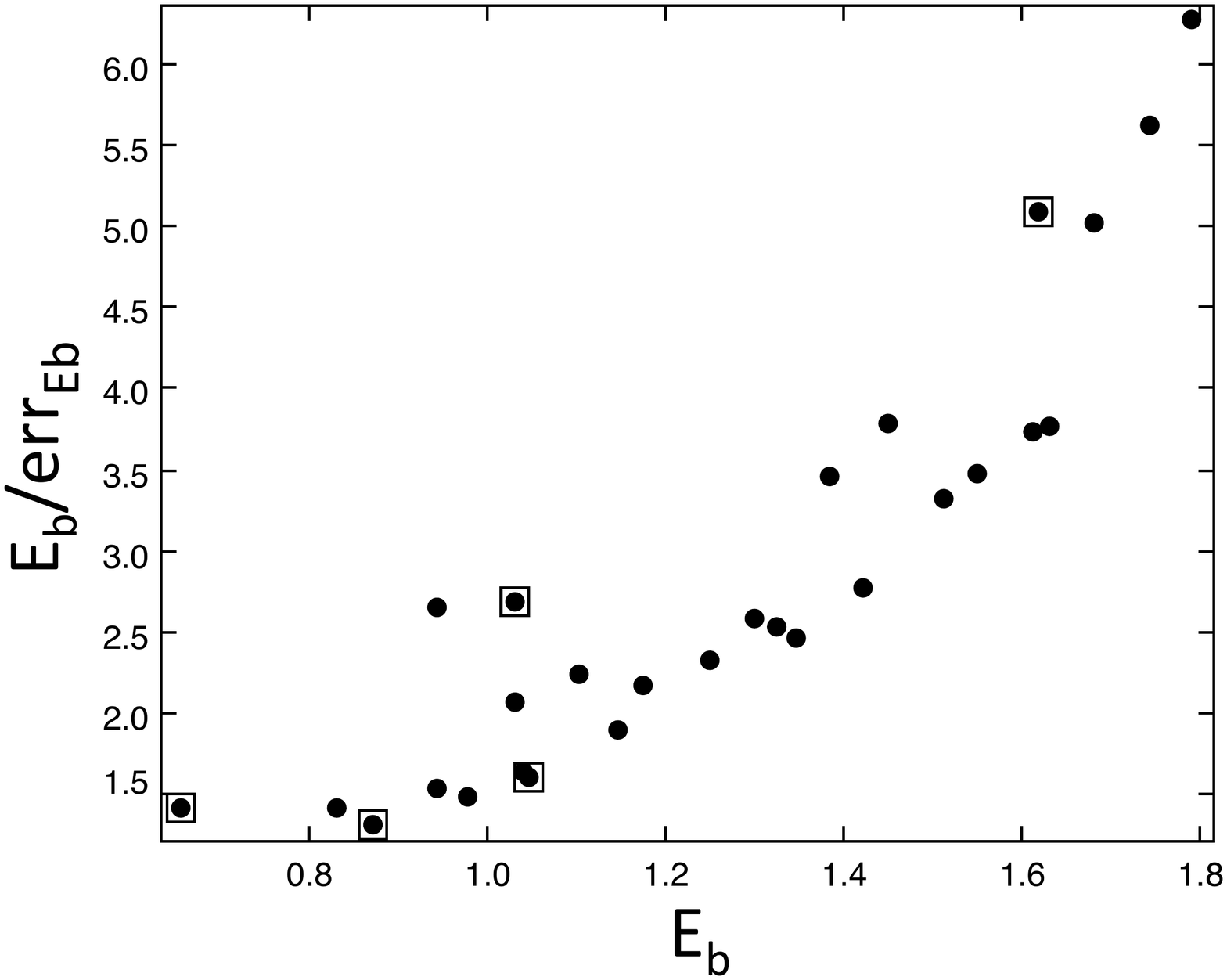}
   \caption{$<E_ {\rm b}>$, the mean value of $E_ {\rm b}$, divided by  its standard deviation $err_{E_{b}}$ versus  $<E_ {\rm b}>$. $E_b/err_{E_b}> 2$ corresponds to $E_{\rm b} > 0$ with a probability of 95$\%$. X-ray galaxies are encircled  with squares. }
              \label{Xray-Eb}%
    \end{figure}

  Extinction curves of AGN and quasars are found to  not display  the 2175 $\AA$ feature and the wavelength dependence of the extinction in these objects remains controversial between a gray or SMC-like extinction  \citep[][and references therein]{li07}. 
   N17 and N25, two of the seven galaxies in our initial sample classified as X-ray emitters by \citet{cardamone09}, are discarded because  of the bad fitting of their SEDs (see discussion in section 3).  The SED fitting process gives amplitudes of the bump for the remaining 5 sources which are different from 0 but in general lower than for non X-ray galaxies (Fig.\ref{bump}). We can quantify this difference: for a Gaussian probability distribution of the parameter $E_b$,  $E_b/err_{E_b}> 2$ corresponds to $E_b > 0$ with a probability of 95$\%$ ($err_{E_b}$ is the standard deviation of the distribution). Only 2 out of 5 X-ray galaxies fulfill this condition compared to  19 of the 23 non X-ray galaxies (Fig.\ref{Xray-Eb}). For the first one, N14 (MUSYCID-30046), $E_b/err_{E_b}=2.7$. This galaxy has a high X-ray luminosity, $\log(L_X/L_{\odot}) = 44.1$ with a hardness ratio equal to 0.63 \citep{cardamone08}. The fit is of average quality with $\chi^2_{min}=4.56$ and the presence of a bump is not obvious from  a  visual inspection of the SED. It is at least partly due to the fact that the best fit model corresponds to a large  width of the bump: 500 $\AA$. The second one is N21 (MUSYCID-55660). This galaxy  (G02:638)  is detected  in the hard X-rays  with a signal to noise ratio of 2 by \citet{giacconi02} only  using  Sextractor and not with a method based on wavelet transform;  \citet{alexander03} did not detect this source,  so the X-ray detection may be spurious.  Its SED shows a clear absorption feature at 2175 $\AA$ with $E_b/err_{E_b}=5.1$ and the SED fitting is good ($\chi^2_{min}=1.71)$. Given the small number of X-ray sources in our sample as well as the uncertainties inherent to the two sources for which a bump is measured above $2\sigma$,  we cannot draw any firm conclusion about the presence or not of a UV bump in the attenuation curve of X-ray galaxies. \\
    The slope of the attenuation curve, $\delta$, is found $\simeq 0$ for two X-ray sources and $\simeq -0.2$ for the 3 other ones. Therefore a  steep attenuation curve is favoured,  not as steep as the   SMC  extinction curve but excluding  a gray extinction which  corresponds to positive values of $\delta$.      \\

    \subsection{Average dust attenuation curve}

\begin{figure}
     \centering
  \includegraphics[width=9cm,angle=-90]{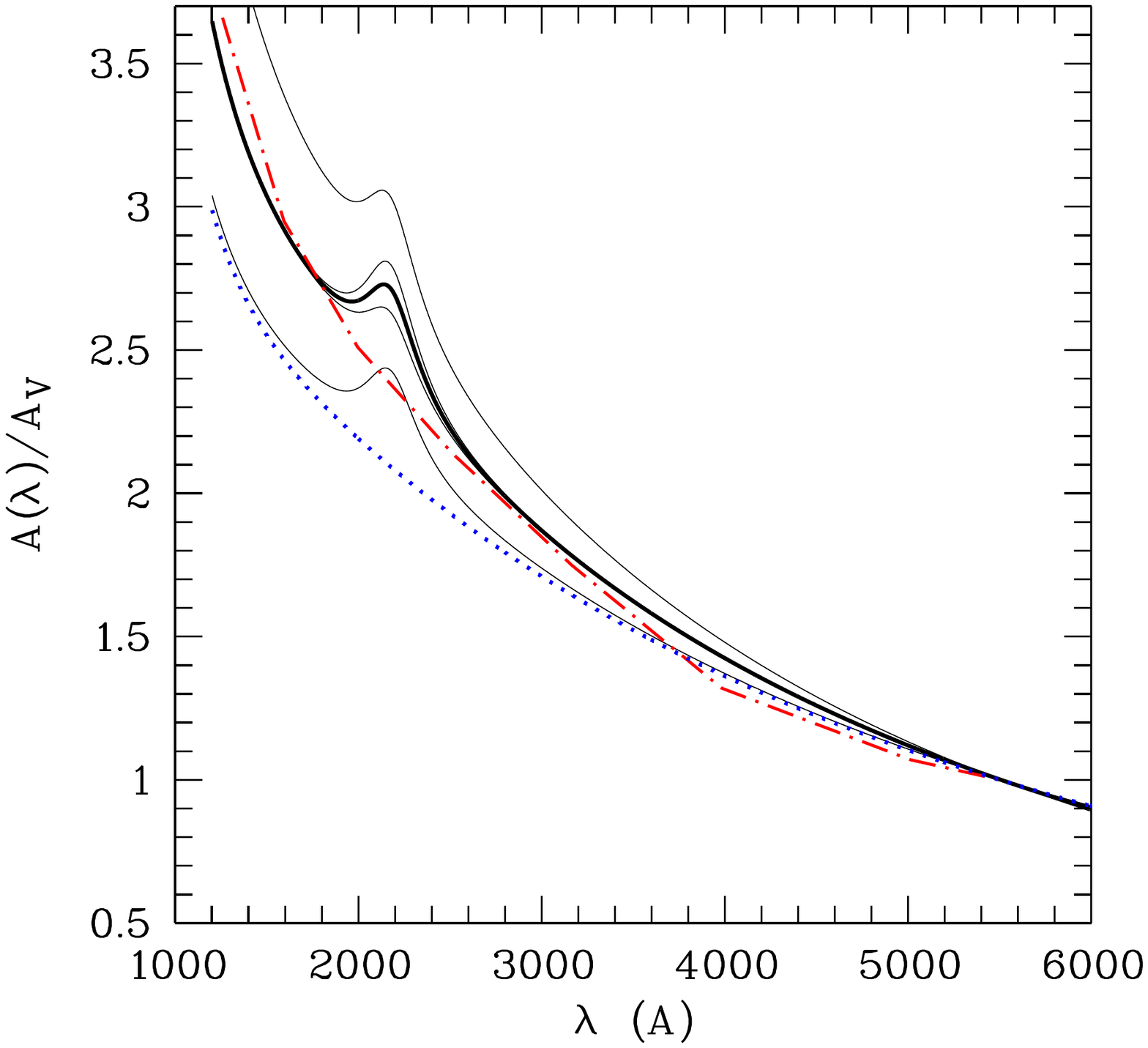}
   \caption{Average dust attenuation curve. The mean curve found in this work is plotted as a thick line.  The thin lines correspond to variations of $\pm 1 \sigma$ of the amplitude of the bump, $E_b$ and the slope of the continuum $\delta$; see text for details. The  \citet{calzetti00} attenuation curve (dotted blue line) and the effective absorption curve of \citet{charlot00} for a starburst age of $3 ~10^8$ years (dot-dashed red line) are shown on the figure for comparison.  }
              \label{att}%
    \end{figure}

       The resulting mean dust attenuation curve obtained for the whole sample is plotted in Fig.~\ref{att}.    All parameters are calculated by   a  Bayesian analysis.      The average dust attenuation curve we obtain can be written:
      \begin{equation}
    { A(\lambda)\over{A_{\rm V}}}= {k'(\lambda)+D_{\lambda_0,\gamma,E_ {\rm b}}(\lambda) \over{4.05}} \left(\lambda \over {\lambda_V} \right)^{-0.13}
    \end{equation}
     where  $\lambda_V = 5500$  $\AA$, $k'(\lambda)$ is given in  \citet{calzetti00} (Eq.4) and 
    
      \begin{equation}
     D_{\lambda_0,\gamma,E_{\rm b}}(\lambda) = {{1.26 \times 356^2 \lambda^2 } \over { (\lambda^2-2175^2)^2+\lambda^2 \times 356^2}}
     \end{equation}
     
     Strictly speaking this average attenuation curve  applies to the young stellar component only (as explained in section 3), the dust attenuation applied to the old component being reduced by a factor 2. In practice, the SEDs of our galaxies appear to be dominated by the young stellar component (modeled as a constant star formation rate over $\sim 10^8$ years), especially in UV. As a result the global attenuation curve is found to be quite similar to that of the young stellar component alone.\\
     
     The uncertainty on the average value of  $\delta$ implies  quite a  large dispersion of the attenuation curve at $\lambda < 2500~ \AA$ (Fig.~\ref{att}) and of the  determination of dust attenuation at UV wavelengths. For a visual attenuation $A_{\rm V} = 1$  mag the difference between the dust attenuation at 1500 $\AA$ obtained by  using  Eq. 3 ($\delta=-0.13$) or  a Calzetti et al.  law ($\delta=0$) reaches 0.5 mag. 
     A value of $\delta \not= 0$ also implies a change in the value of the effective total obscuration ${\rm R_V= A_{\rm V}/E_{B-V}}$. \citet{calzetti00} obtain $\rm {R_V=4.05\pm 0.80}$ by comparing the predicted and observed absorbed emission (i.e. by performing an energetic budget). Adopting a value of $\delta=-0.13$ leads to $\rm {R_V= A_V/E_{B-V}=3.7}$, still consistent with the value found by \citet{calzetti00}. \\
The introduction  of a bump of moderate amplitude in the dust attenuation curve (Eq. 3)  leads to an increase of  $A(\lambda)/A_{\rm V}$  at $\simeq 2000~ \AA$ of $\simeq 0.35$ mag as compared to the same curve with  $D_{\lambda_0,\gamma,E_{\rm b}}(\lambda) = 0$.

\begin{figure}
     \centering
  \includegraphics[width=9cm,angle=-90]{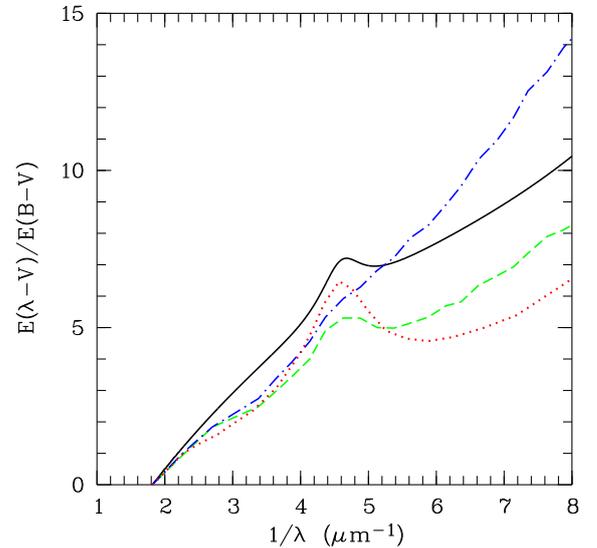}
   \caption{$E(\lambda-V)/E(B-V)$.   The mean curve found in this work is plotted as a thick line  together with the extinction curve of the MW (red dotted line) from \citet{whittet03}, LMC2 supershell (green short dashed line) and SMC (blue dot-dashed line) from \citet{gordon03}.}
              \label{att1}%
    \end{figure}

   We can compare our average dust attenuation curve to other attenuation curves in the literature.   \citet{charlot00} propose a recipe consisting in attenuating the  stellar emission by a factor proportional to $\lambda^{-0.7}$, and in reducing the normalization  of the attenuation for stars older than $10^7$ years; the resulting effective absorption curve they obtain for a constant star formation rate over $3~10^8$ years (their Fig. 5) is reported in Fig.~\ref{att}. A period of $3~10^8$ years  is  close to  the mean duration of the burst of star formation we obtain for the young stellar population ($10^8$ years, cf. section 3). Apart from the presence of the bump, not considered by \citet{charlot00}, our mean attenuation curve and that of \citet{charlot00} exhibit similar shapes and steepening for decreasing wavelengths.\\
    Although we are dealing with attenuation and not extinction, we also   compare our resulting curve with the extinction curves found in the Milky Way and the Magellanic Clouds. The relative extinctions  are compared in Fig. \ref{att1}. The general shape of the extinction curve of the LMC2 supershell is consistent with our attenuation curve, whereas the MW and SMC extinction curve show   flatter and steeper variation, respectively, at short wavelengths.\\
   The exact shape of the extinction curve (slope in the UV range and amplitude of the bump) cannot be constrained with the present data since the attenuation curve we derive strongly depends  on the distribution of both stars and dust.  \citet{witt00} show that the shape of the attenuation function also depends on the total amount of dust attenuation. We do not observe  any variation of $E_{\rm b}$ and $\delta$ with $A_{\rm UV}$: it can be explained by  the homogeneity of our sample in terms of dust attenuation associated to the large uncertainties on the determination of these parameters. 
    Radiative transfer models in opaque disks with various configurations of stars and dust (especially in a clumpy medium) are able   to reduce the amplitude of the  UV bump in the effective attenuation  with a  dust   composition similar to that of the MW or the average LMC, the  attenuation curve becoming grayer  when   the opacity  increases \citep[e.g.][]{witt00,pierini04,inoue06,panuzzo07}. These models also predict a flattening of the attenuation curve in UV (as an effect of the gray attenuation) as compared to the original extinction curve. The average  attenuation curve we deduce increases  steeply at short wavelengths: it seems to  favour the scenario of  a deficiency of  bump carriers with respect to the case of MW or  average LMC.   We will  see in section 5.2 that  our sources roughly follow the $A_{UV}$-$\beta$ relation found by \citet{Meurer99} for local starbursts  which is well explained by a  clumpy shell geometry (\citet{witt00,calzetti01}). Such a dust/star configuration favours the interpretation of a moderate UV bump in the extinction curve itself in agreement with  our conclusion \citep[e.g.][]{witt00,noll09a}.
       

    \section{Slope of the UV continuum from broad band photometry  and  dust attenuation }
       \begin{figure}
   \centering
  \includegraphics[width=9cm]{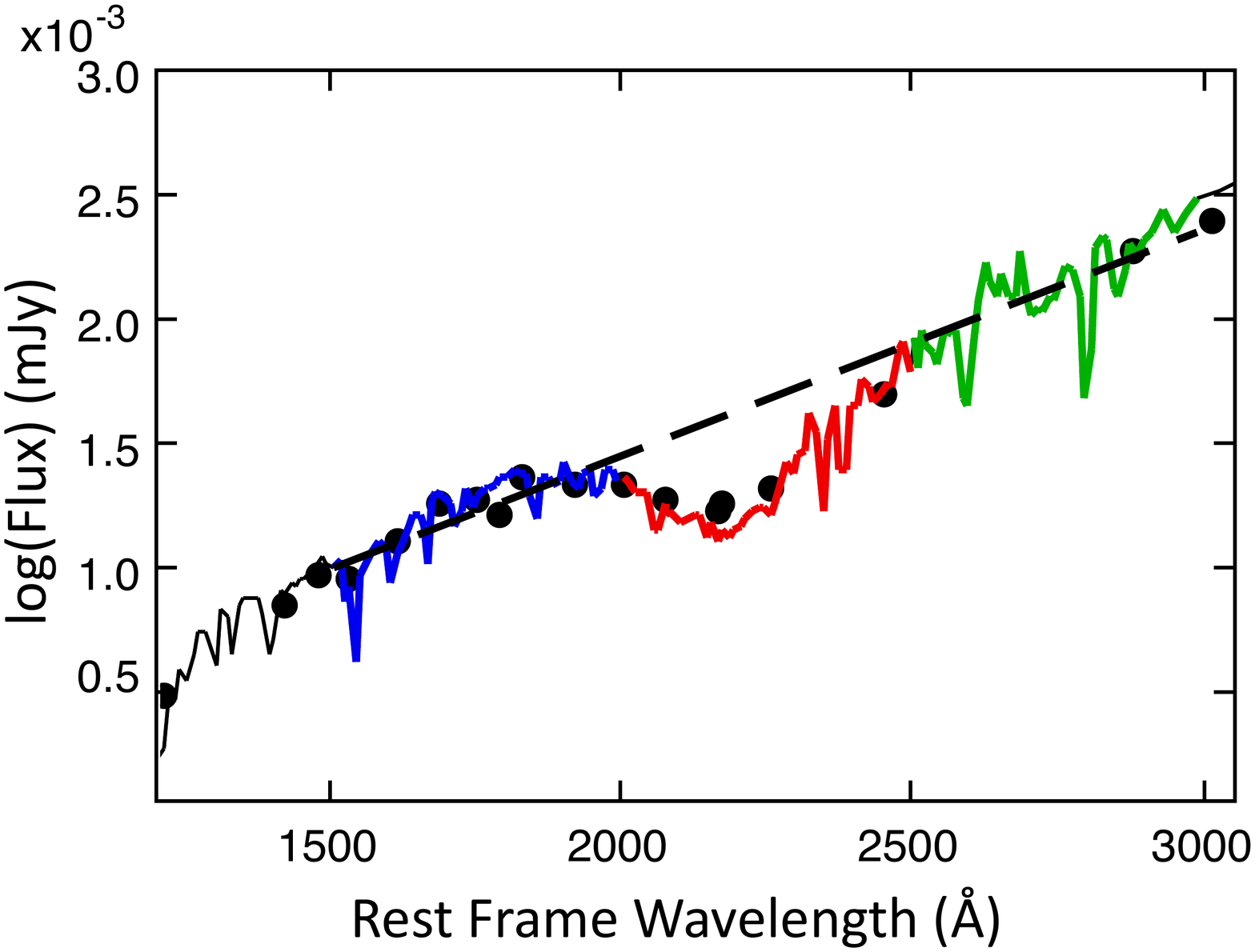}
   \caption{Definition of 3 bands for ID 2733. The best model found for this galaxy is plotted and the range of wavelengths corresponding to B1, B2 and B3 is in blue, green and red respectively. The black dots are the observed fluxes.}
              \label{1096}%
    \end{figure}

       \begin{figure}
   \centering
 \includegraphics[width=9cm]{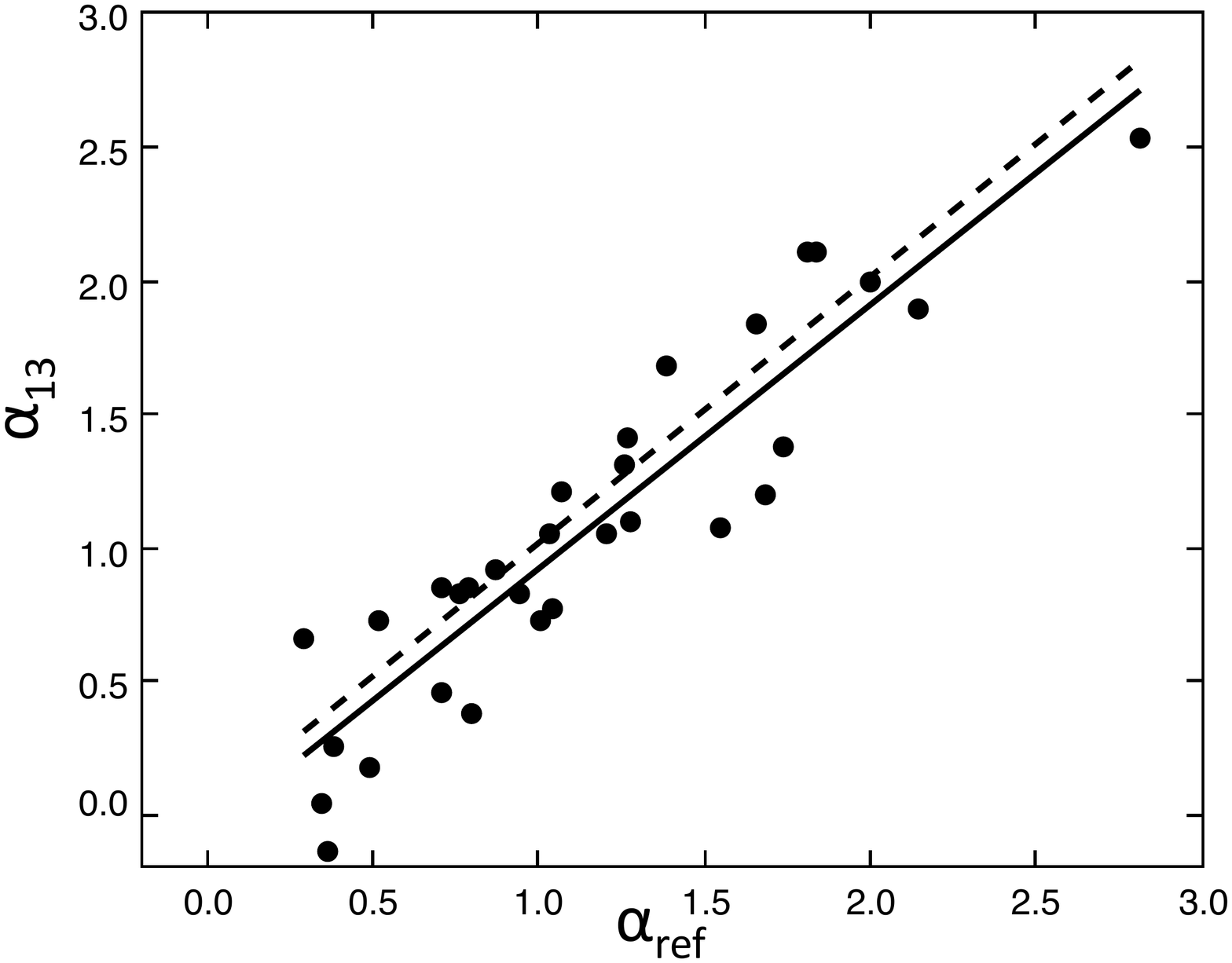}
  \includegraphics[width=9cm]{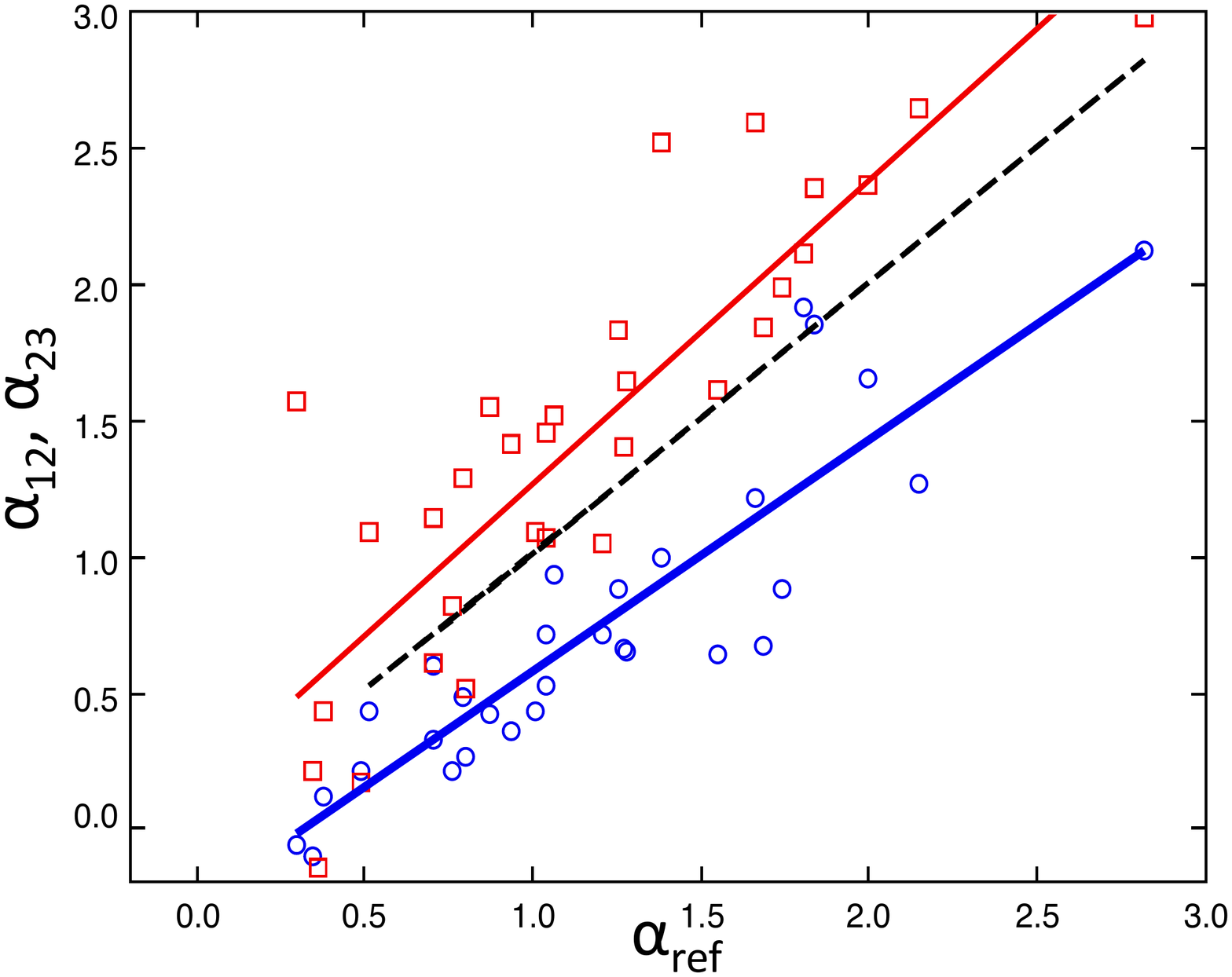}
  \includegraphics[width=9cm]{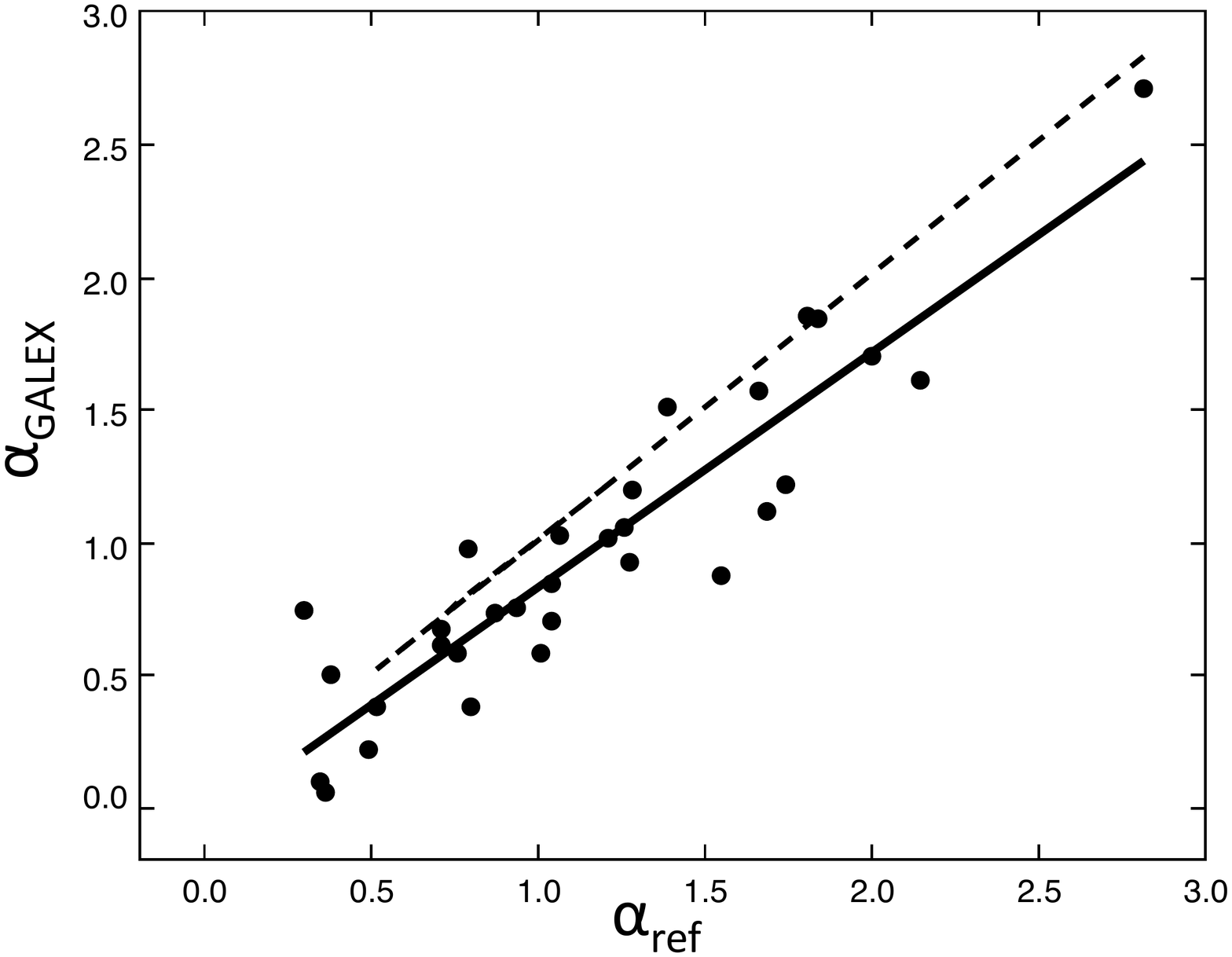}
   \caption{Various estimates of $\alpha$ compared to the reference value $\alpha_{\rm ref}$ measured on the full SED ,  top:  $\alpha_{13}$ versus $\alpha_{\rm ref}$; middle:  $\alpha_{12}$  (blue empty circles) and $\alpha_{23}$ (red empty squares) versus $\alpha_{\rm ref}$;  bottom: $\alpha_{\rm GALEX}$  calculated with  the rest-frame FUV and NUV fluxes versus $\alpha_{\rm ref}$. On each figure, the dashed lines represent the 1:1 line, the solid lines are  the results of linear regressions.}
              \label{alpha}%
    \end{figure}
    \subsection{ Determination of the UV slope}
    \subsubsection{Influence of the bump}
With the imprint of the UV bump clearly visible in our sample, the assumption of a simple power-law for the entire UV continuum is no longer valid, and its influence in the measured slopes can be quantified. A single power law might be  valid if we exclude the spectral region of the bump (Fig.\ref{SED-bump}).  Since our fluxes are defined in Jansky we define $\alpha$ as  $f_{\nu} \propto \lambda^{\alpha}$ with $\alpha=\beta+2$ where $\beta$ corresponds to the original definition  $ f_\lambda ({\rm erg ~cm^{-2} s^{-1}  \AA^{-1}}) \propto \lambda^\beta$. From Fig.~\ref{SED-bump} it is clear that  we can extend the model  of a power-law  up to 3000 $\AA$. Therefore  we calculate the slope of the UV continuum in several ways. First  we use  all the data between 1200 and 3000 $\AA$, avoiding the bump range  (1975-2375 $\AA$) and perform a linear regression. The resulting value of $\alpha$ will be taken as our reference $\alpha_{\rm ref}$. Then  we define  3 wavelength ranges 1500-2000 $\AA$ (B1) , 1500-2500 $\AA$  (B2) and 2500-3000 $\AA$ (B3) which mimic broad band filters and combine two of these bands to estimate $\alpha$. The definition of these 3 bands is illustrated in Fig.\ref{1096} for a galaxy exhibiting a strong bump.  B2  overlaps the bump, therefore we expect some effects when the slope is calculated by considering it. 
    The slopes of the UV continuum are calculated following the prescription of  \citet{kong04} adapted to our notation:
    \begin{equation}
    \alpha = {{\log(f_{\nu_i}/f_{\nu_j})}\over{\log(\lambda_i/\lambda_j)}}
    \end{equation}
     where $f_{\nu_i}$ and $f_{\nu_j}$ are the mean flux densities   per unit frequency through the filters i and j  whose effective wavelengths are  $\lambda_i$ and $\lambda_j$.
    In Fig.\ref{alpha}  the $\alpha$ values obtained by combining  B1 and B2 ($\alpha_{12}$), B2 and B3 ($\alpha_{23}$) and B1 and B3 ($\alpha_{13}$) are compared to  $\alpha_{\rm ref}$ measured on the entire UV SED. As expected $\alpha_{13}$, calculated avoiding the bump is found to correlate very well with $\alpha_{\rm ref}$ (Fig~\ref{alpha} top panel, $R=0.92$). The combination of B1-B2 leads to bluer slopes whereas the slopes calculated with B2 and B3 are redder than $\alpha_{\rm ref}$. These effects are expected since the presence of  the bump increases the dust  attenuation in the B2 band. The amplitude of the shifts are $<\alpha_{12}-\alpha_{\rm ref}> = -0.46 \pm 0.26$ and $<\alpha_{23}-\alpha_{\rm ref}> = 0.28 \pm 0.40$ whereas $<\alpha_{13}-\alpha_{\rm ref}> = -0.09 \pm 0.25$.
        \subsubsection{ GALEX-like UV slope}
    The FUV and NUV  bands of  GALEX are commonly used to measure the slope of the UV continuum in the nearby universe \citep[e.g.][]{seibert05}. Since the NUV band (1771-2831 $\AA$) \citep{morissey05} overlaps the UV bump we might expect some underestimation  of the slopes.  $\alpha_{\rm GALEX}$ is an output of CIGALE: the spectral energy distribution of each model generated by the code is integrated in the FUV and NUV (rest-frame) filter bands and $\alpha_{\rm GALEX}$ calculated with the above formula. Then the expected value of $\alpha_{\rm GALEX}$  for each galaxy of our sample is estimated through a Bayesian calculation \citep{giovannoli10}. The correlation between $\alpha_{\rm GALEX}$ and $\alpha_{\rm ref}$ is very good ($R=0.91$). As expected $\alpha_{\rm GALEX}$ is typically lower (bluer) than $\alpha_{\rm ref}$ but the systematic difference remains small  with $<\alpha_{\rm GALEX}-\alpha_{\rm ref}> = -0.19 \pm 0.24$.  This moderate shift, smaller than the one  found considering $\alpha_{12}$,  is probably due to the wide large bandpass of the NUV filter  (1060 $\AA$,   \citep{morissey05}), much wider than the bump itself. Therefore, as long as galaxies exhibit bumps in their attenuation curve similar to those found in this study (1/3 that of the Milky Way), the slopes measured with the GALEX data are found reliable. \citet{burgarella05} showed that Milky Way-like bumps have a large influence on the determination of the UV slope with GALEX filters.
    
    \subsubsection{Impact of photometric errors}
    It is worth noting that the determination of the UV slope through two  broad band filters is very sensitive to photometric errors as well as to the interval between the effective wavelengths of both filters i.e.  the factor $\log(\lambda_i/\lambda_j)$ in Eq.(3). We can re-write Eq.(3) in terms of magnitudes instead of fluxes:
     \begin{equation}
    \alpha =\beta+ 2 = -0.4{{(m_i-m_j)}\over{\log(\lambda_i/\lambda_j)}}
    \end{equation}
   Using the FUV and NUV filters of GALEX the relation becomes $\alpha = -2.22~ (m_i-m_j)$ and a typical error of 0.1 mag in FUV and NUV translates to an error of 0.3 on $\alpha$. It is obvious from Eq.6 that increasing the wavelength difference between the two filters reduces the uncertainty on the determination of $\alpha$ and $\beta$. For the domain of validity of the power law adopted in this work ($\lambda < 3000 \AA$, cf. section 5.1.1), using filters with the largest  possible interval between  rest-frame wavelength, i.e. 1300 $\AA$ (to avoid the flattening of the attenuation curve for $\lambda \le 1200 \AA$ \citep{leitherer02,buat02} and 3000 $\AA$ reduces the error by  a factor $\sim$ 2 (this time $\alpha = -1.11 (m_i-m_j)$) as compared to the use of  GALEX filters.\\
To summarize, given the uncertainty on the presence and strength of a UV bump in a galaxy, one must avoid as much as possible the region affected by the bump when measuring the UV slope using broad band filters and  use filters with effective wavelengths that are  as  widely separated as possible from each other. If  the region of the bump cannot be avoided,  the  attenuation curve presented in Eq.3 might be used to interpret the data.

    \subsection{Dust attenuation corrections}
      \begin{figure}
  \includegraphics[width=9.5cm]{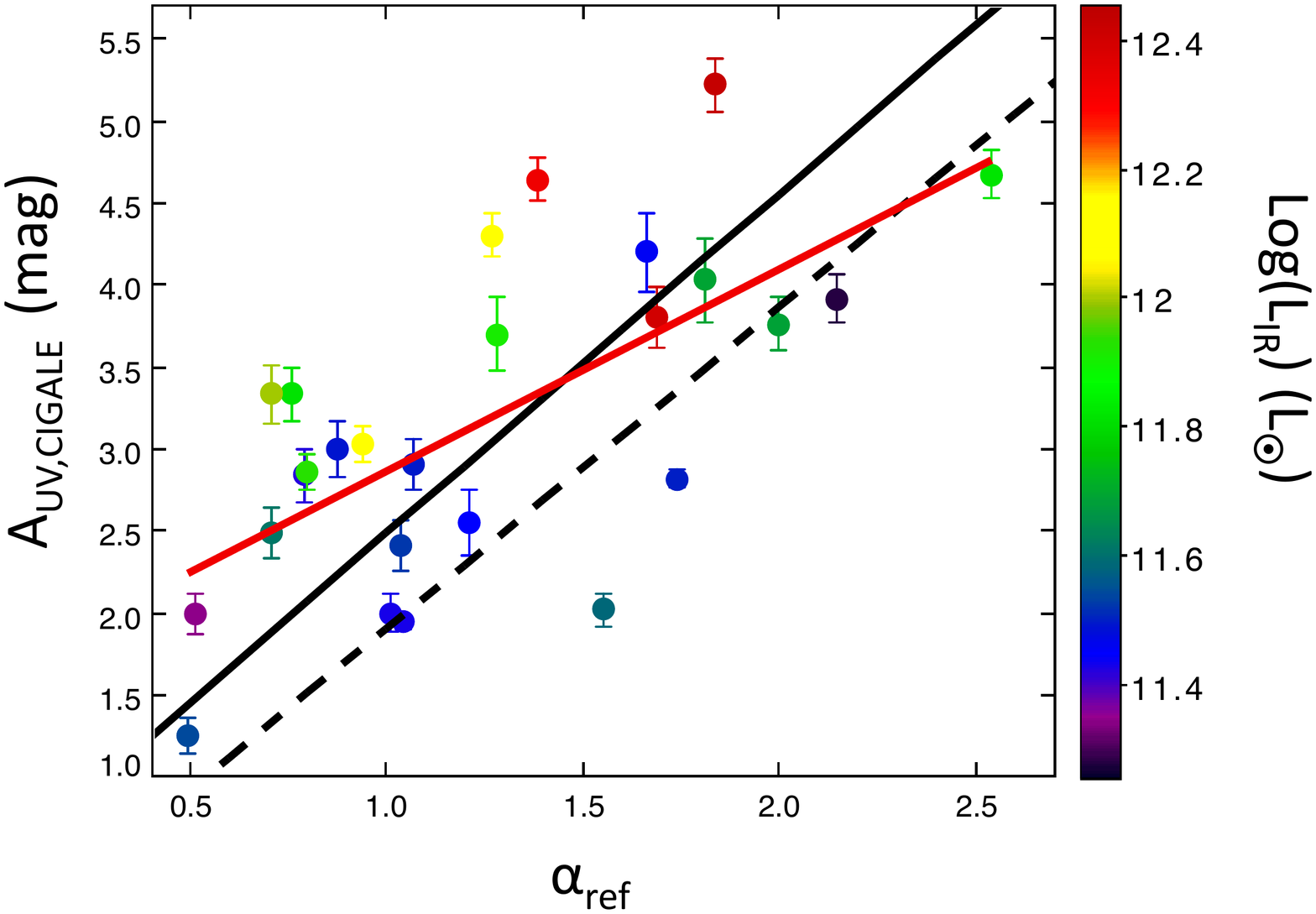}
  \caption{UV dust attenuations from CIGALE plotted against the slope of the UV continuum calculated with the full SED. The total IR luminosity  is colour-coded. The solid and dashed lines correspond to the inner and total relations of \citet{overzier10} respectively. The error bars come from CIGALE (Bayesian estimates of the parameters and their associated errors). The red line  is the result of the linear regression }
                \label{auv-alpha}%
    \end{figure}
   
 \citet{Meurer99} first  deduced a relation between $A_{\rm UV}$ and  $\beta$ for local starbursts, avoiding the bump area on the spectra obtained with IUE. Several calibrations have been proposed since this pioneering work \citep{calzetti00,seibert05,overzier10}. One issue of the preliminary analyses was that the UV data came from IUE whose aperture sampled only the inner part of the galaxies. \citet{overzier10} re-investigate the topic by using GALEX images defining an 'inner' and 'total' relation, the first one based on UV fluxes measured into an IUE-like aperture , the second one corresponding to the total UV fluxes. These relations are:
 $$A_{\rm UV} = 4.54+2.07 \beta= 4.54+2.07 (\alpha-2)$$ for the inner relation and $$A_{\rm UV} = 3.85+1.96 \beta = 3.85+1.96 (\alpha-2)$$ for the total relation.
 In Fig.\ref{auv-alpha} these relations are compared to the values of $\alpha_{\rm ref}$ and  $A_{\rm UV,CIGALE}$ (the dust attenuation obtained from the SED fitting process and a Bayesian calculation). Whereas LIRGs seem to  follow more or less  the local starburst relations,  the ULIRGs are found above the local relations which is in  agreement with previous studies at low and high  redshift  \citep{takeuchi10, howell10,burgarella07, reddy08}. \citet{buat10}  performed a  selection at 250 $\mu$m at $z<0.3$ and found that their sample (including LIRGs) lie below the local starburst relation. This different behavior is likely to be due to a  selection effect biased towards more  quiescent galaxies  when selected at $250 \mu$m than in  the mid-IR or around 60 $\mu$m and to the use of the GALEX colour  to measure the UV slope at $z>0$.  \\
 A linear regression on the whole sample gives: 
 \begin{equation}
 A_{\rm UV, CIGALE} = 1.46 ~(\pm 0.21) ~\alpha_{\rm ref}+1.50~ (\pm 0.29)
 \end{equation}
with a correlation coefficient $R=0.79$. The correlation is rather weak resulting in a large dispersion with a RMS scatter of 0.73 mag. This error directly affects any estimate of $L_{\rm IR}$ based on  relations between $L_{\rm IR}/L_{\rm UV}$, a quantity tightly related to $A_{\rm UV}$ \citep[][and references therein]{buat11}, and the slope of the UV continuum.

     \begin{figure}
   \centering
  \includegraphics[width=9cm]{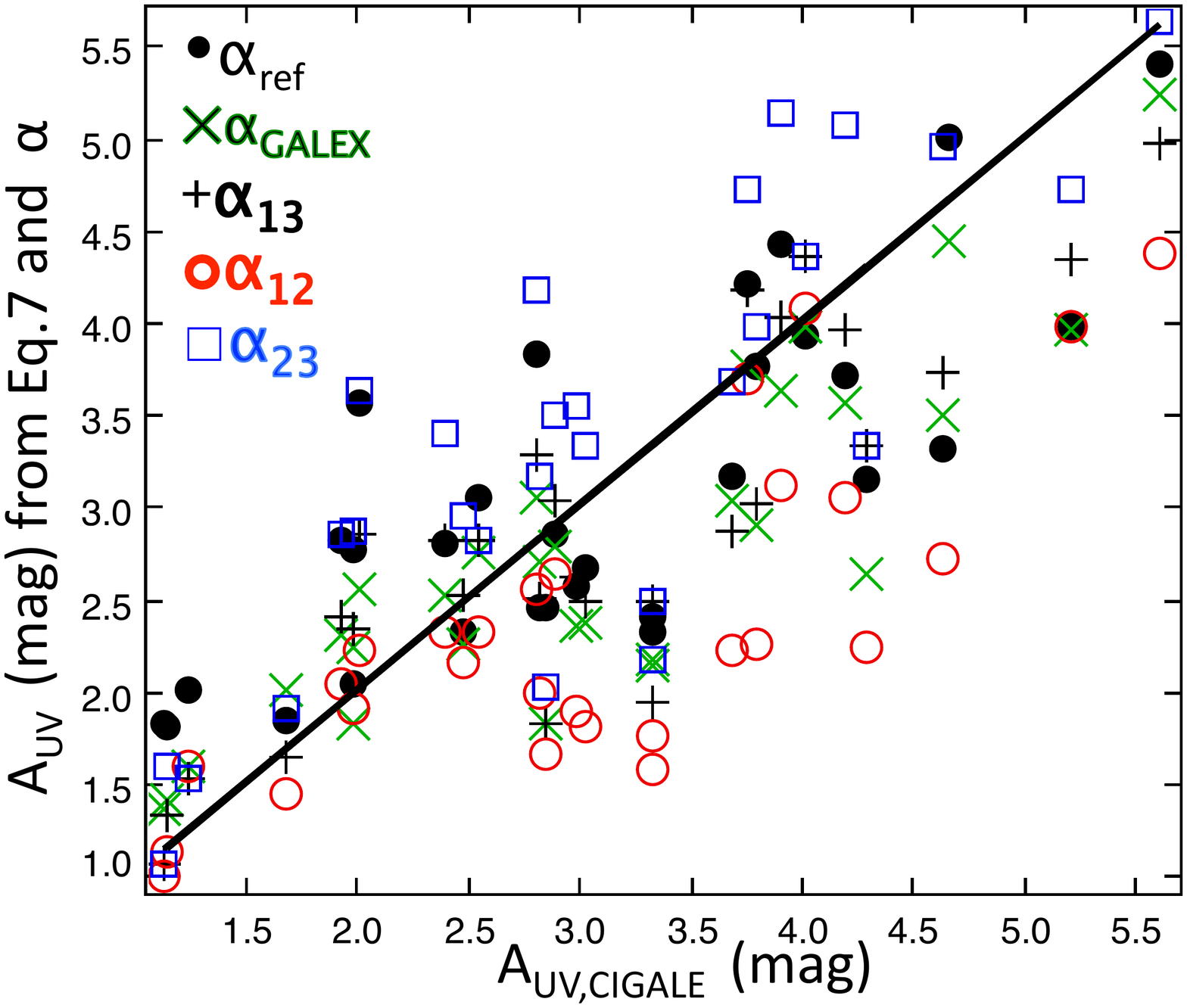}
       \caption{ Comparison of  UV dust attenuations: the values of $A_{\rm UV}$  directly obtained from the SED fitting with  CIGALE are plotted  on the x-axis and the estimates of $A_{\rm UV}$ from different UV slopes and Eq. 7 on the y-axis .}

              \label{auv-alphas}%
    \end{figure}

 We can also compare dust attenuation corrections  obtained by applying Eq. 7 to    the various estimates of $\alpha$  defined in section 5.1.1 and 5.1.2.  The results are reported in Fig.~\ref{auv-alphas} and compared to the dust attenuation  given by CIGALE: $A_{\rm UV, CIGALE}$.
 A large dispersion is found, as a consequence of the loose relation found initially between $A_{\rm UV}$ and $\alpha_{\rm ref}$ (Fig~\ref{auv-alpha}):    $<A_{\rm UV, \alpha_{\rm ref}}- A_{\rm UV,\rm CIGALE} > = -0.005 \pm 0.707$ mag. When using $\alpha_{13}$  the systematic shift remains moderate $<A_{UV, \alpha_{13}}- A_{UV,\rm CIGALE} > = -0.18\pm 0.58$ mag.  $\alpha_{\rm GALEX}$, $\alpha_{12}$ and $\alpha_{23}$ give worse estimates with substancial shifts and large dispersions:   $<A_{\rm UV, \alpha_{\rm GALEX}}- A_{UV,\rm CIGALE} > = -0.32\pm 0.58$ mag, $<A_{\rm UV, \alpha_{12}}- A_{\rm UV, CIGALE} > = -0.73\pm 0.70$ mag, $<A_{\rm UV, \alpha_{23}}- A_{\rm UV,\rm CIGALE} > = 0.37\pm 0.71$ mag.\\
 Therefore, using UV slopes derived from broad band colours  leads to large uncertainties in the derivation of dust attenuation ($\sim 1 $ mag) even if the bump area is avoided and good photometry is available. This is due to the intrinsic dispersion found between the slope of the UV continuum and the amount of dust attenuation. The presence of a bump in one of the filters may add  a systematic shift which can reach 0.7 mag for the galaxies of our sample which is likely to be representative of star forming galaxies at redshifts between 1 and 2.  These results hold for galaxies with a moderate bump in their dust attenuation curve ($35\%$ of the mean  amplitude found in the MW) and we expect larger effects for larger bumps. 
    
  \section{Conclusions}
 We have identified the presence of a UV bump in  a sample of  galaxies of the CDFS at a redshifts between $0.9 < z < 2.2$, selected to be observed in intermediate and broad band optical filters as well as at mid-and far-IR wavelengths  by {\it Spitzer} and $Herschel$. This study demonstrates the capabilities of intermediate-band photometry to give details on SED of galaxies, and to derive physical properties. \\ 
The average dust attenuation curve we deduce is well described by a modified Calzetti et al. law slightly steeper than the original one and with a UV bump at 2175 $\AA$ whose amplitude is $\sim$ 35$\%$ of the MW one. We propose an  analytical expression of the average attenuation curve which can  be used to correct the SEDs of  galaxies for  dust attenuation.The moderate amplitude of the bump together with the substantial slope of our average attenuation curve in the UV argue for a deficit of UV bump carriers in our sample galaxies as compared to   the MW.\\
Our sample contains seven X-ray galaxies. Five of them are reasonably fitted: a steep attenuation curve is favoured excluding a gray extinction. The sample is too small to draw any  firm conclusion about the presence of a UV bump. In any case the amplitude of the bump, if any, is smaller than that found for non X-ray galaxies.\\
The presence of a bump in the mean  dust attenuation curve has implications for the derivation of  the slope of the UV (rest-frame) continuum of galaxies from broad band data alone.  The power law modeling of the UV spectral distribution of our galaxies can be extended up to 3000 $\AA$ but the bump area must be avoided. When a broad band filter overlaps the bump, despite its moderate amplitude, the error on the determination of the slope may reach $\sim$0.5.  The use of GALEX  filters leads to an underestimation  of the slope  of only  0.2, even if the NUV filter overlaps the bump. This is likely to be due to the very large bandpass of this filter. However, the proximity of the two central wavelengths of the GALEX filters implies a large uncertainty in the determination of the slope.  It is stressed that a larger wavelength baseline would give a  better measure. \\
Dust attenuation estimated with  UV slopes, even when the latter are reliable,  remains uncertain, with an RMS error of 0.7 mag. \\
\onlfig{1}{
 \begin{figure}
  \centering
\includegraphics[width=5.8cm,angle=-90]{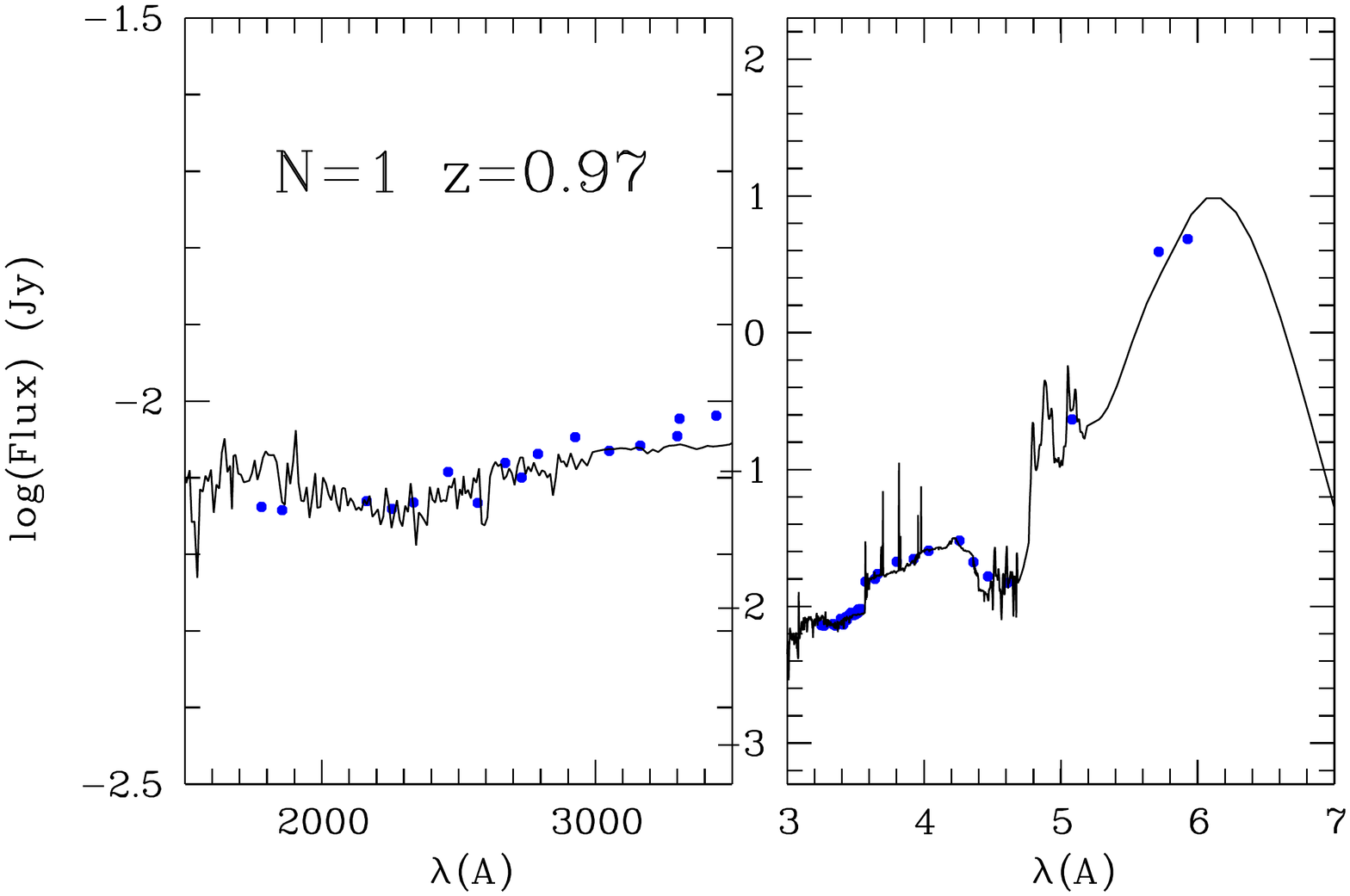}
\includegraphics[width=8cm]{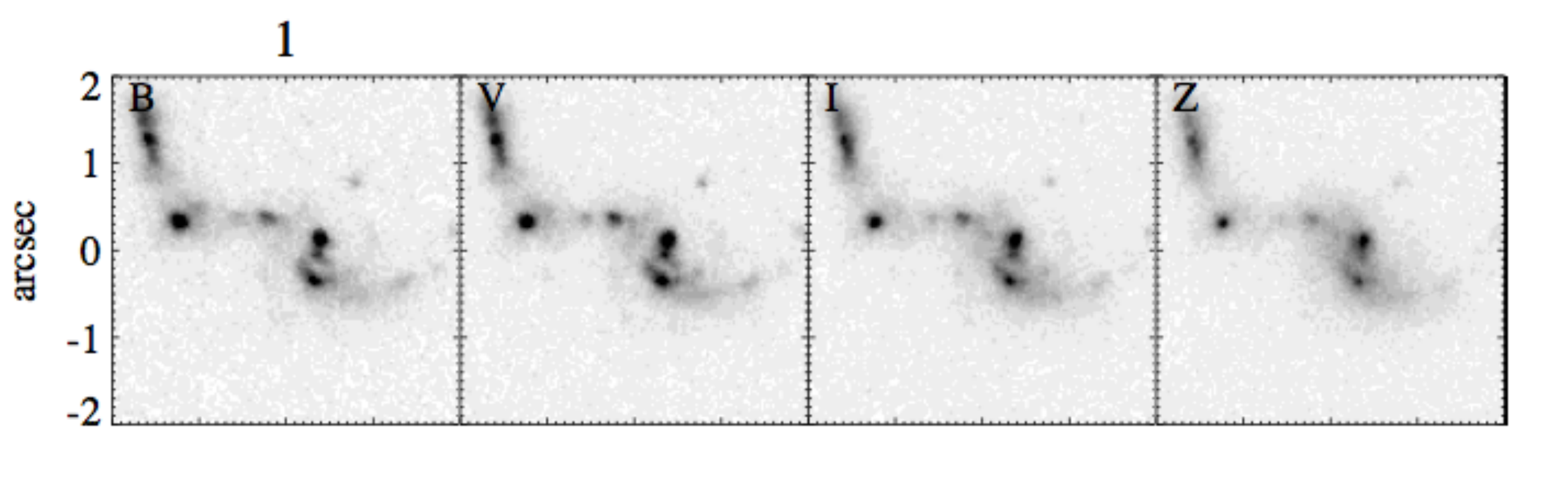}
\includegraphics[width=5.8cm,angle=-90]{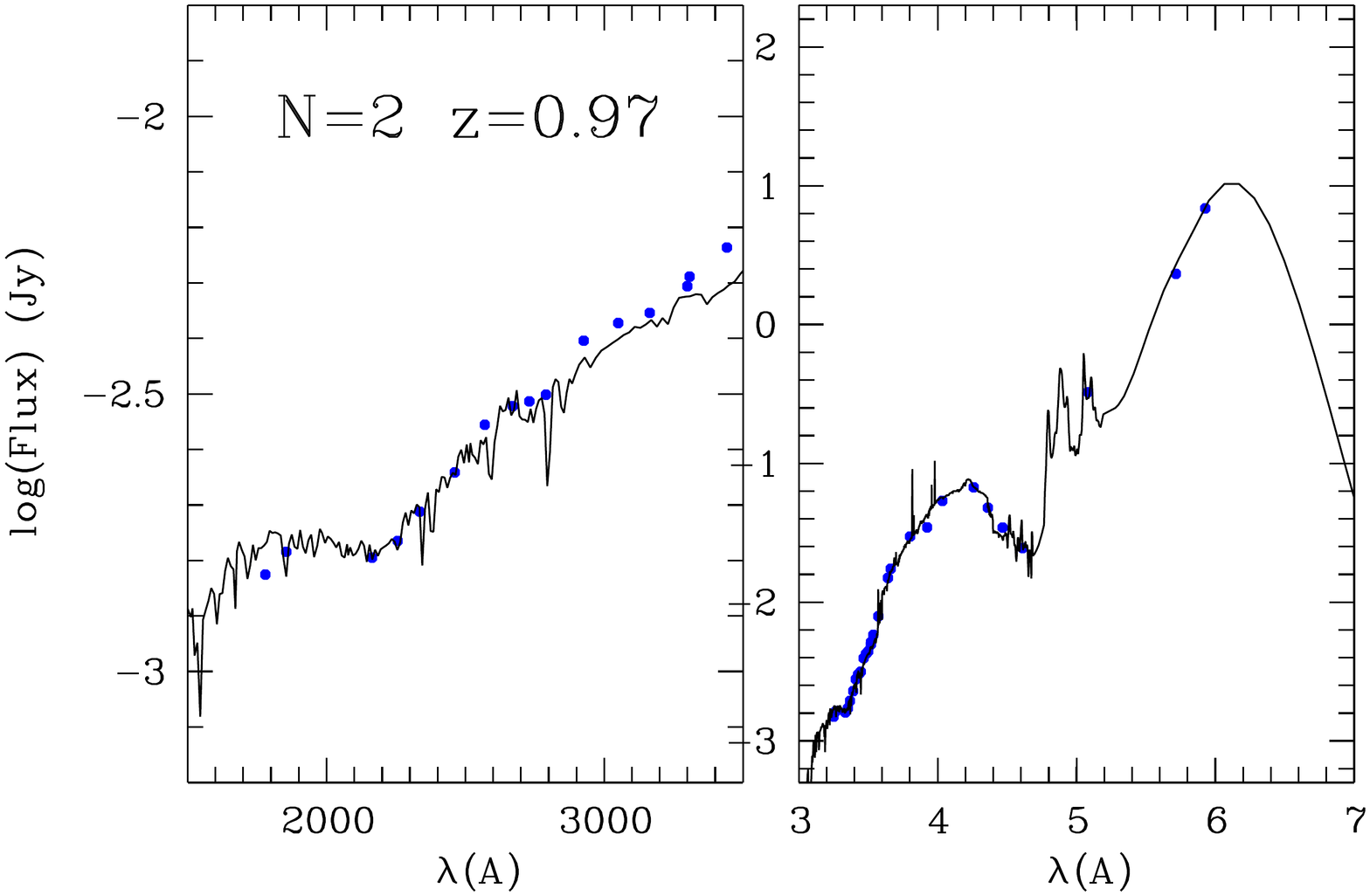}
\includegraphics[width=8cm]{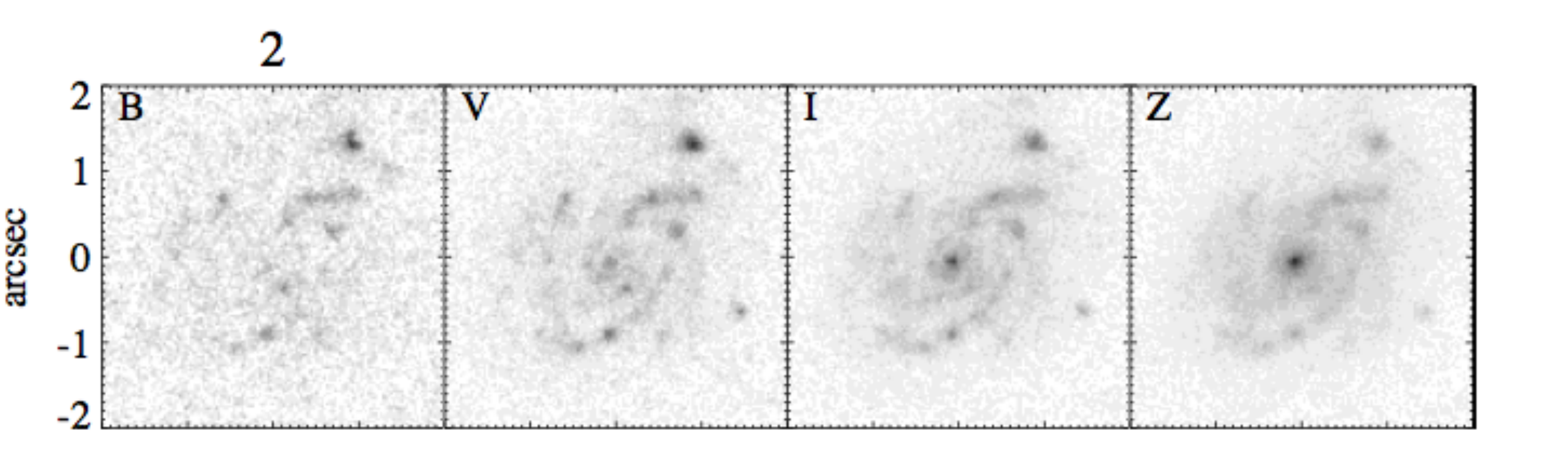}
\includegraphics[width=5.8cm,angle=-90]{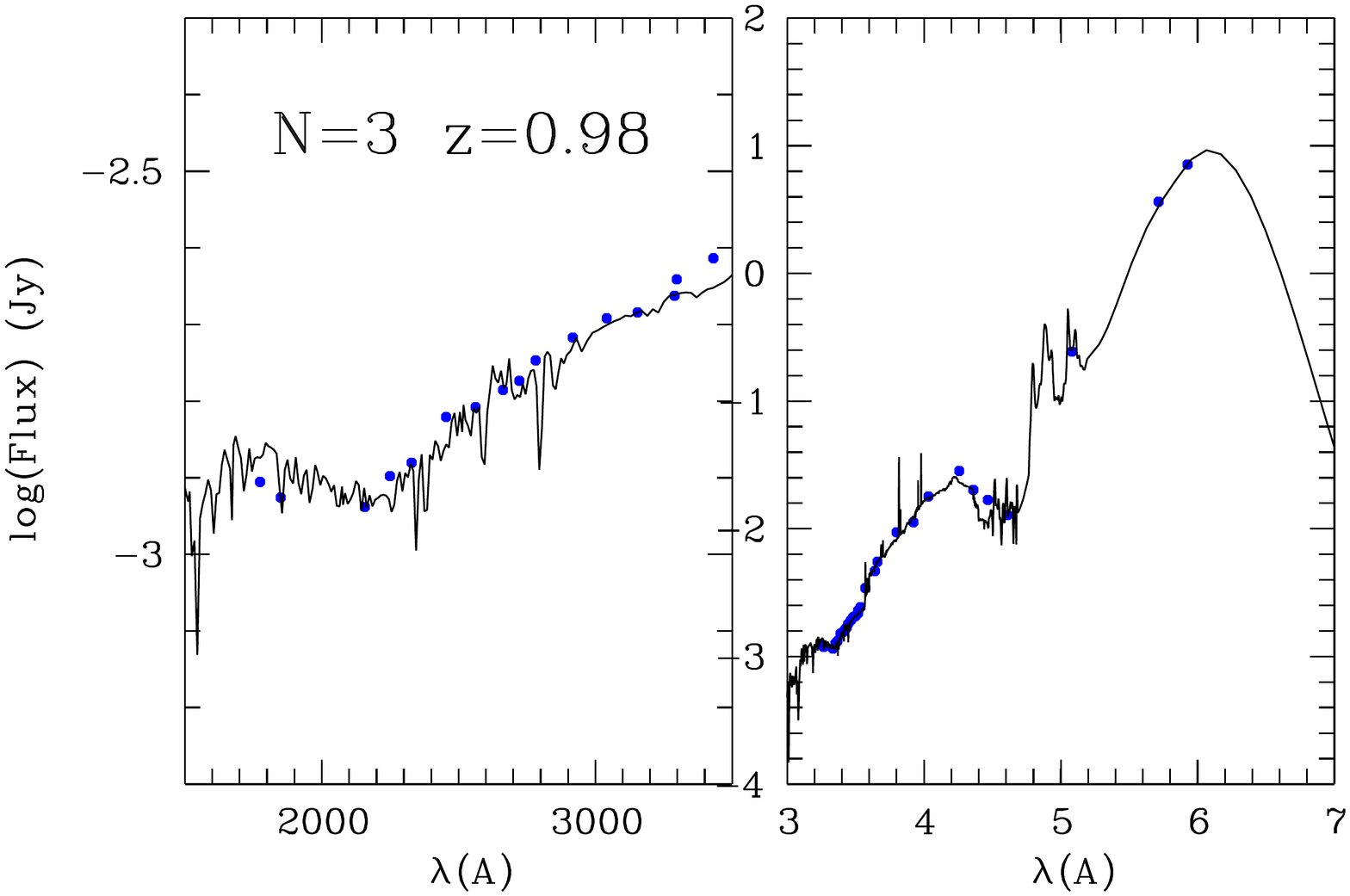}
\includegraphics[width=8cm]{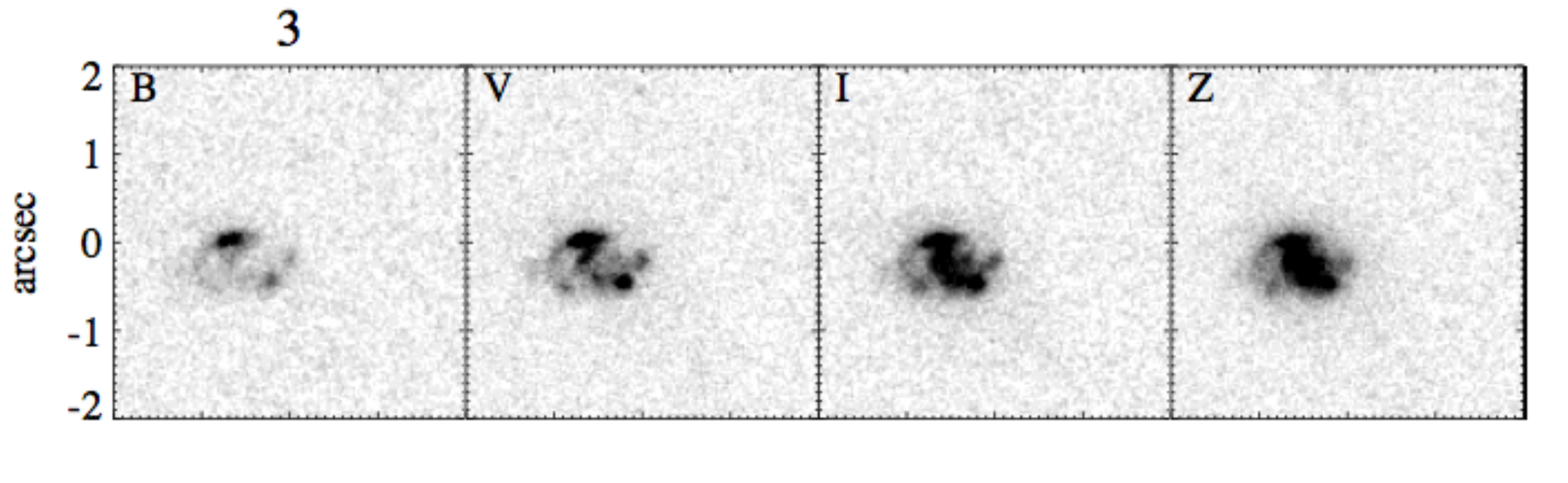}

\caption{Spectral energy distributions for the full sample and HST images obtained from the GOODS cutout service V0.2. The best fit is represented with a solid line. The rest-frame wavelength is plotted on the x-axis and the fluxes in Jy on the y-axis.  }
\end{figure}
    \begin{figure}
    \ContinuedFloat
  \centering
  \includegraphics[width=5.8cm,angle=-90]{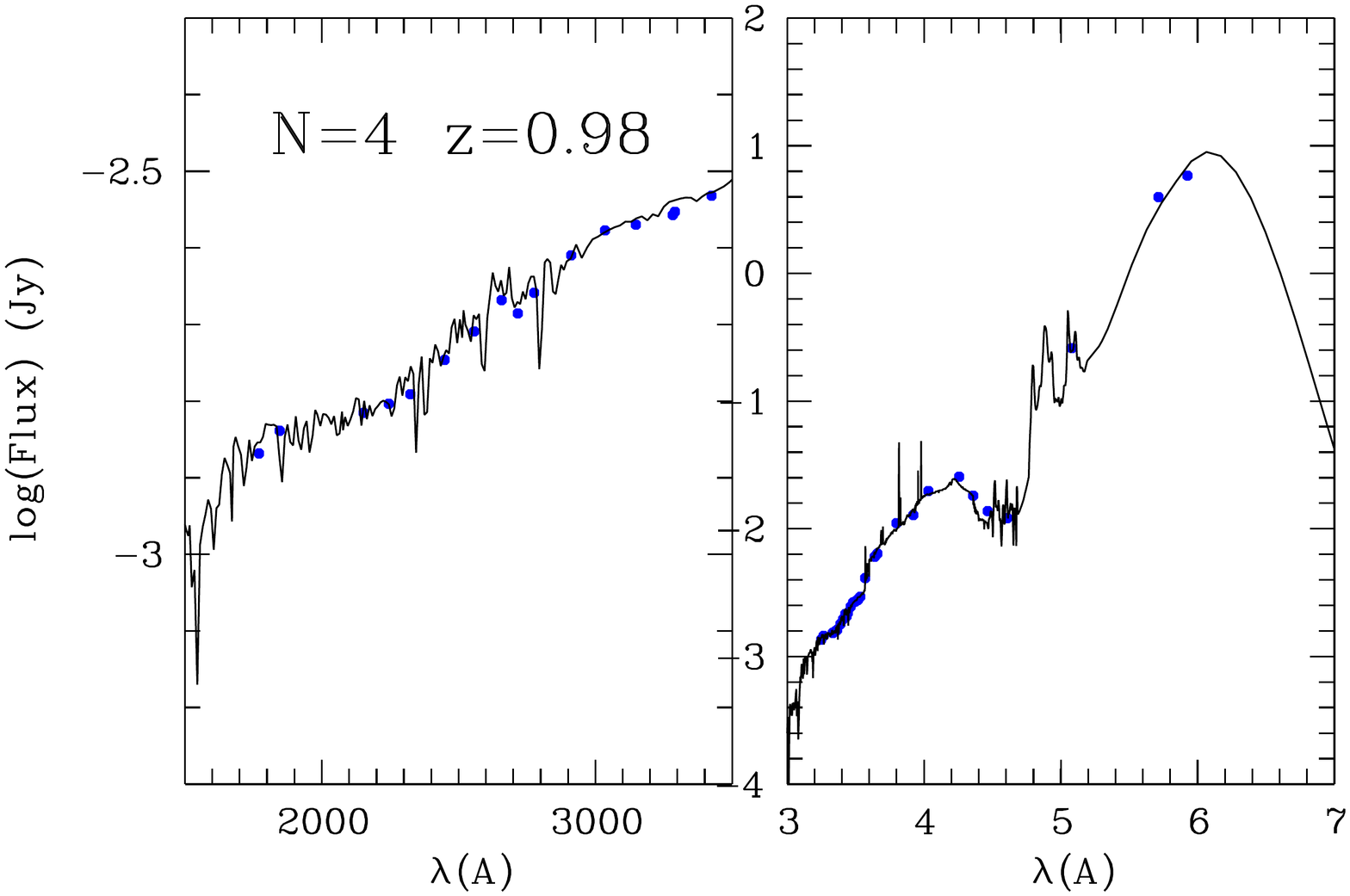}
 \includegraphics[width=8cm]{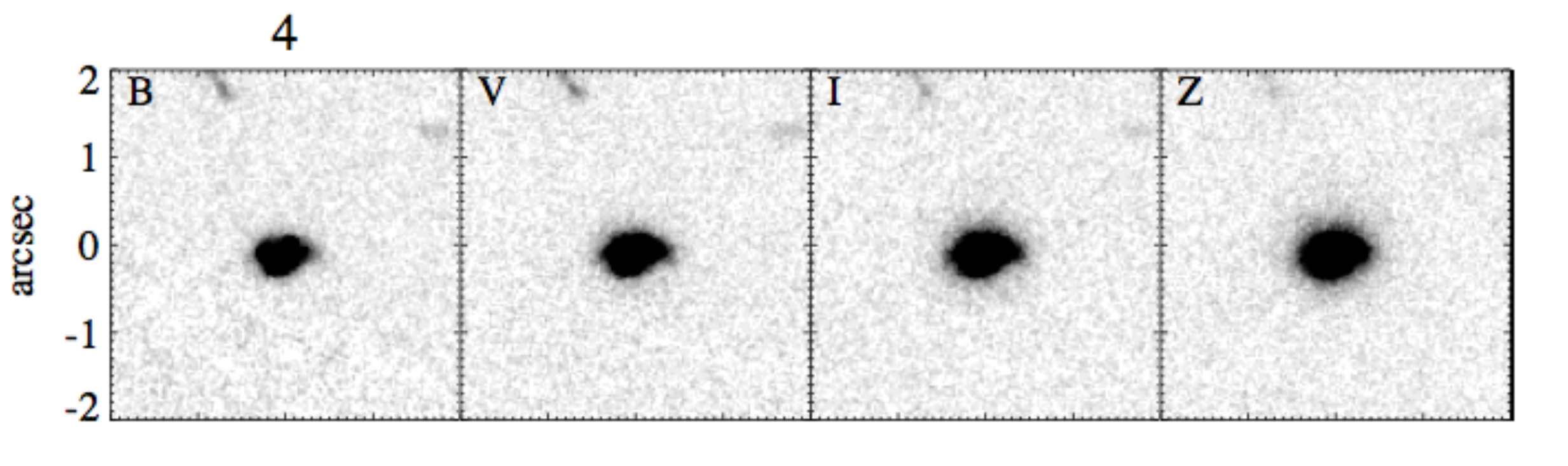}
\includegraphics[width=5.8cm,angle=-90]{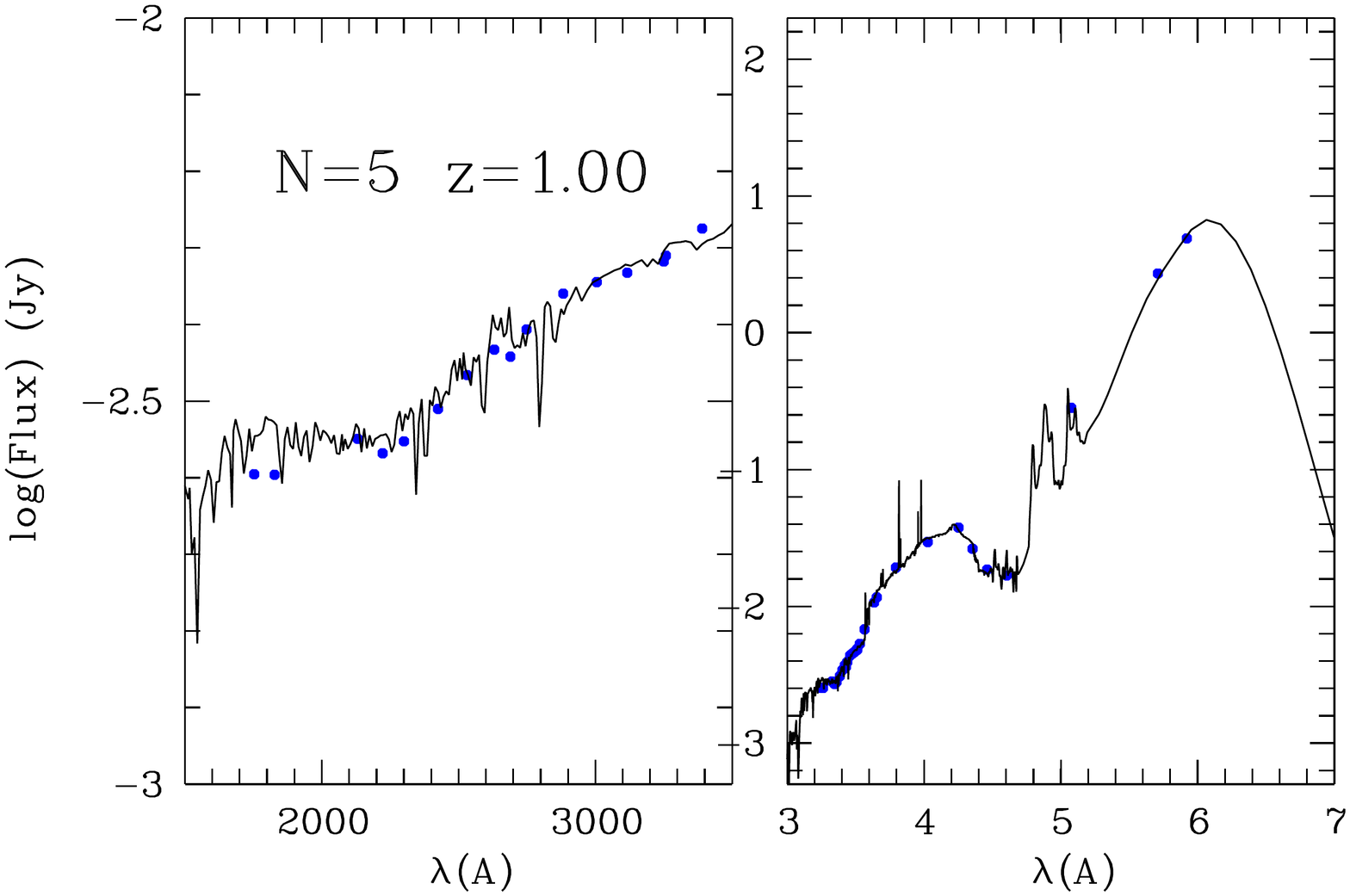}
\includegraphics[width=8cm]{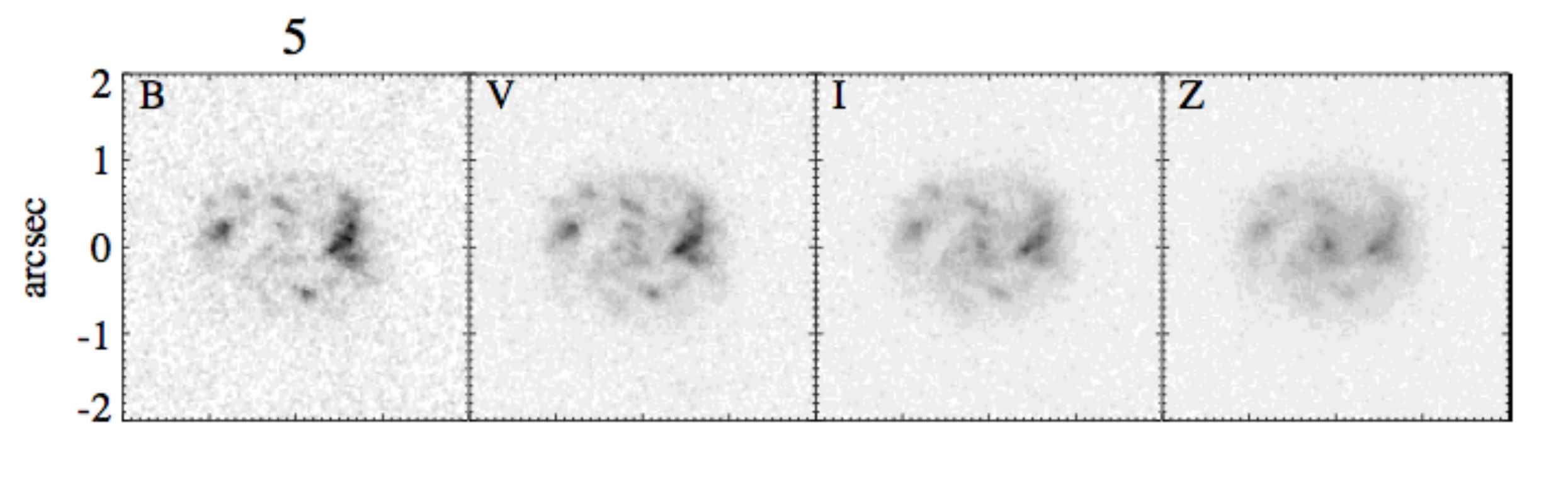}
\includegraphics[width=5.8cm,angle=-90]{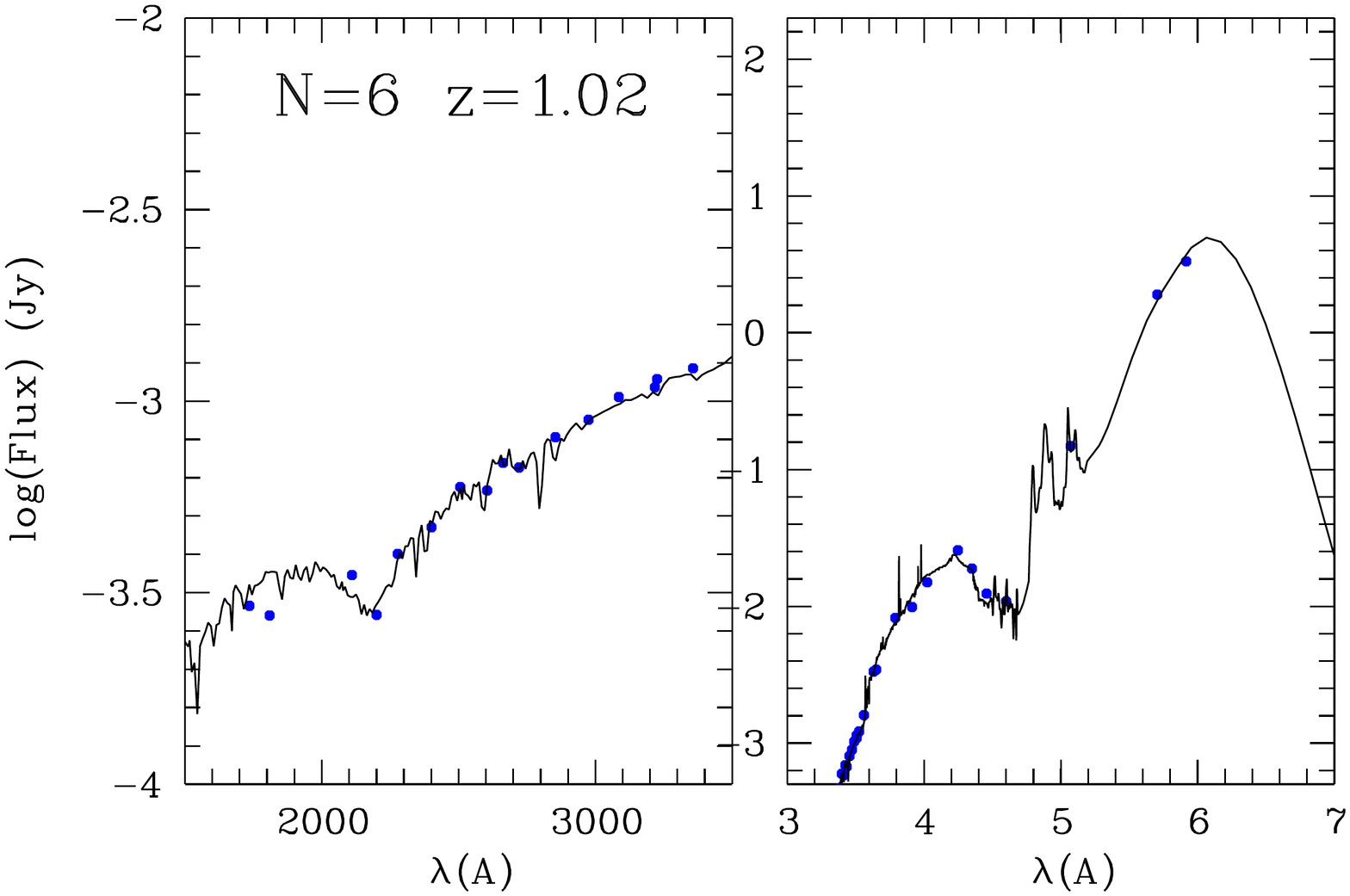}
\includegraphics[width=8cm]{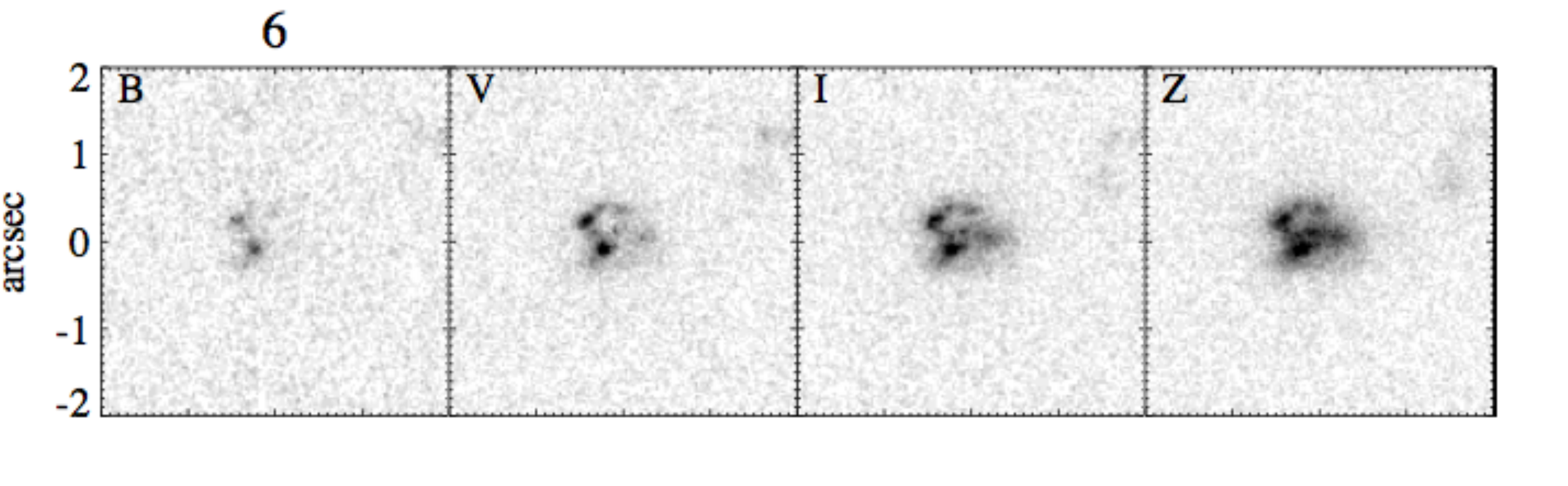}
\caption{ cont'd}
\label{fig:continued:second}
      \end{figure}
    \begin{figure}
    \ContinuedFloat
  \centering
  \includegraphics[width=5.8cm,angle=-90]{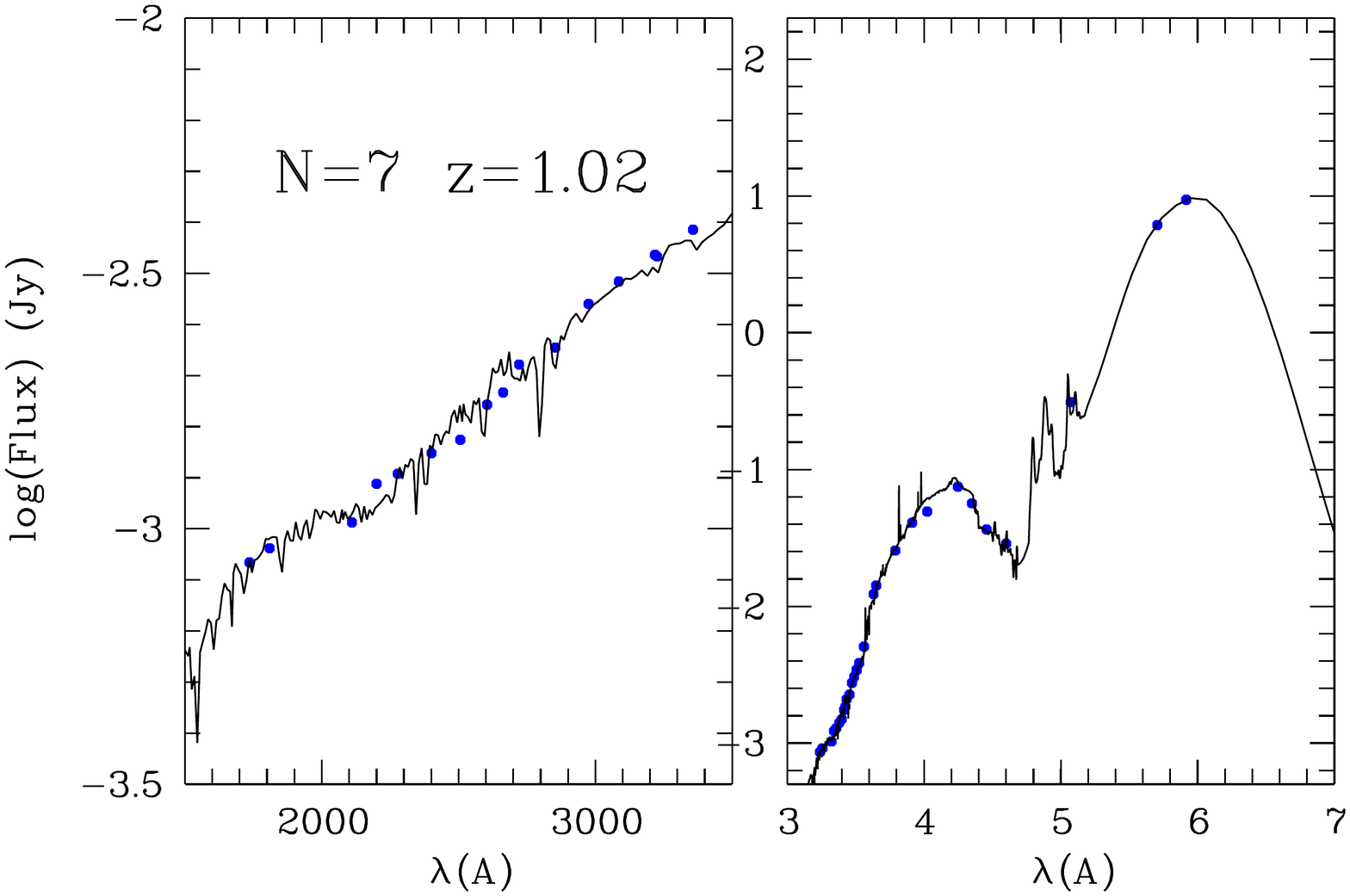}
  \includegraphics[width=8cm]{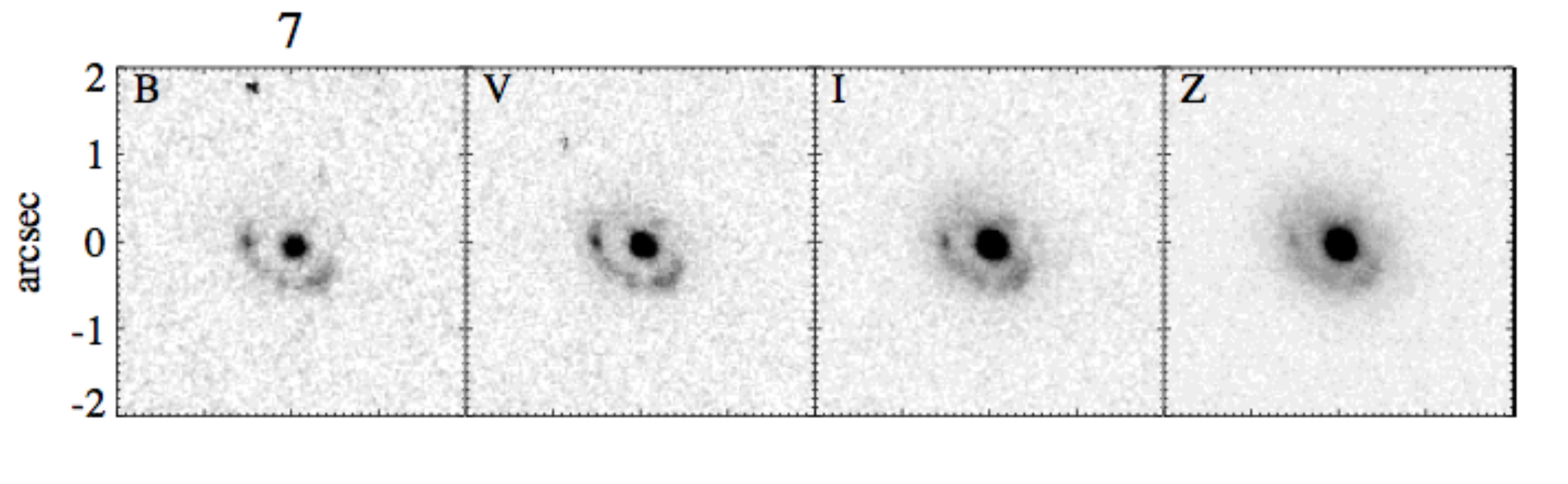}
\includegraphics[width=5.8cm,angle=-90]{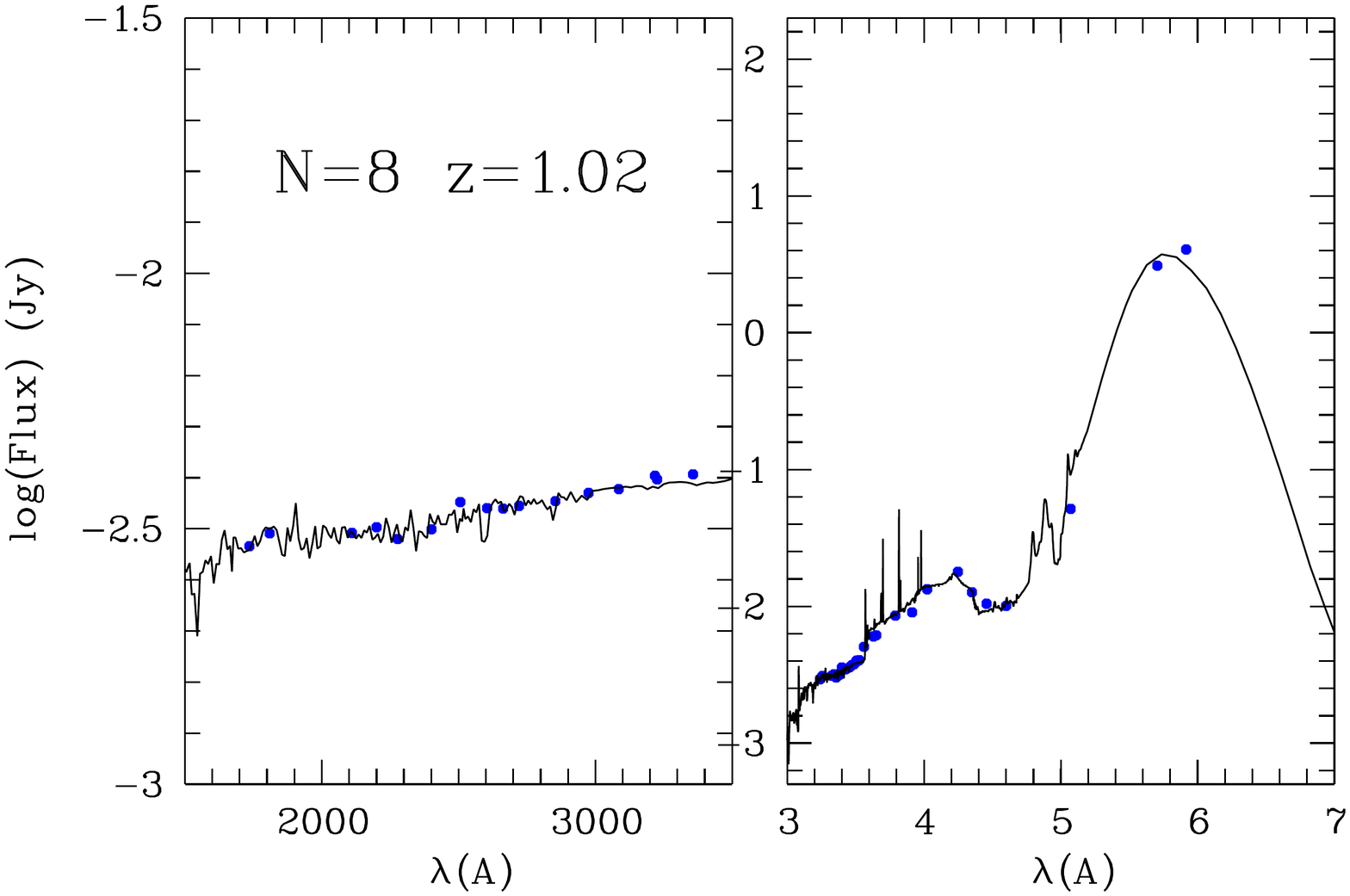}
\includegraphics[width=8cm]{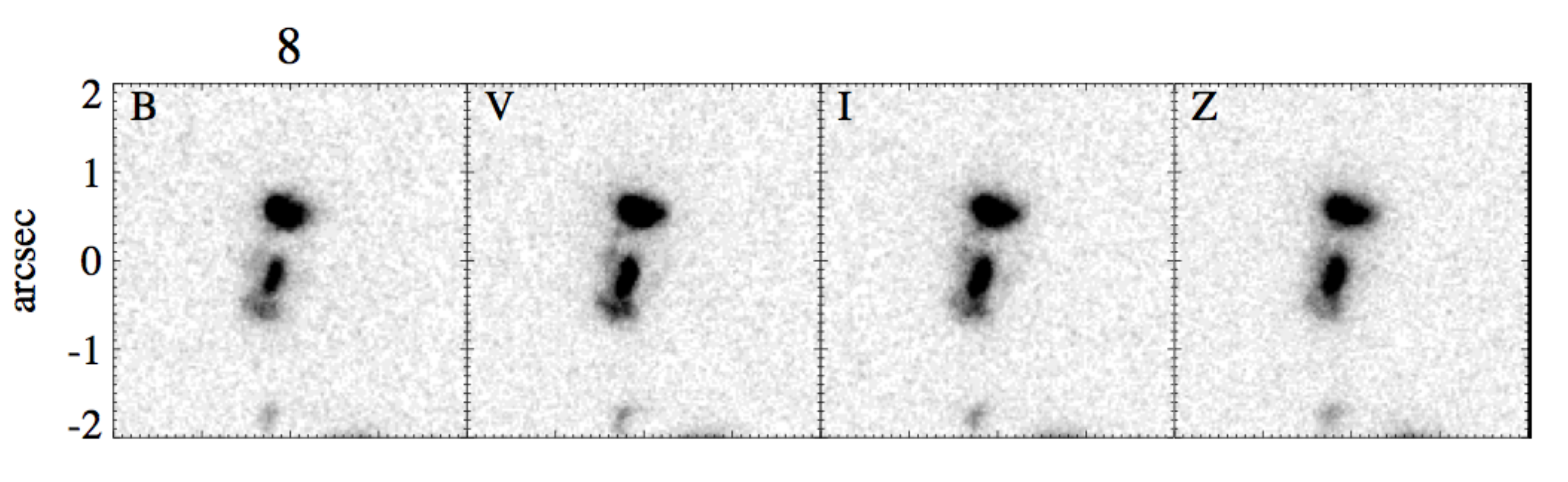}
\includegraphics[width=5.8cm,angle=-90]{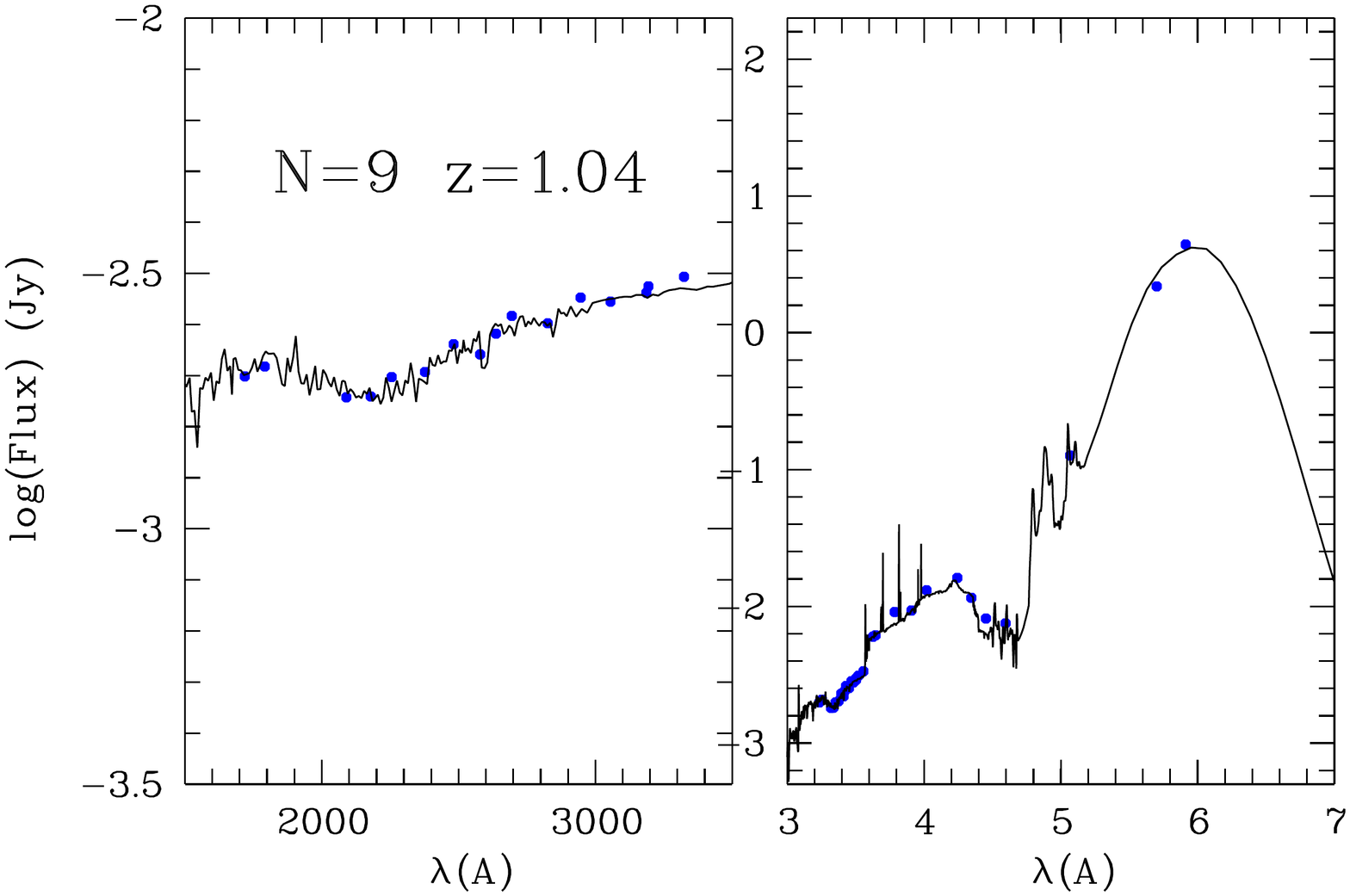}
\includegraphics[width=8cm]{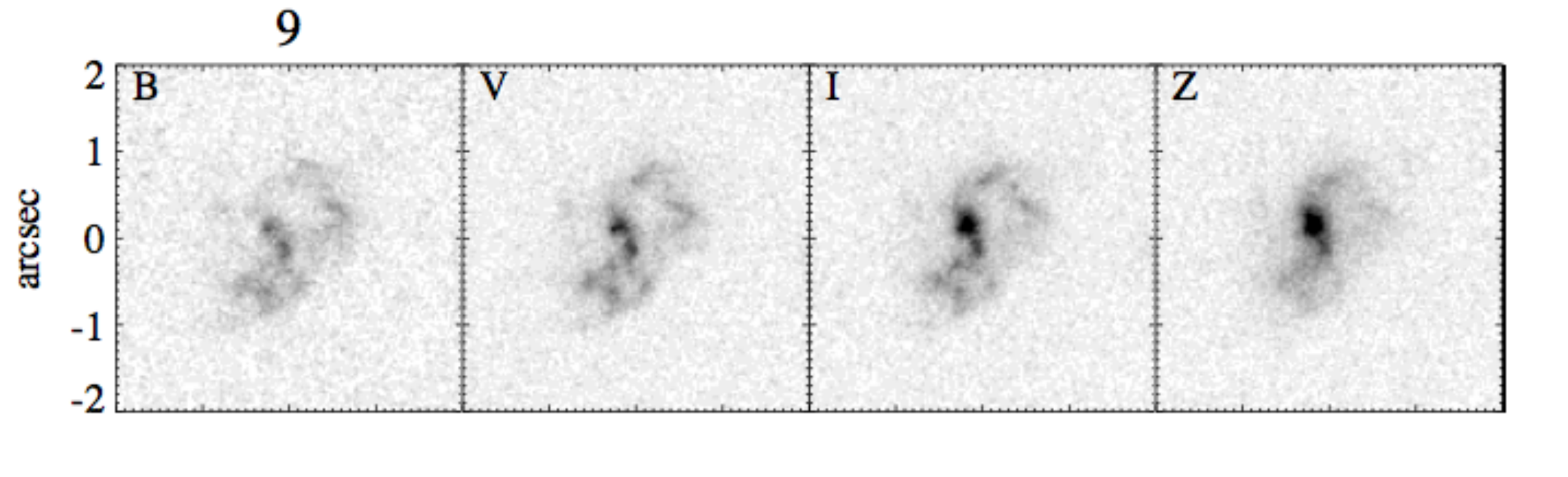}
\caption{ cont'd}
      \end{figure}
        \begin{figure}
    \ContinuedFloat
  \centering
\includegraphics[width=5.8cm,angle=-90]{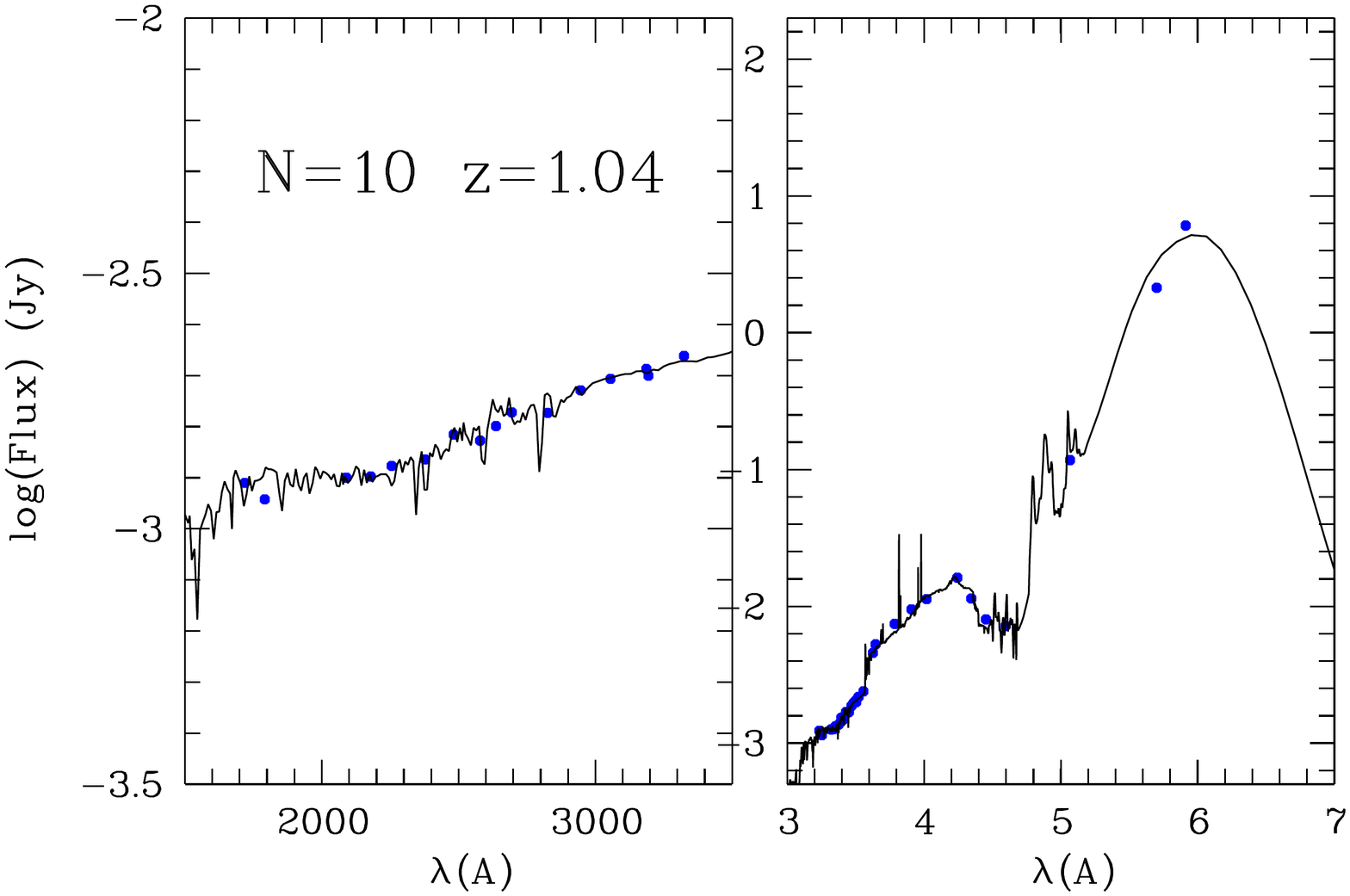}
\includegraphics[width=8cm]{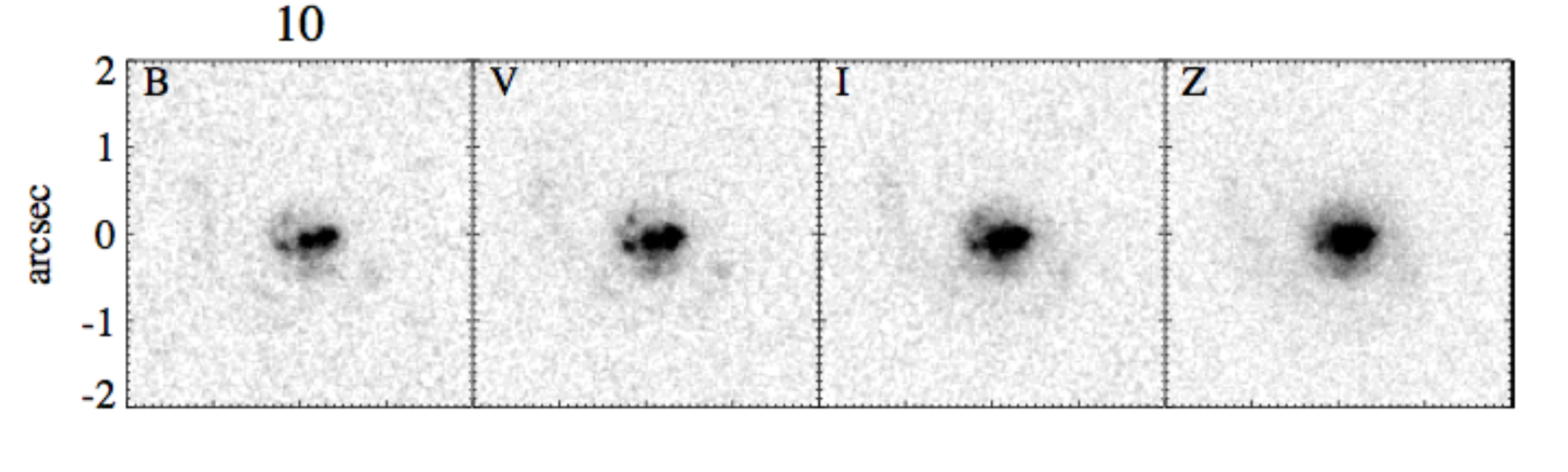}
\includegraphics[width=5.8cm,angle=-90]{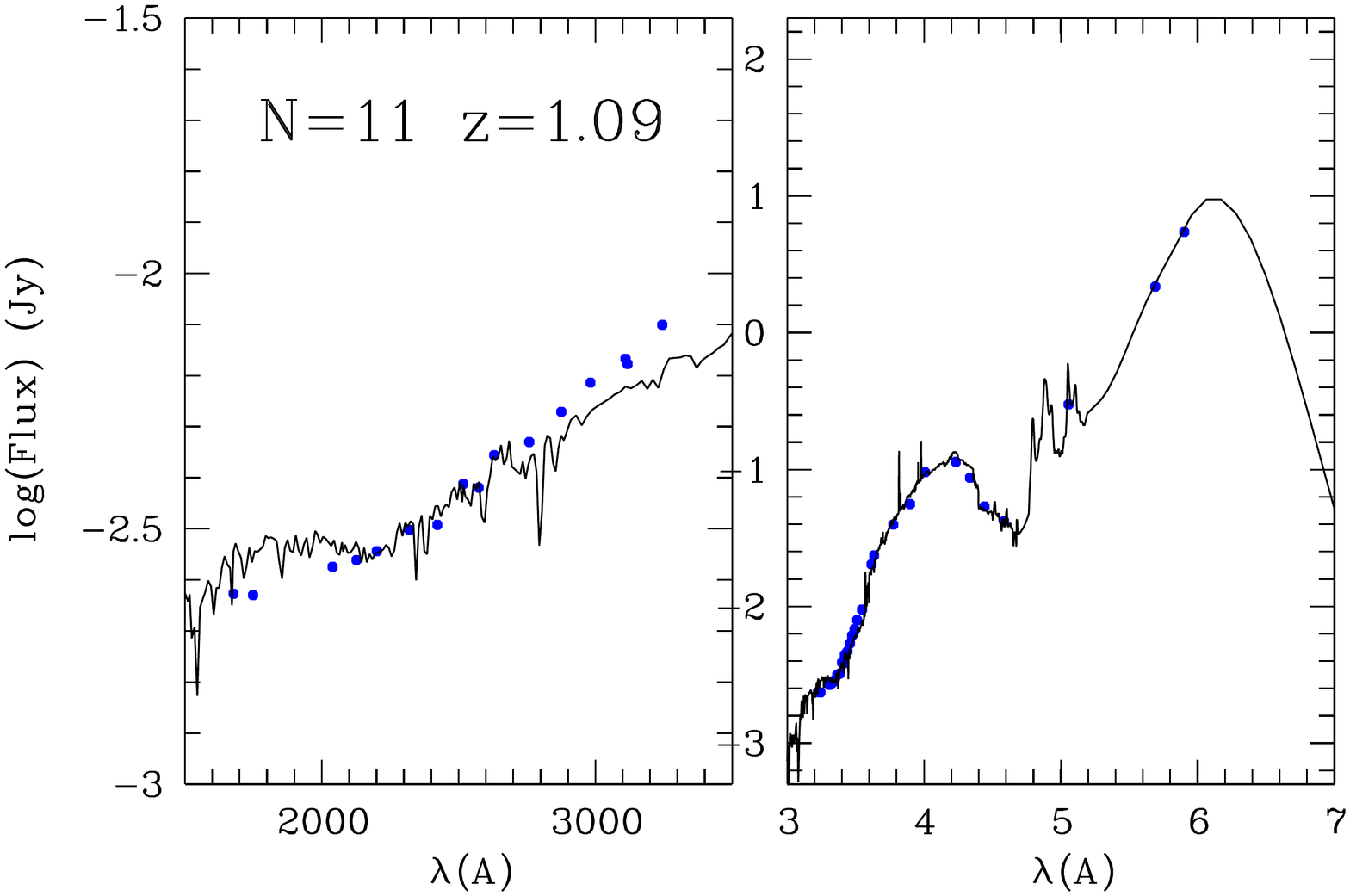}
\includegraphics[width=8cm]{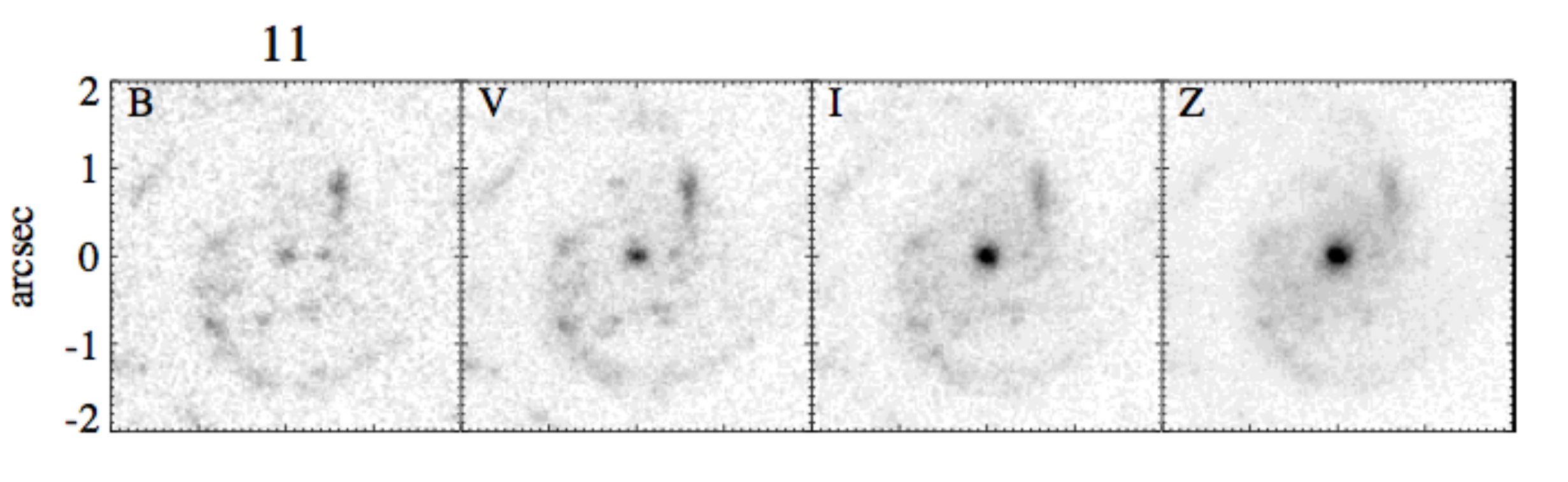}
\includegraphics[width=5.8cm,angle=-90]{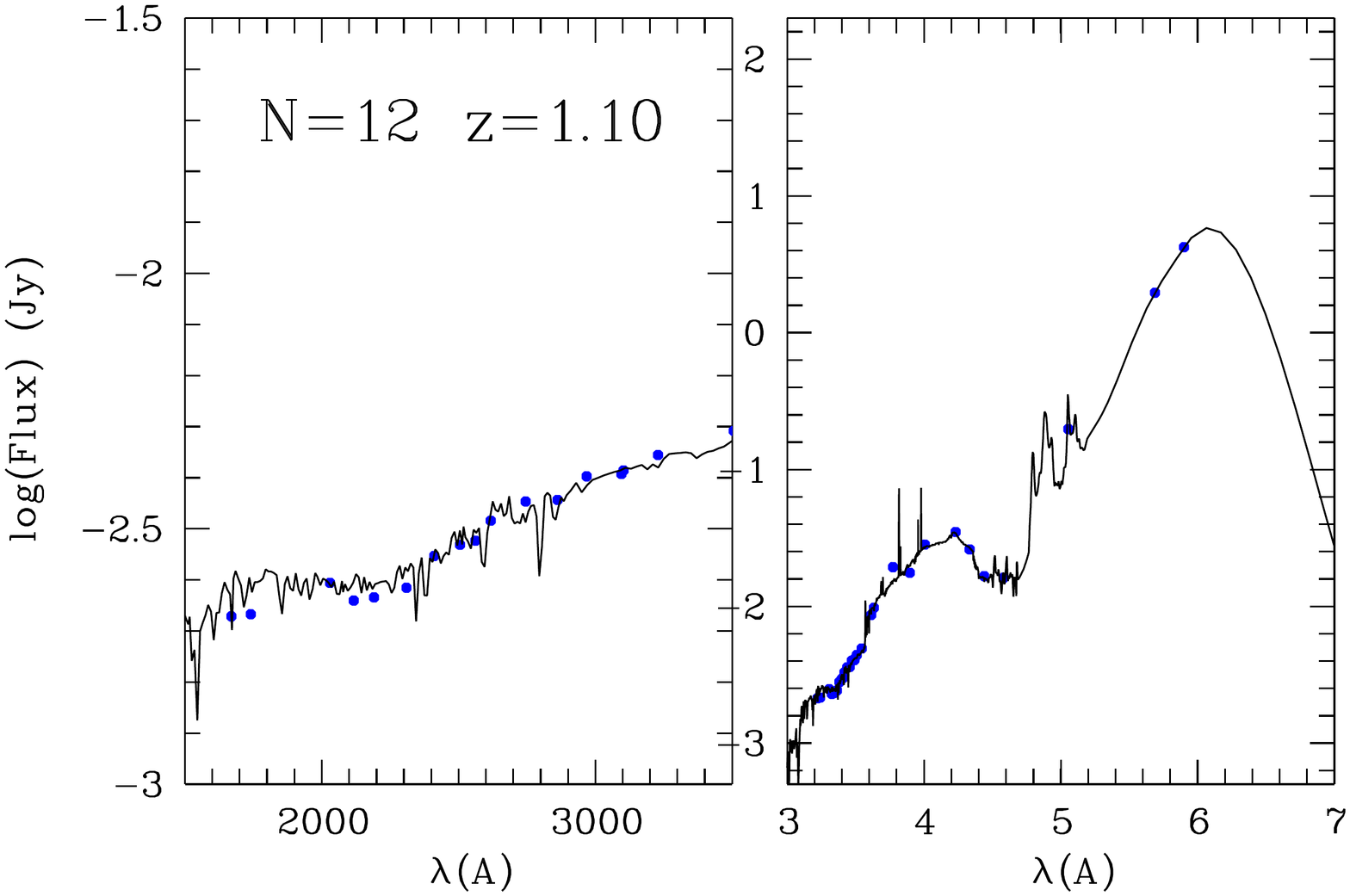}
\includegraphics[width=8cm]{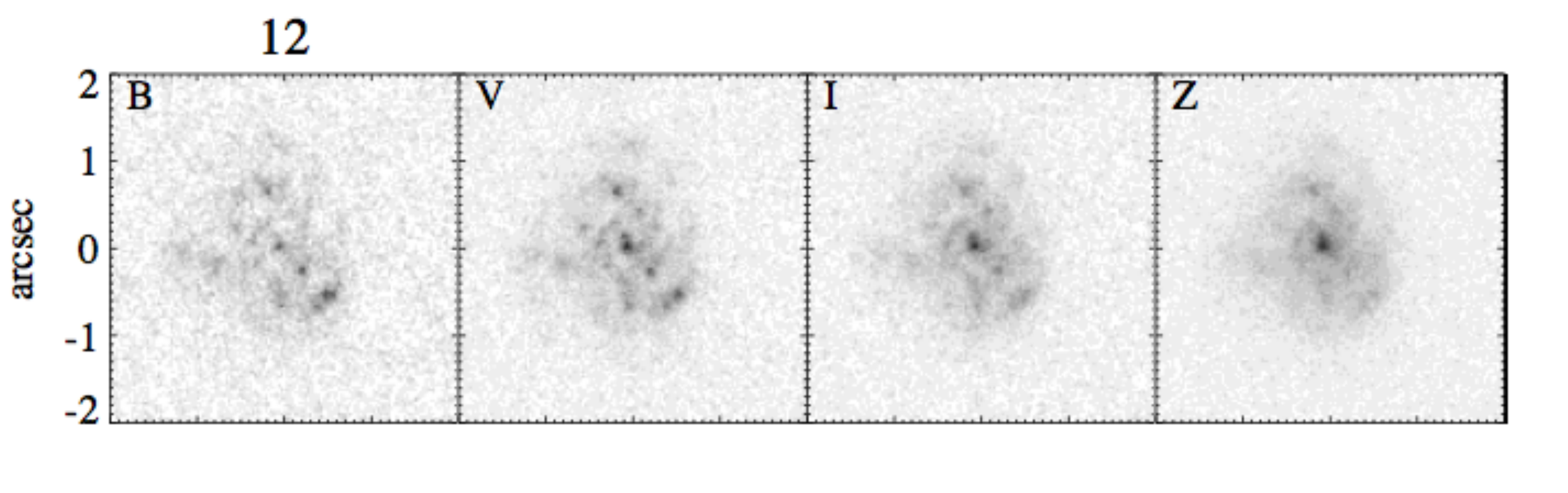}
\caption{ cont'd}
      \end{figure}
        \begin{figure}
    \ContinuedFloat
  \centering
\includegraphics[width=5.8cm,angle=-90]{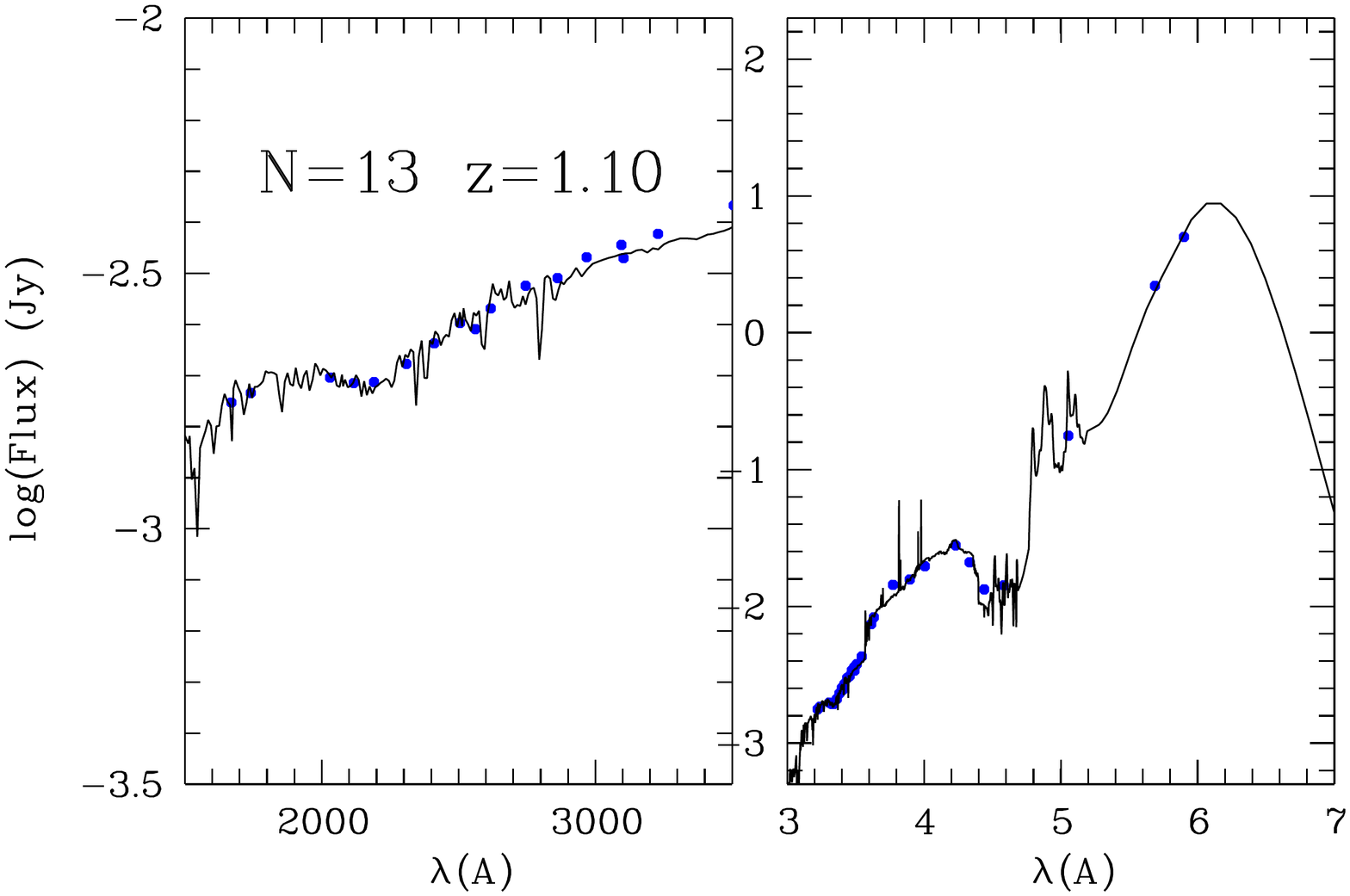}
\includegraphics[width=8cm]{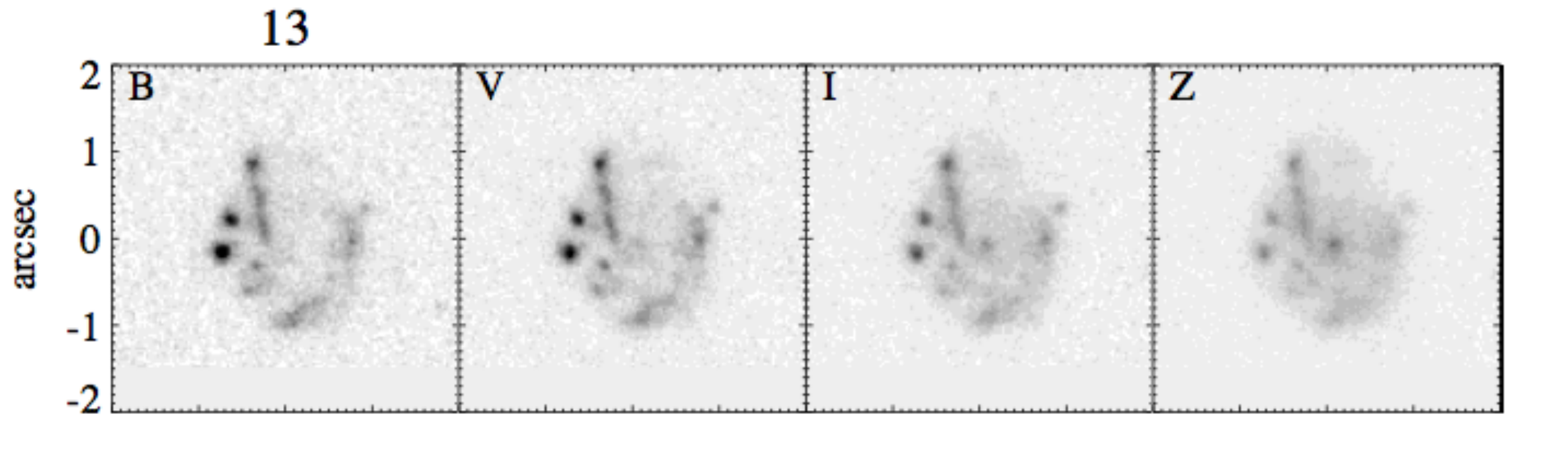}
\includegraphics[width=5.8cm,angle=-90]{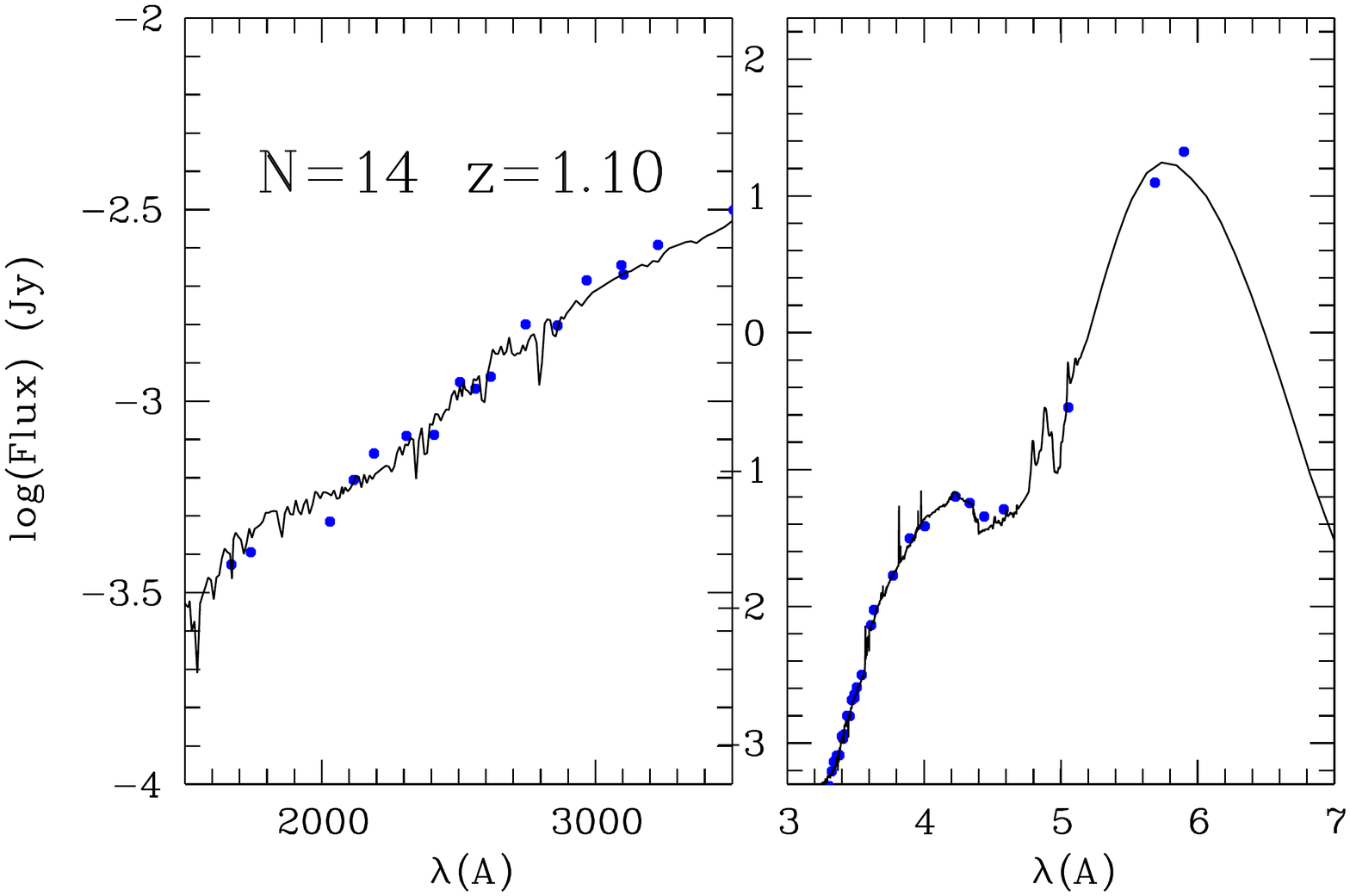}
\includegraphics[width=8cm]{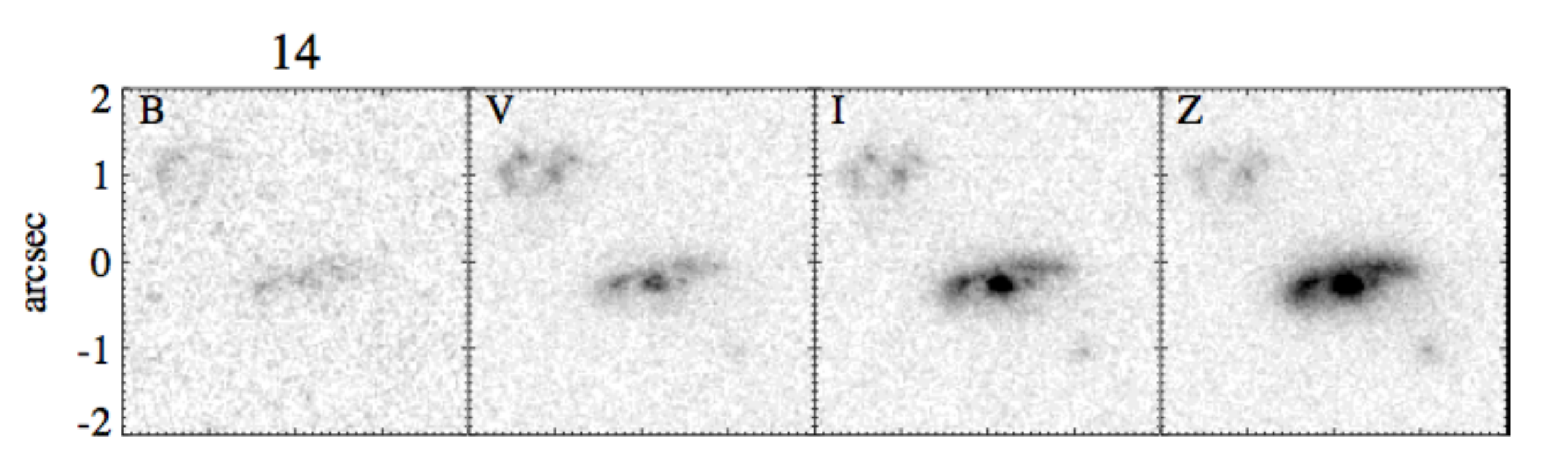}
\includegraphics[width=5.8cm,angle=-90]{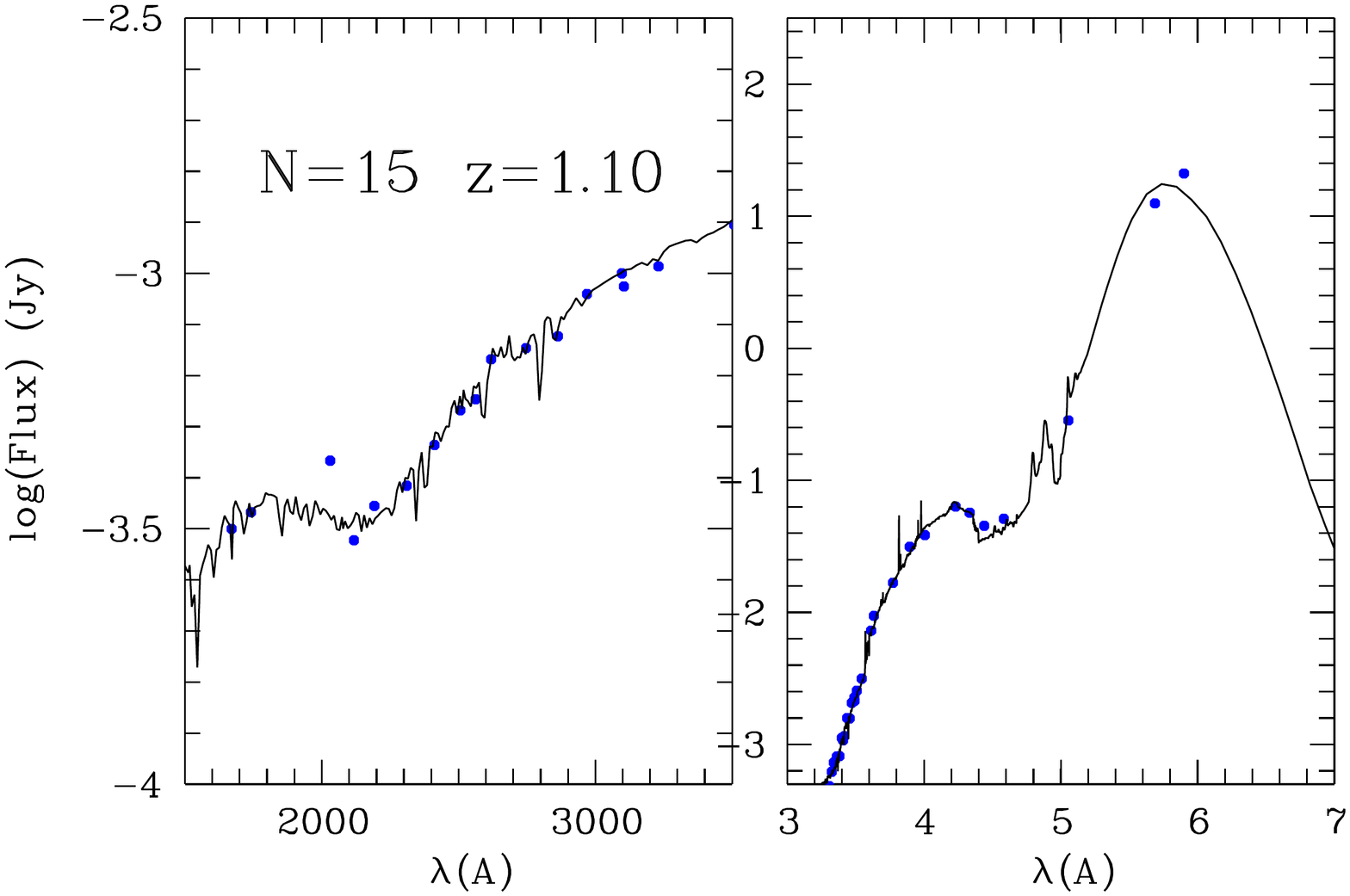}
\includegraphics[width=8cm]{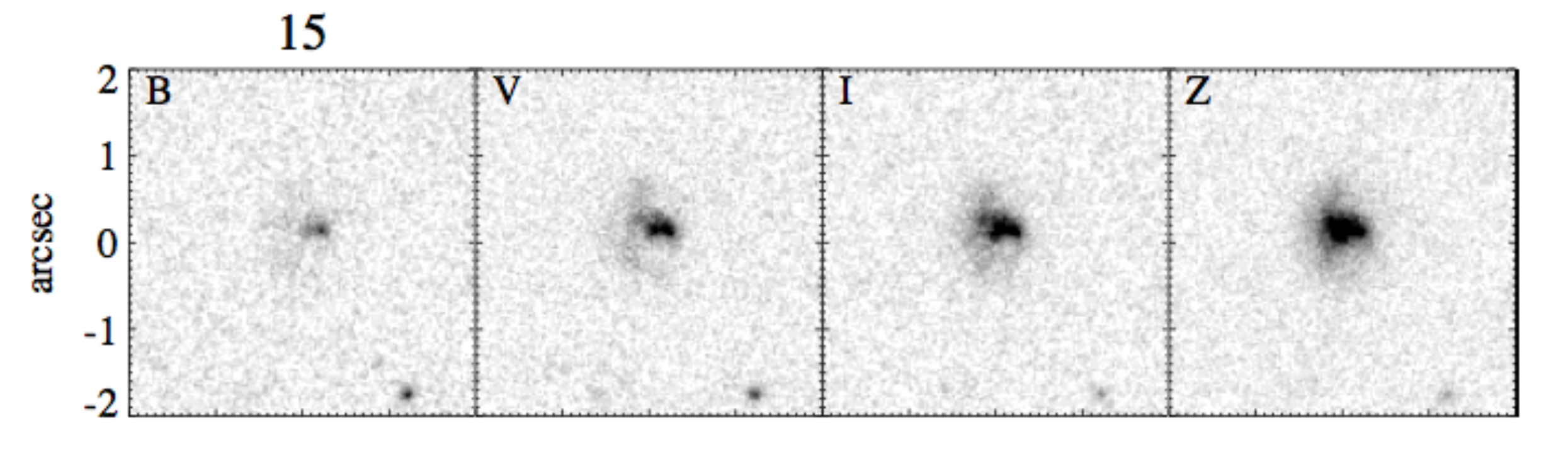}
\caption{ cont'd}
      \end{figure}
  \begin{figure}
    \ContinuedFloat
  \centering
  \includegraphics[width=5.8cm,angle=-90]{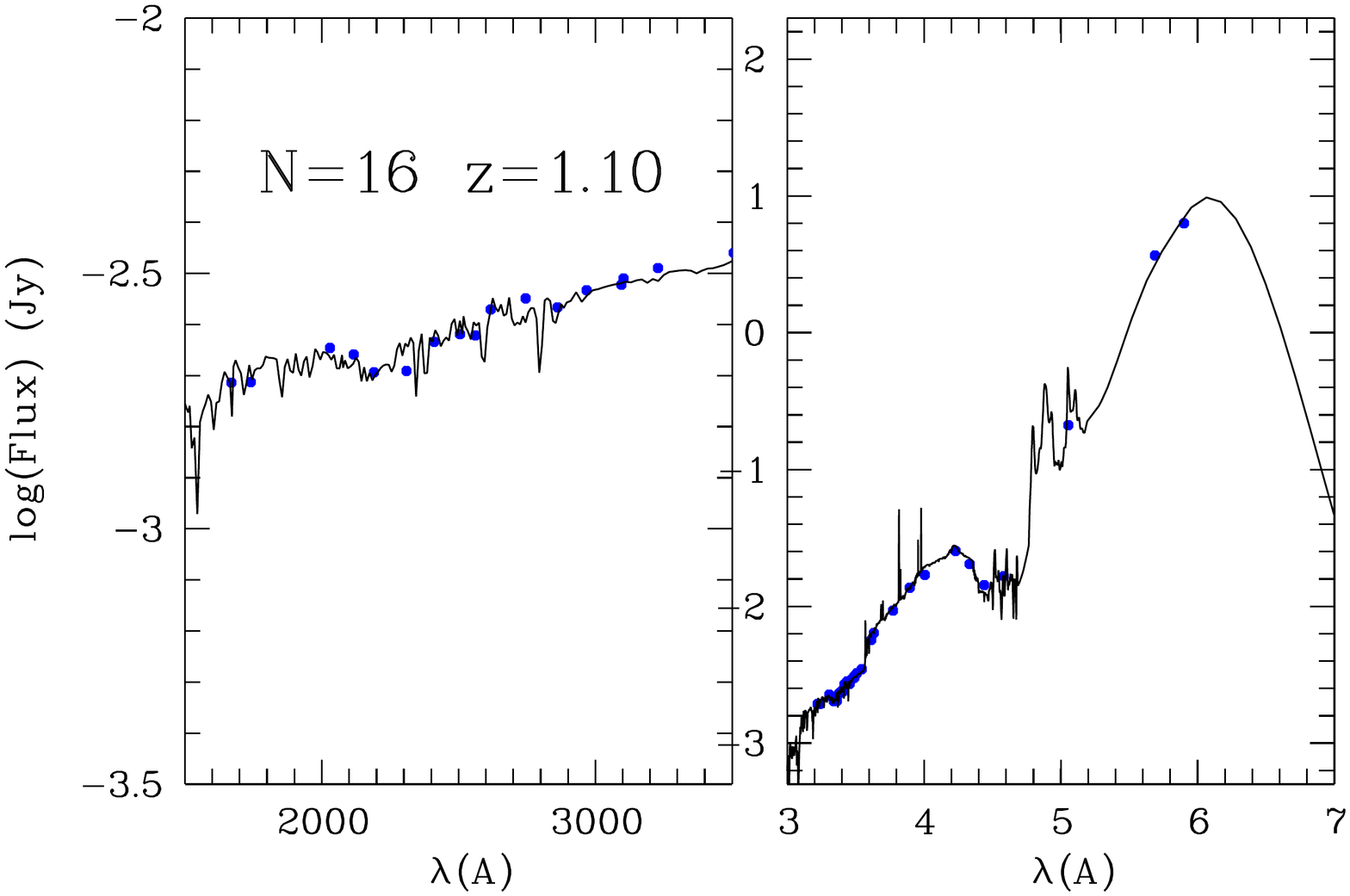}
  \includegraphics[width=8cm]{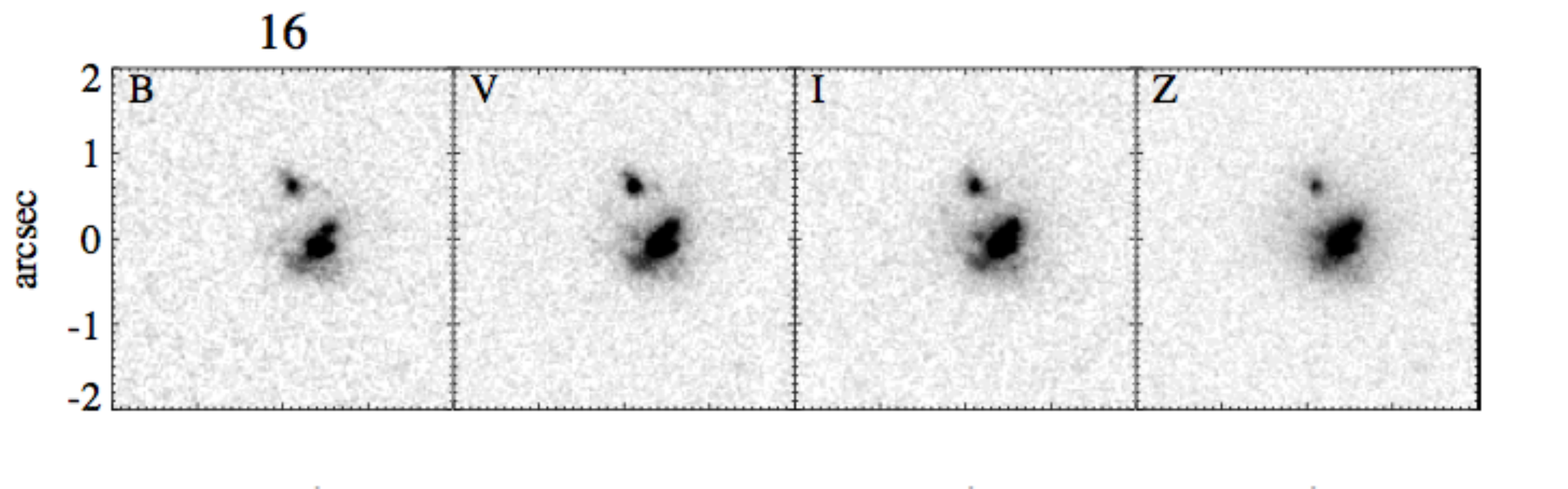}
\includegraphics[width=5.8cm,angle=-90]{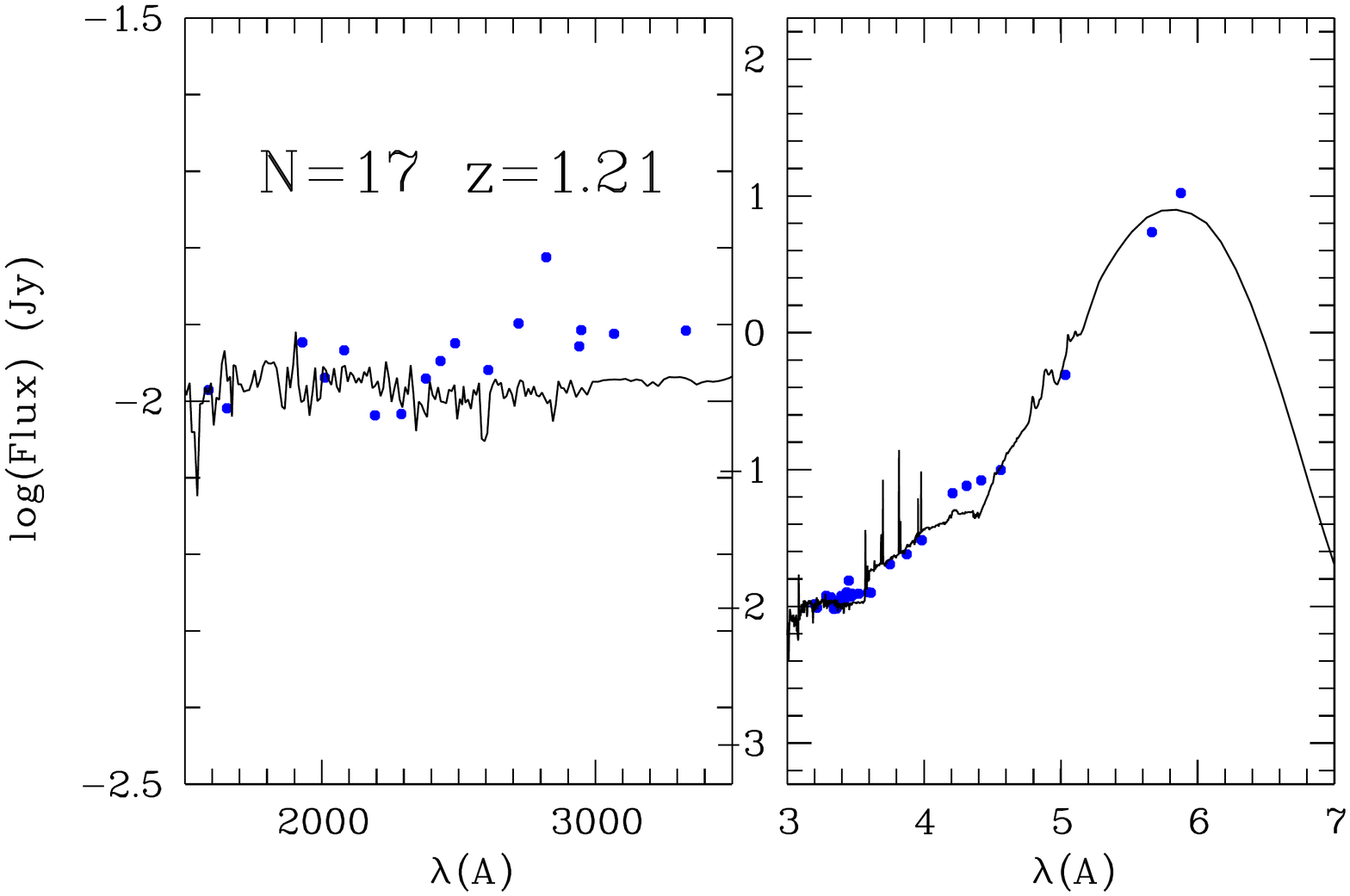}
\includegraphics[width=8cm]{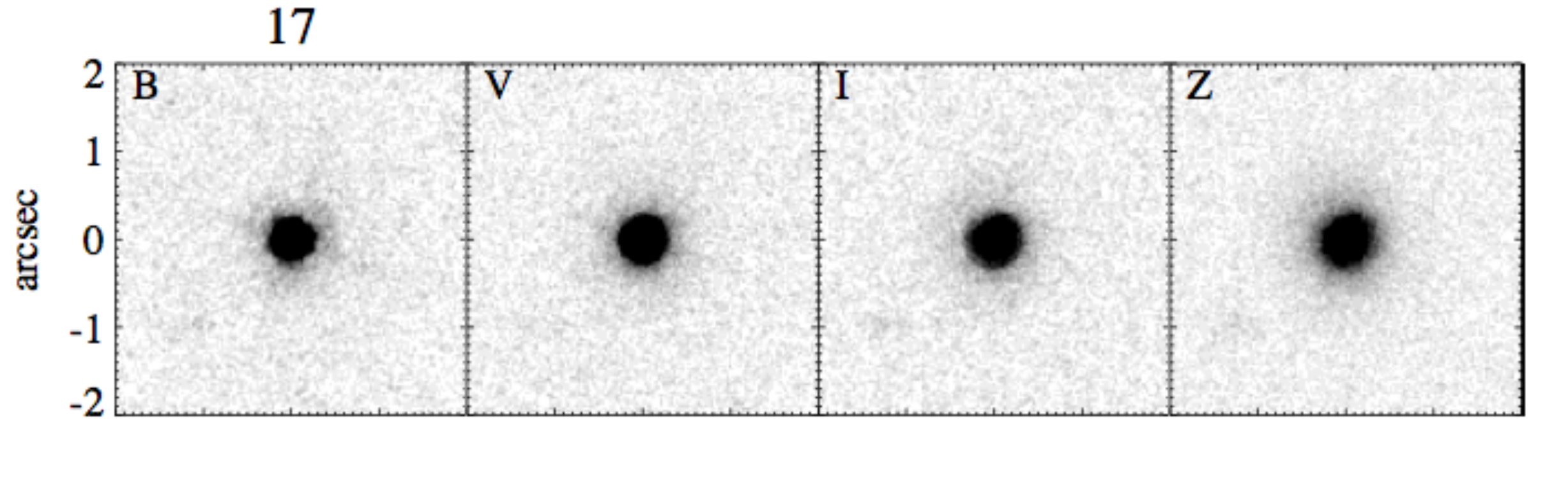}
\includegraphics[width=5.8cm,angle=-90]{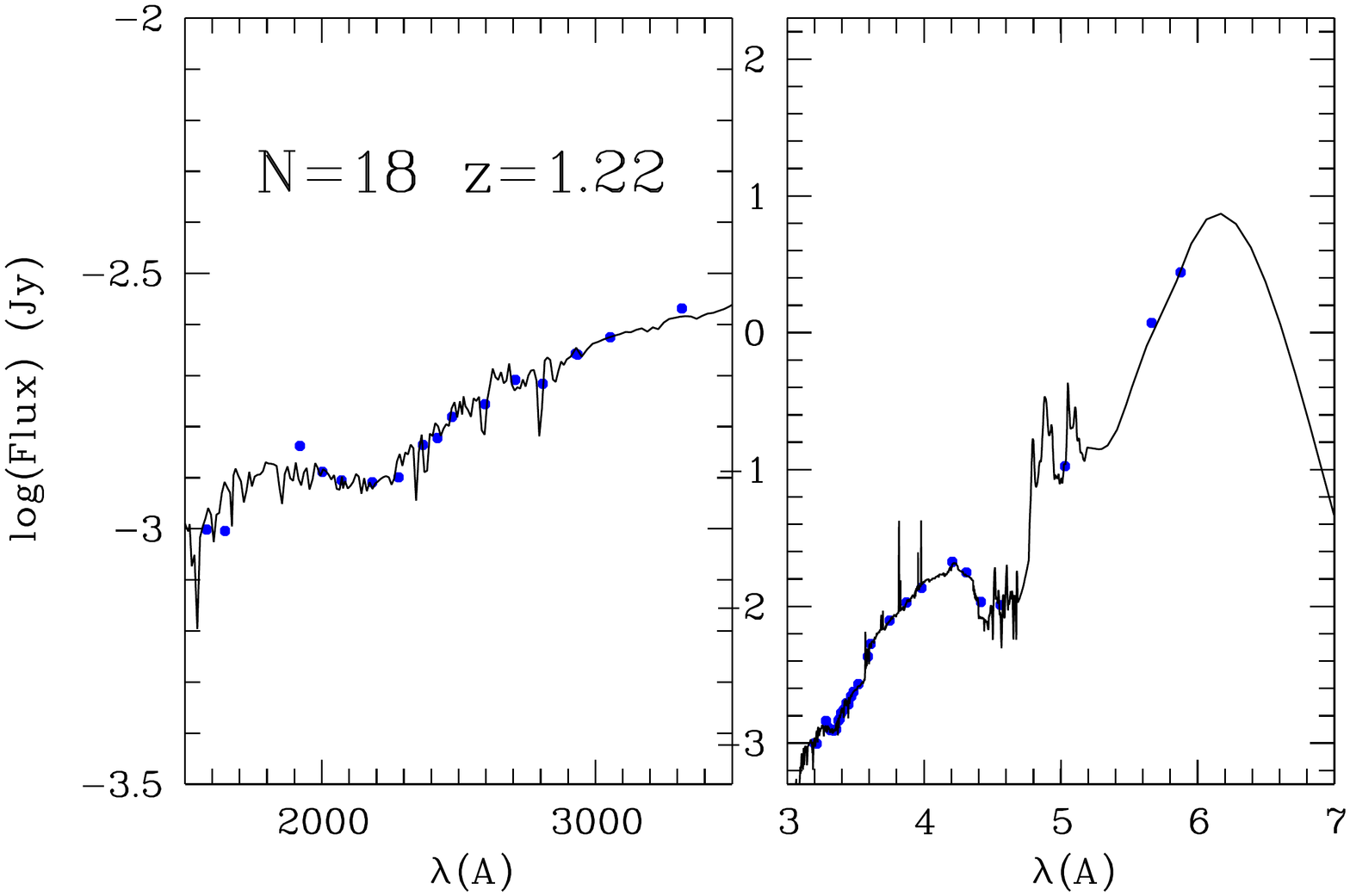}
\includegraphics[width=8cm]{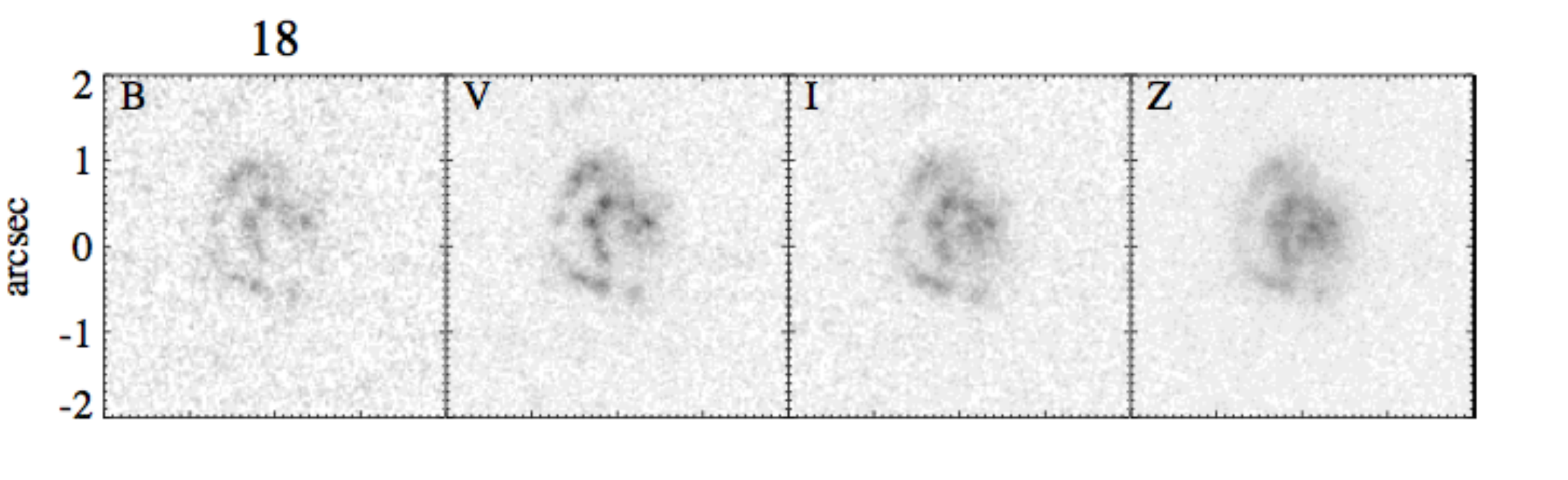}
\caption{ cont'd}
      \end{figure}
  \begin{figure}
    \ContinuedFloat
  \centering
  \includegraphics[width=5.8cm,angle=-90]{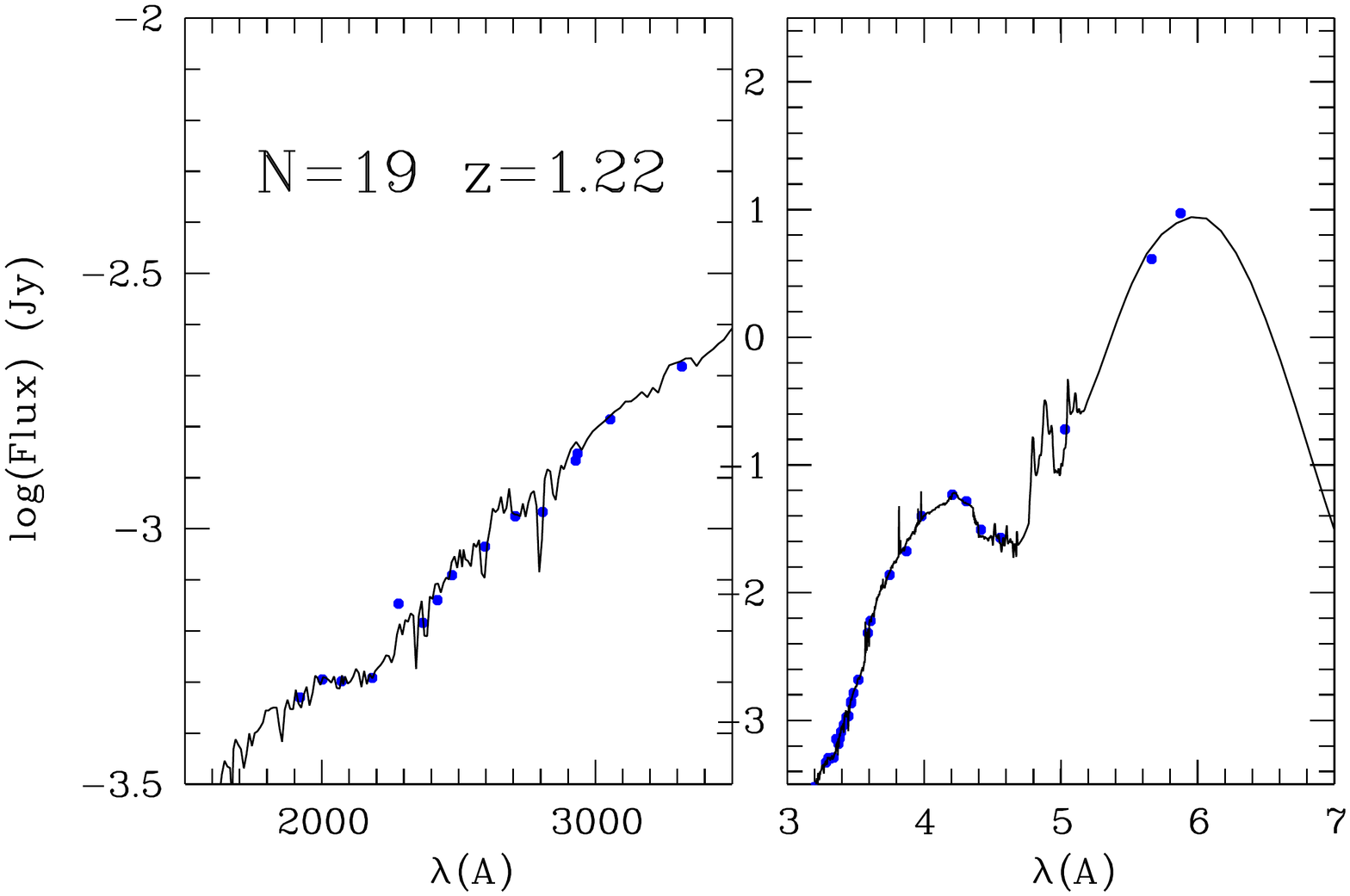}
  \includegraphics[width=8cm]{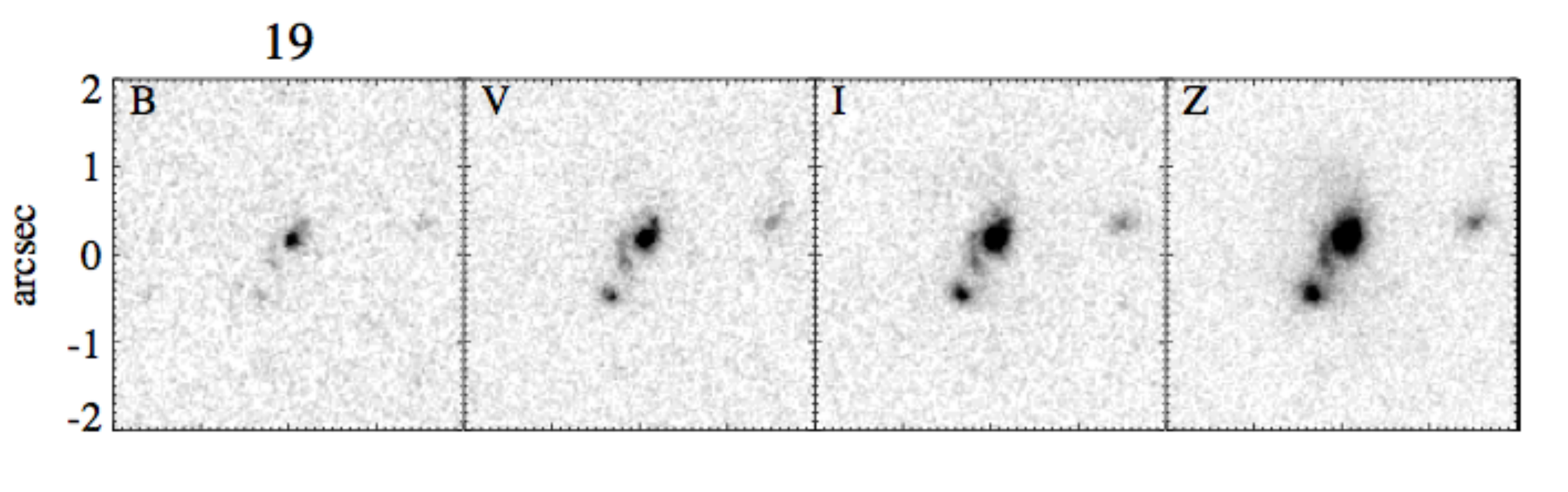}
\includegraphics[width=5.8cm,angle=-90]{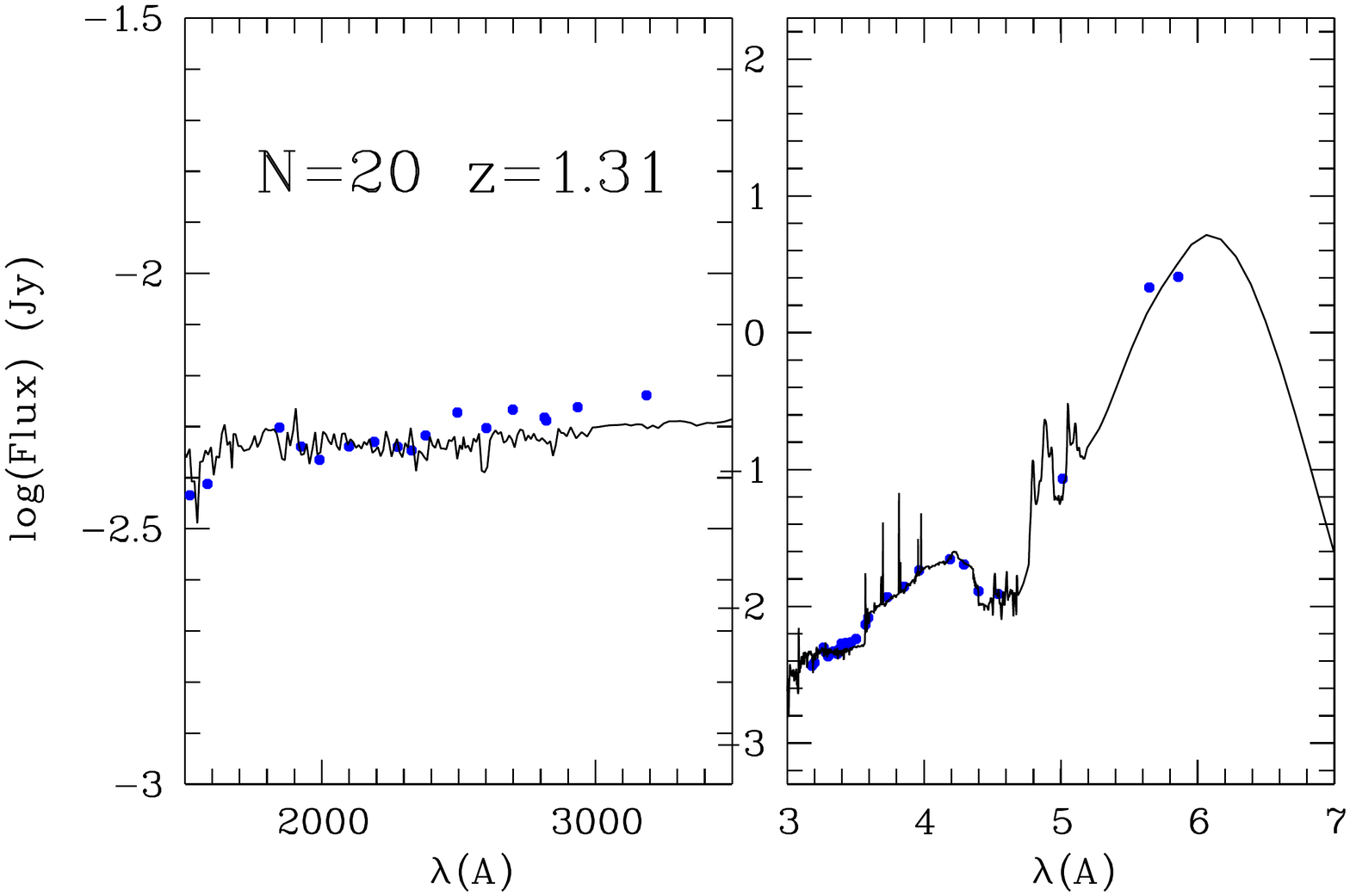}
\includegraphics[width=8cm]{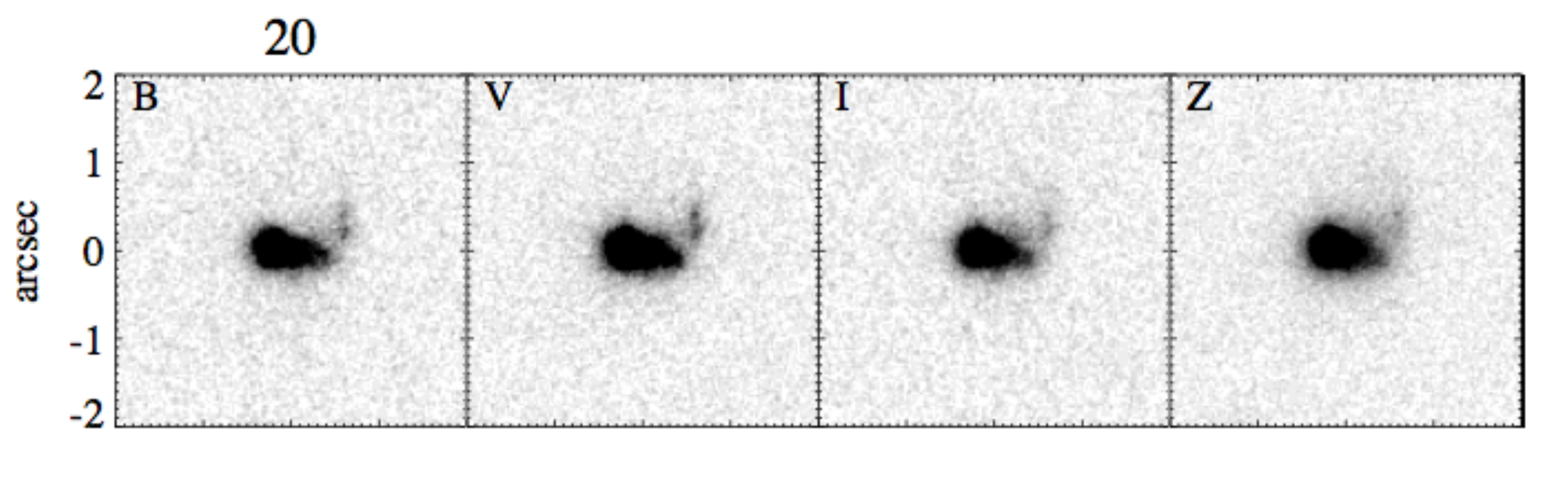}
\includegraphics[width=5.8cm,angle=-90]{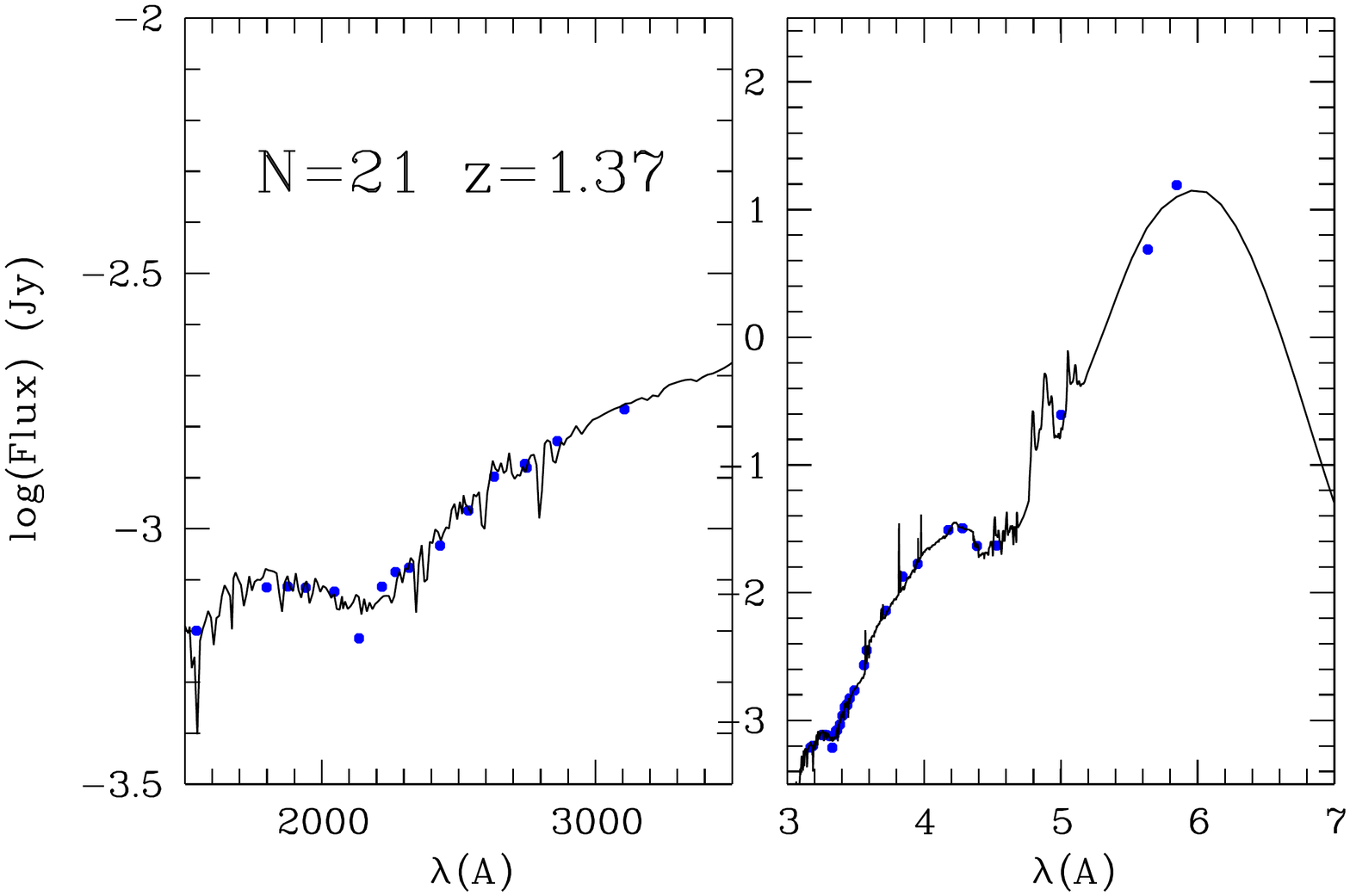}
\includegraphics[width=8cm]{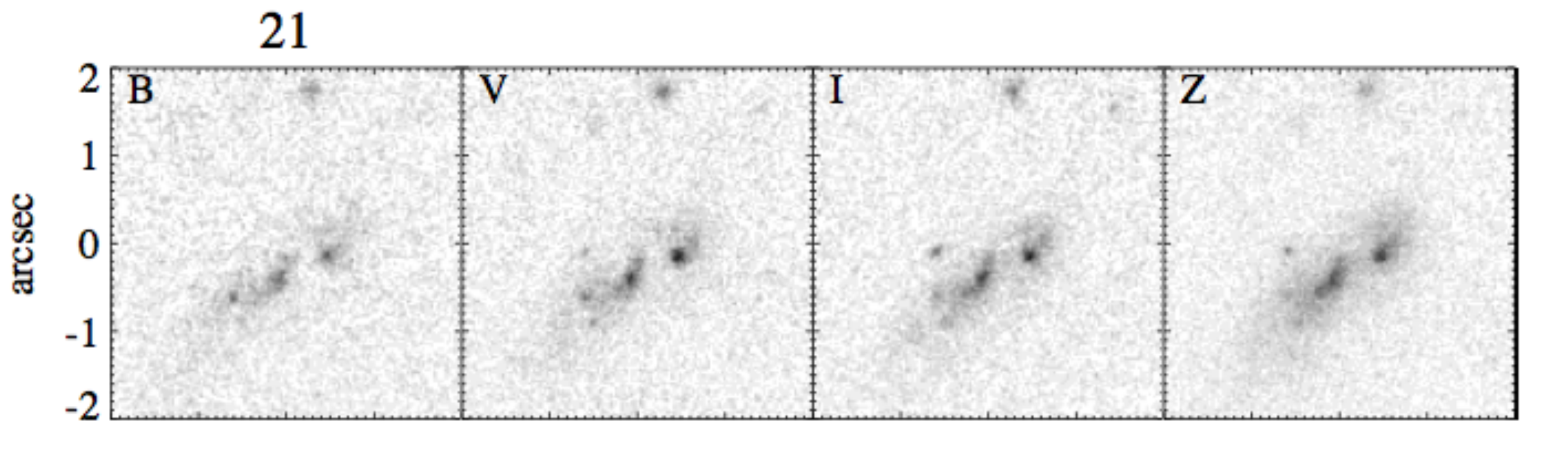}
\caption{ cont'd}
      \end{figure}
  \begin{figure}
    \ContinuedFloat
  \centering
\includegraphics[width=5.8cm,angle=-90]{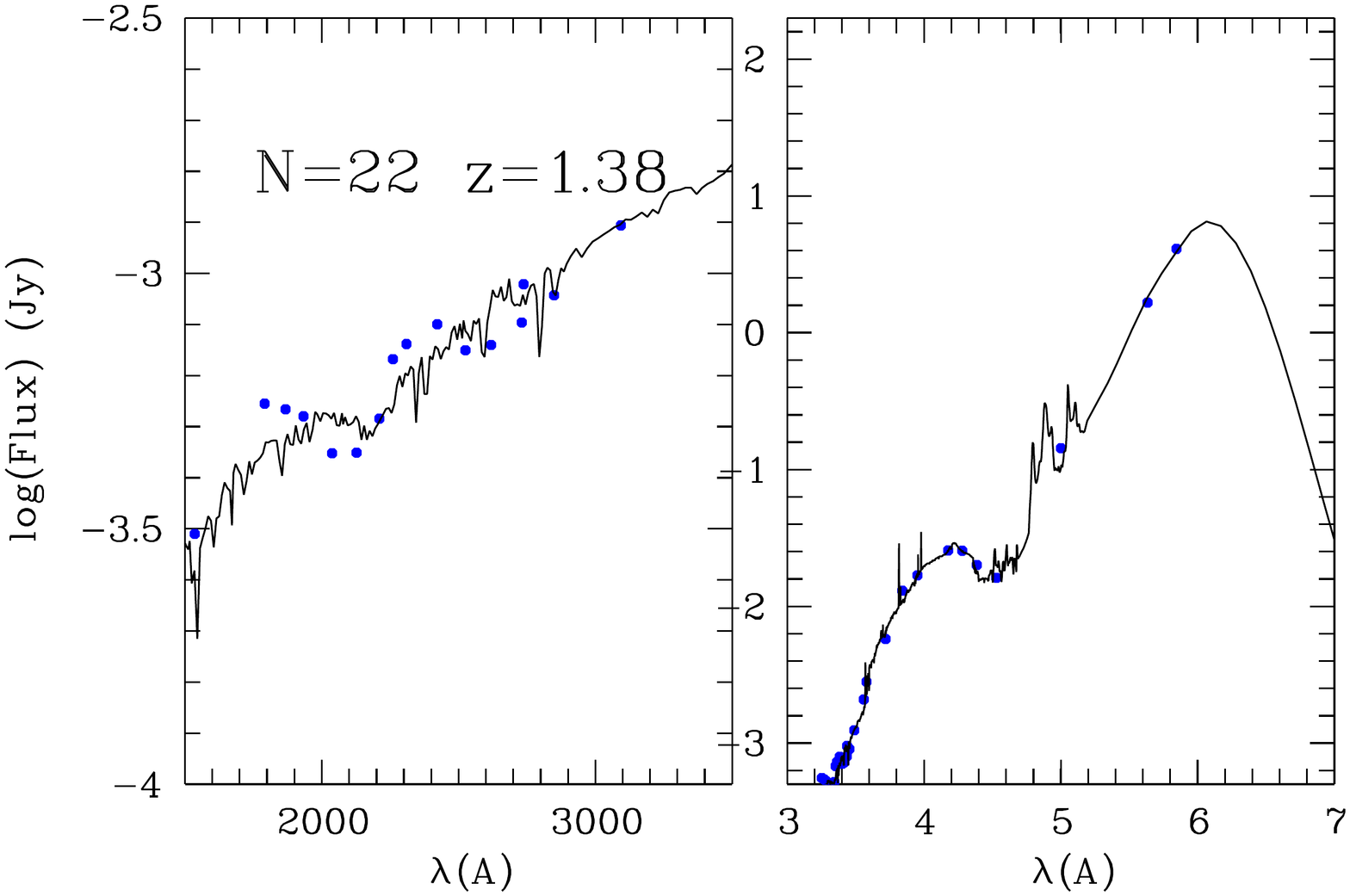}
\includegraphics[width=8cm]{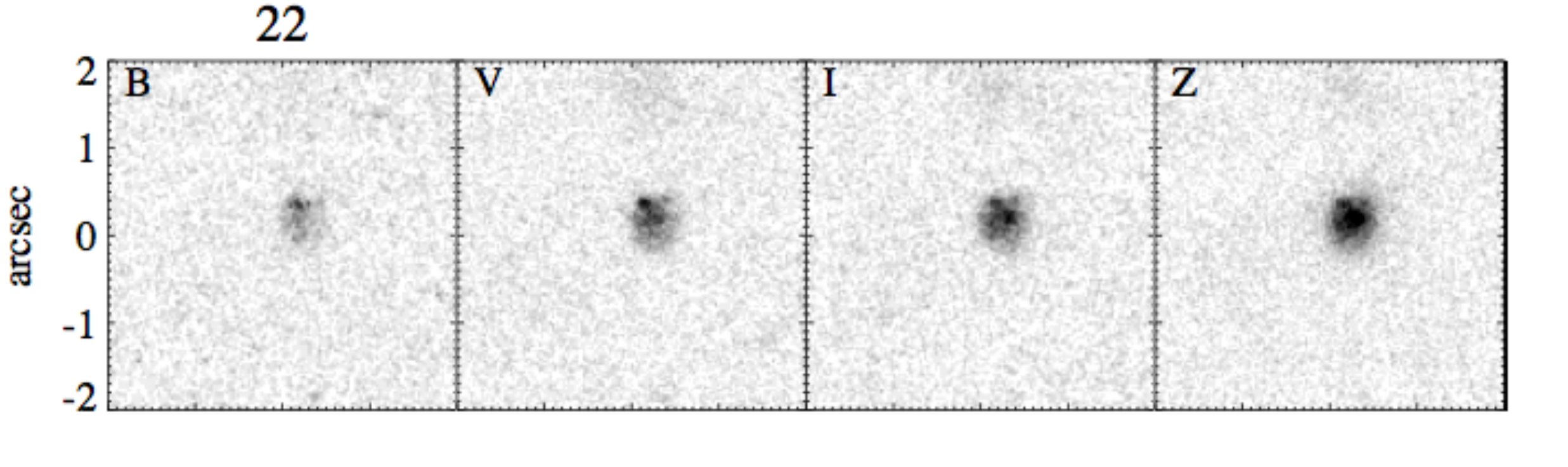}
\includegraphics[width=5.8cm,angle=-90]{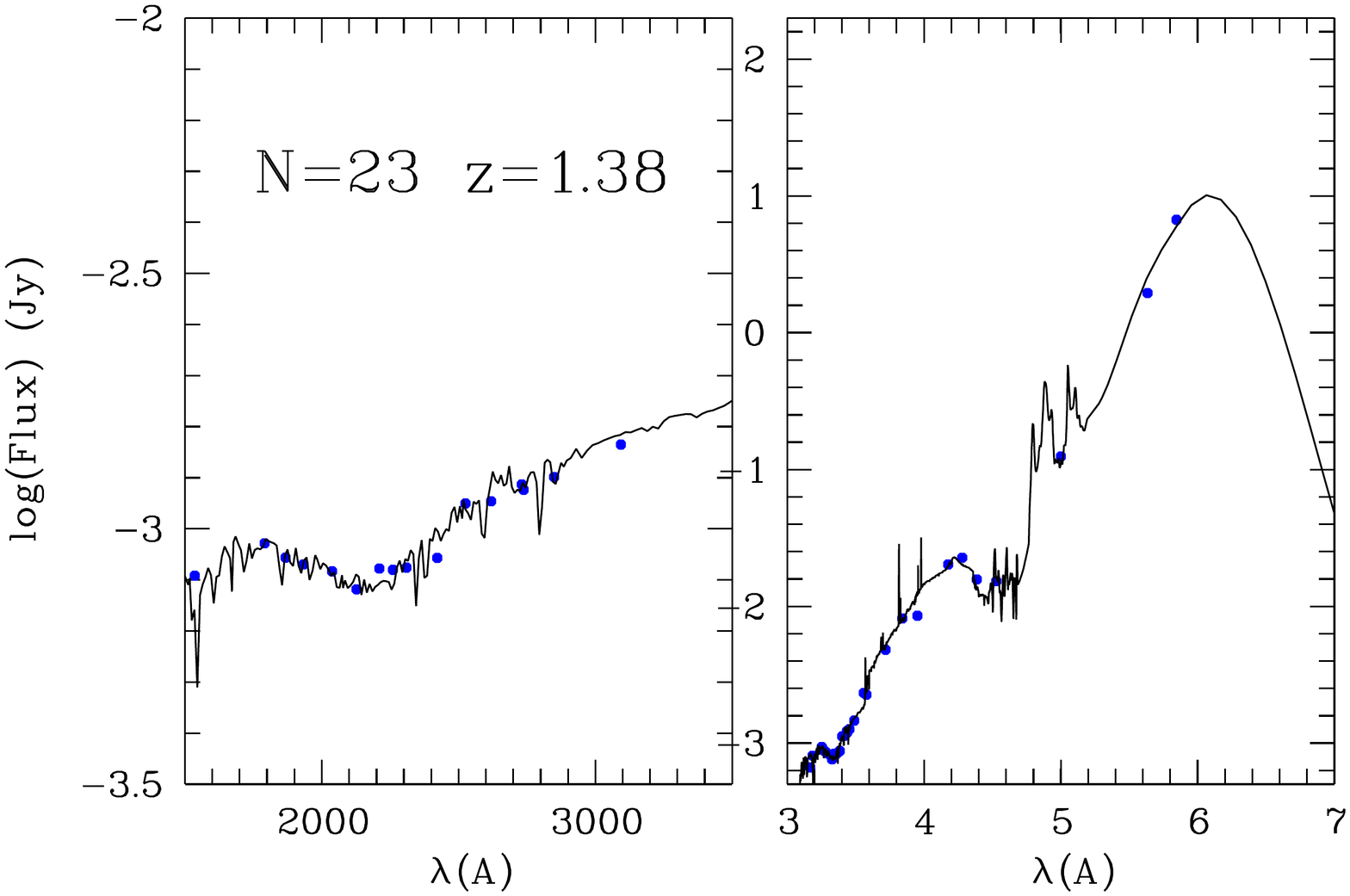}
\includegraphics[width=8cm]{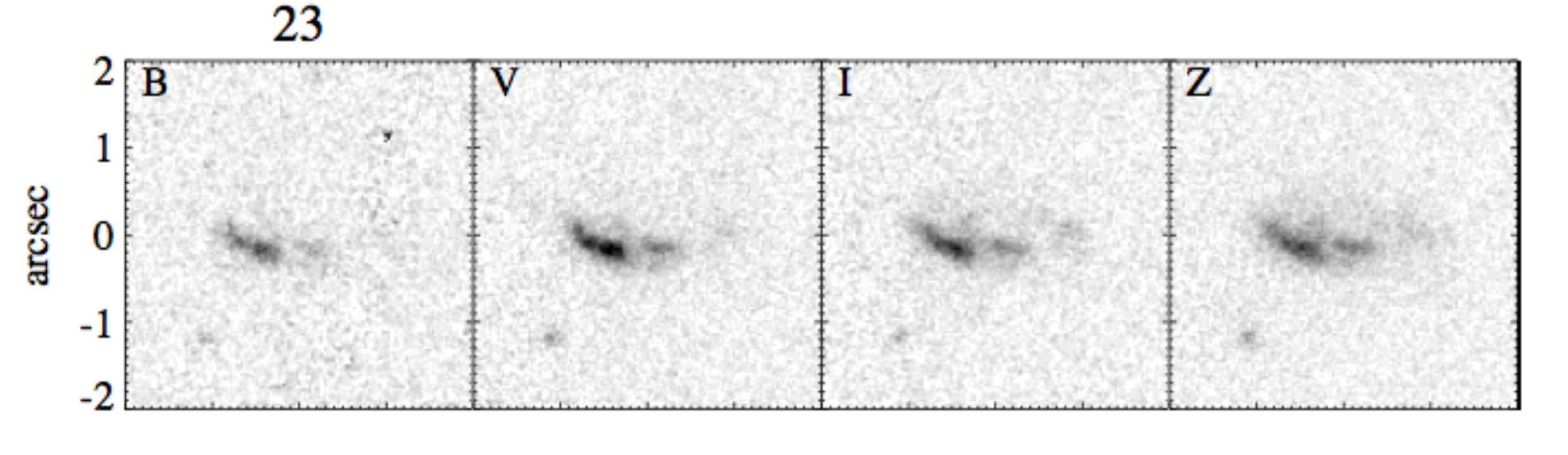}
\includegraphics[width=5.8cm,angle=-90]{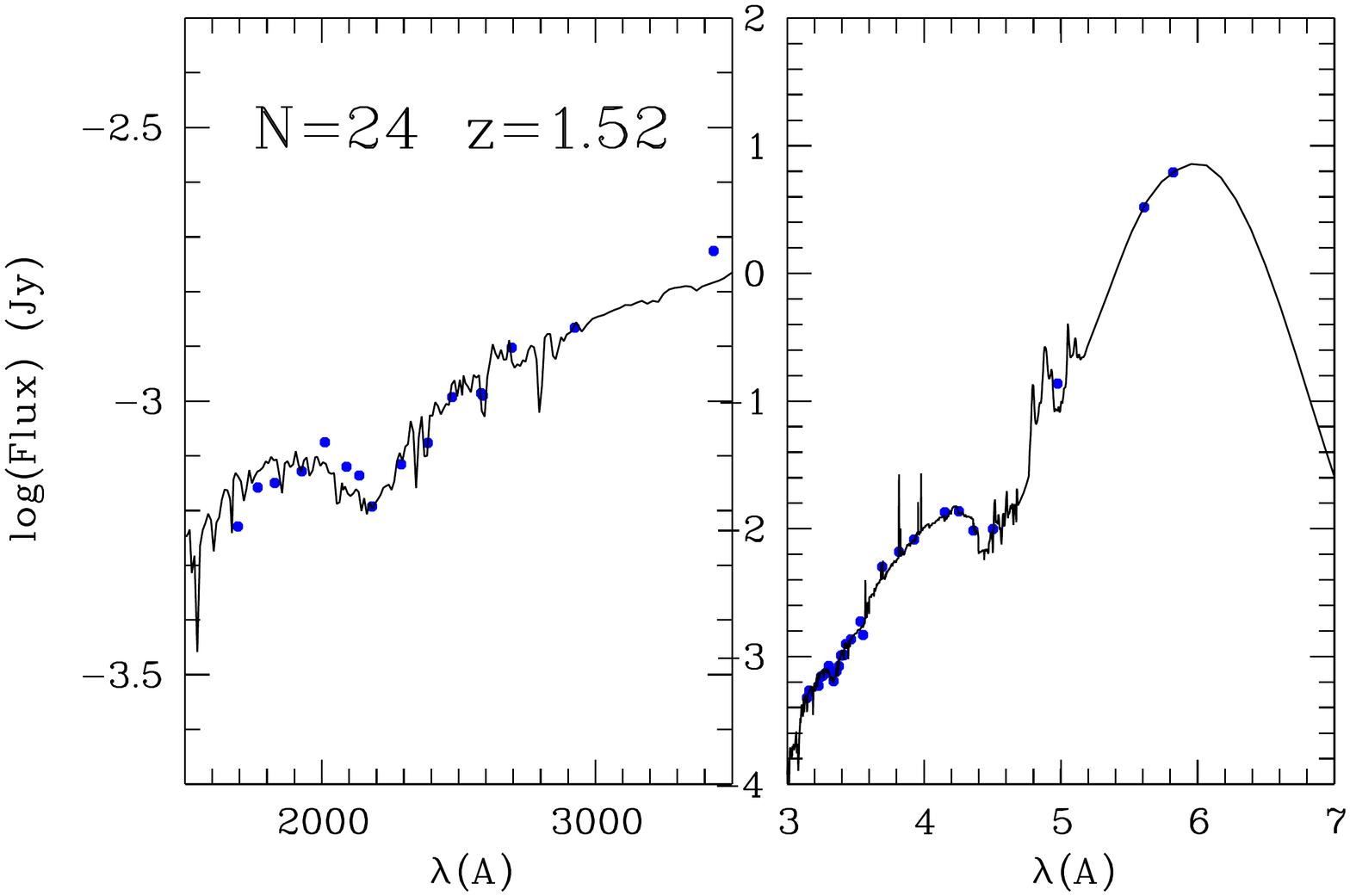}
\includegraphics[width=8cm]{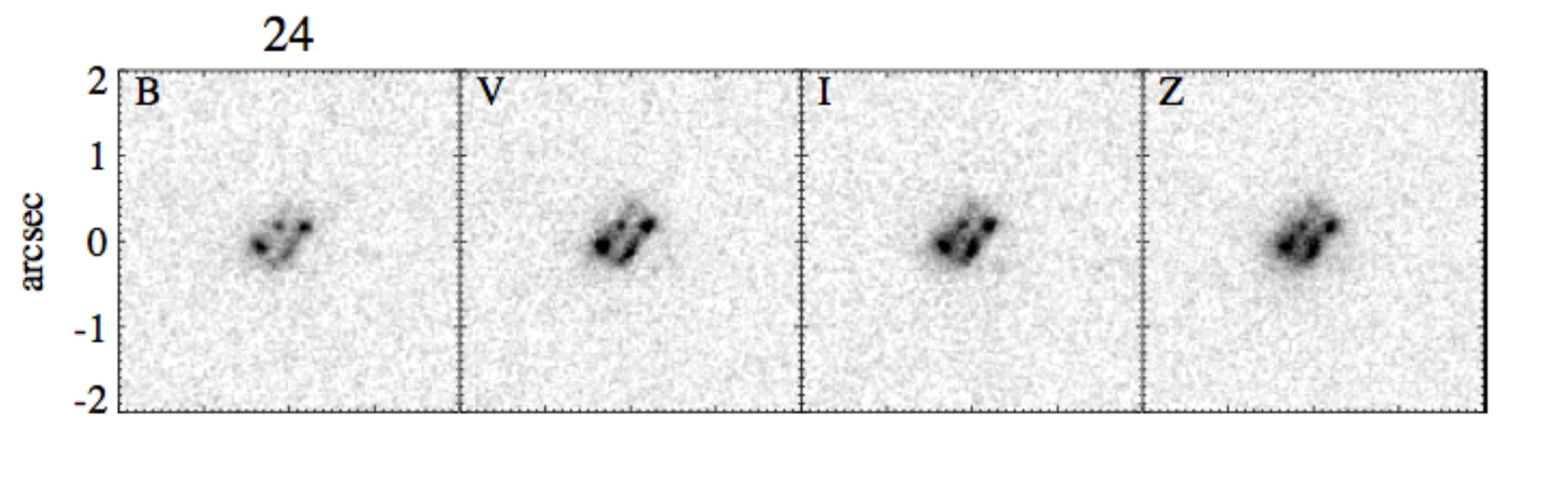}
\caption{ cont'd}
      \end{figure}
  \begin{figure}
    \ContinuedFloat
  \centering
\includegraphics[width=5.8cm,angle=-90]{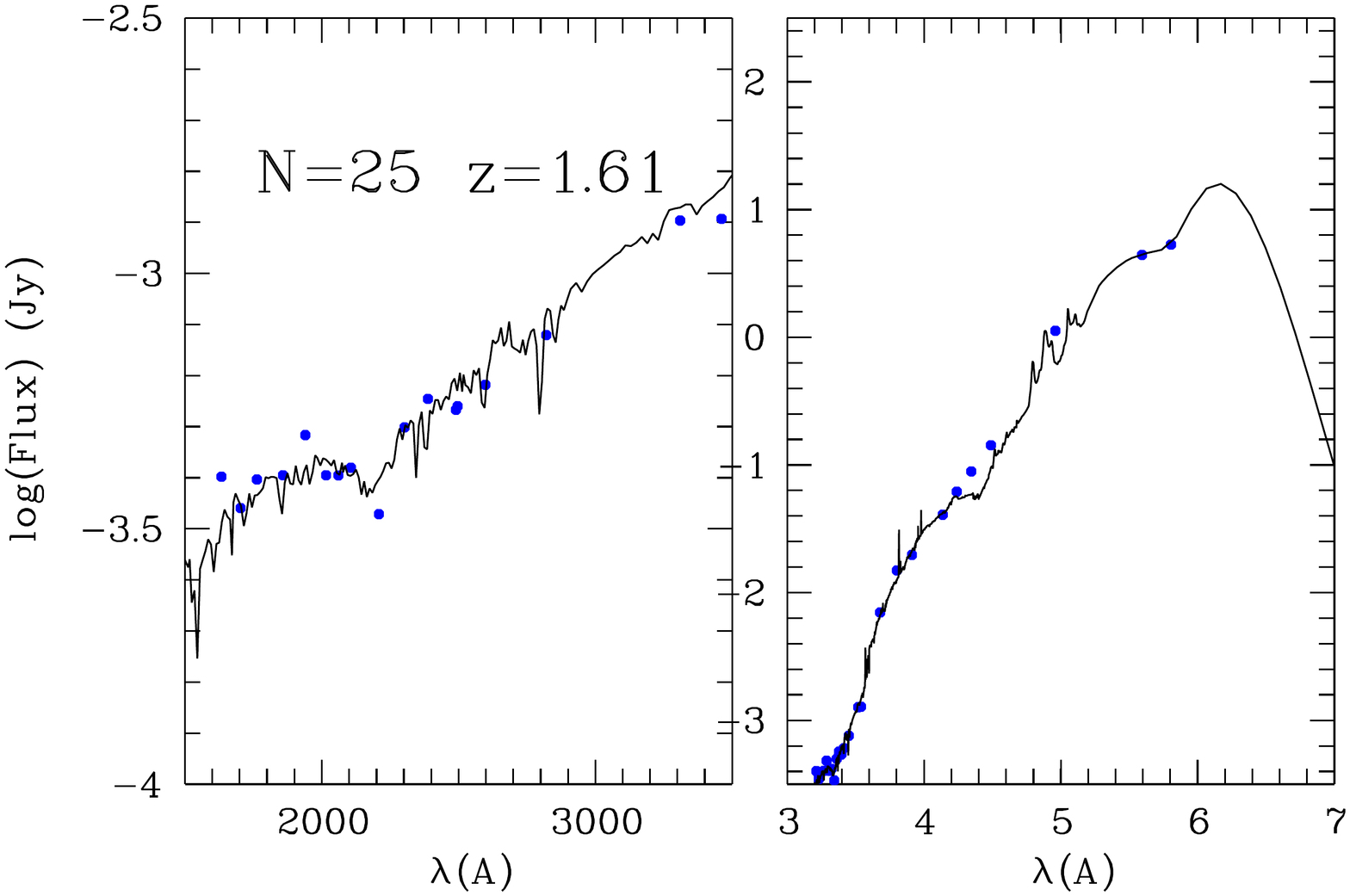}
\includegraphics[width=8cm]{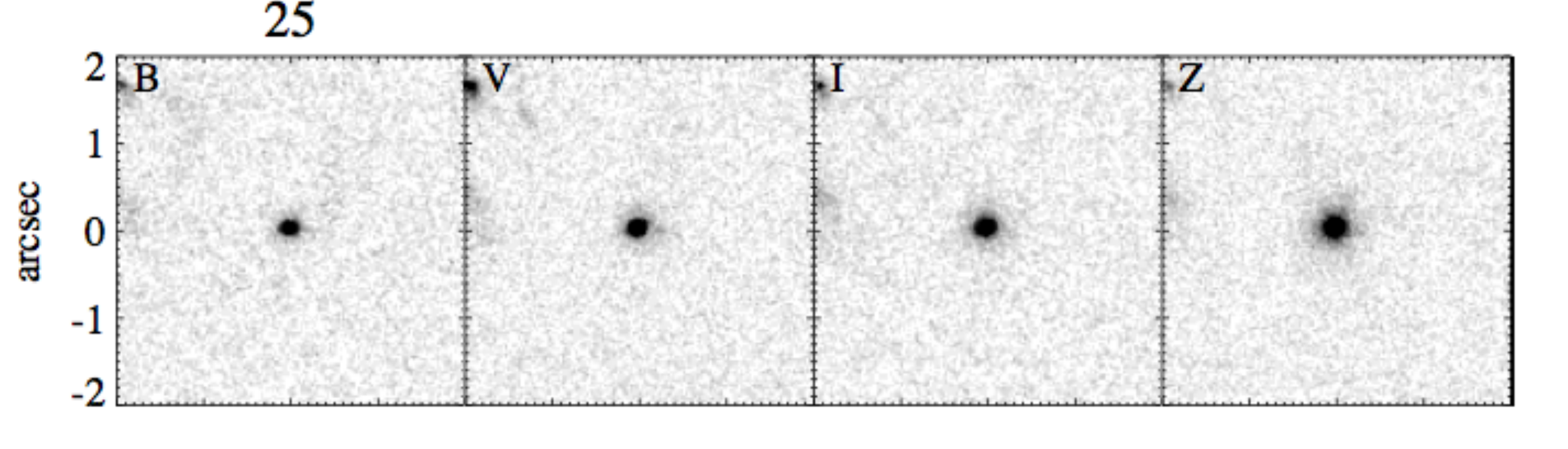}
\includegraphics[width=5.8cm,angle=-90]{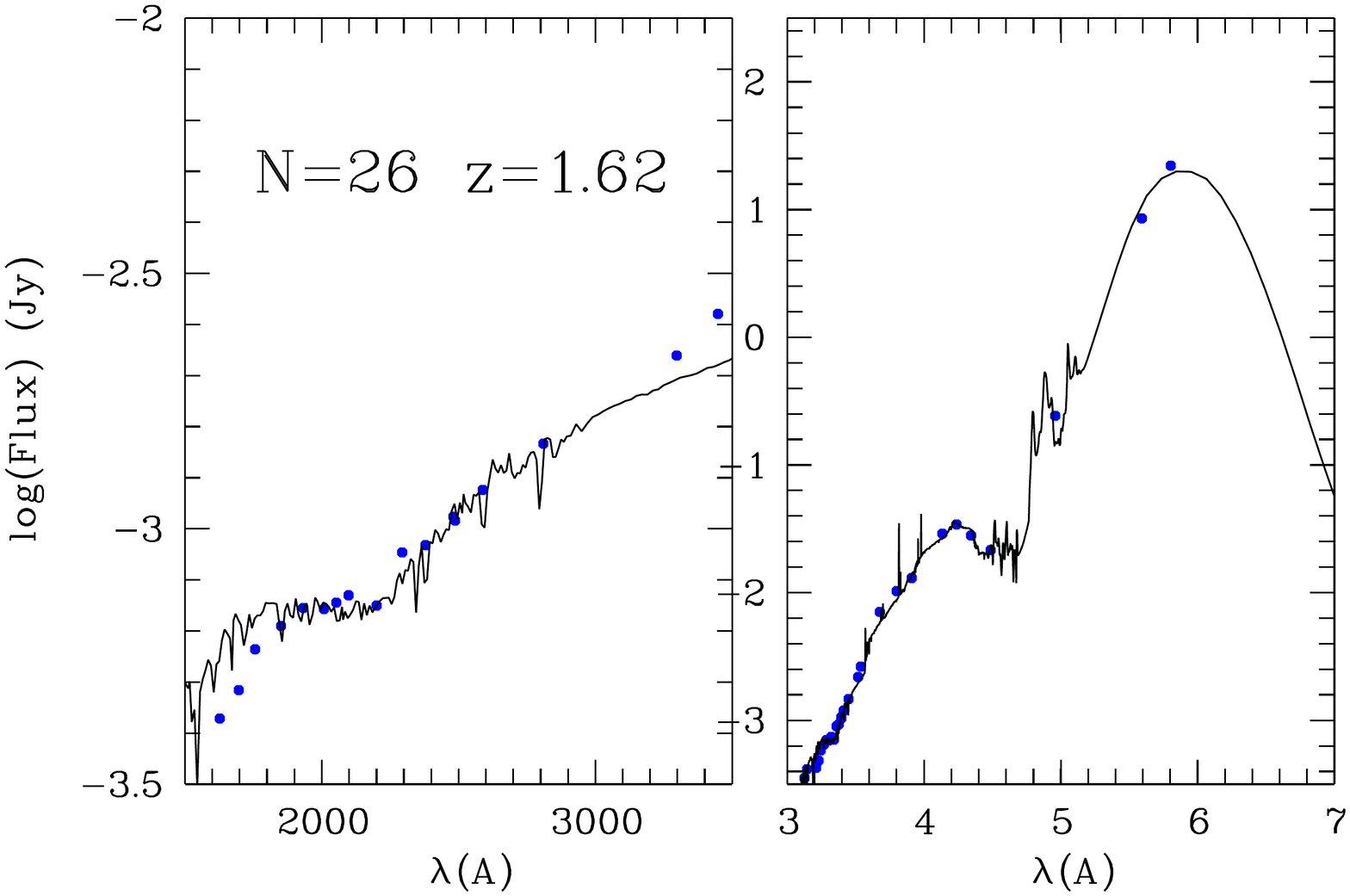}
\includegraphics[width=8cm]{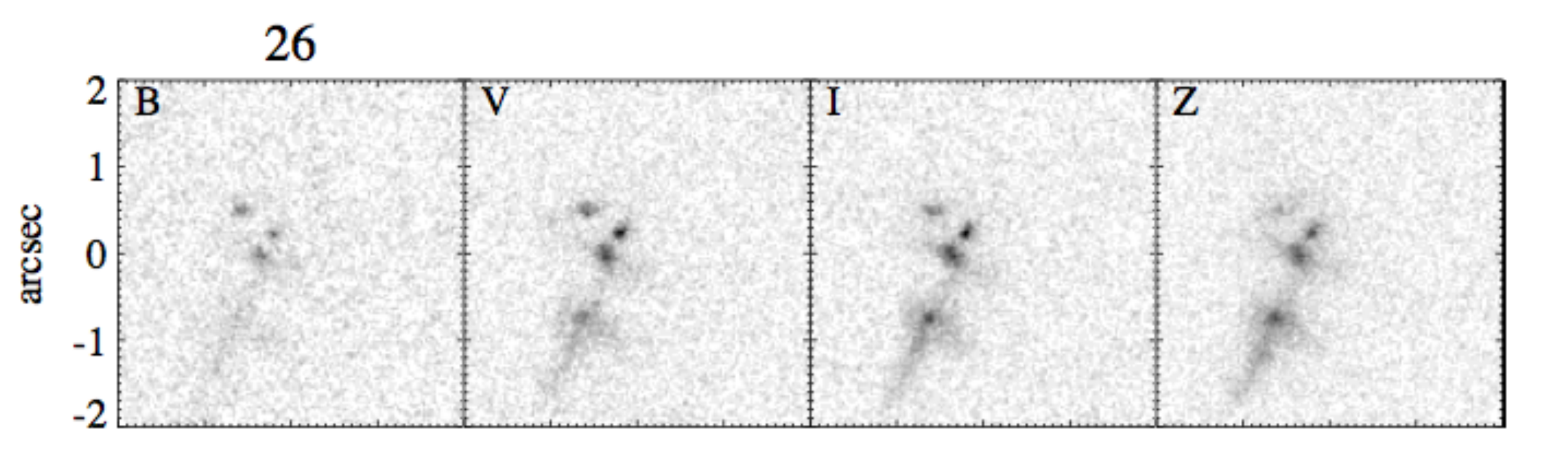}
\includegraphics[width=5.8cm,angle=-90]{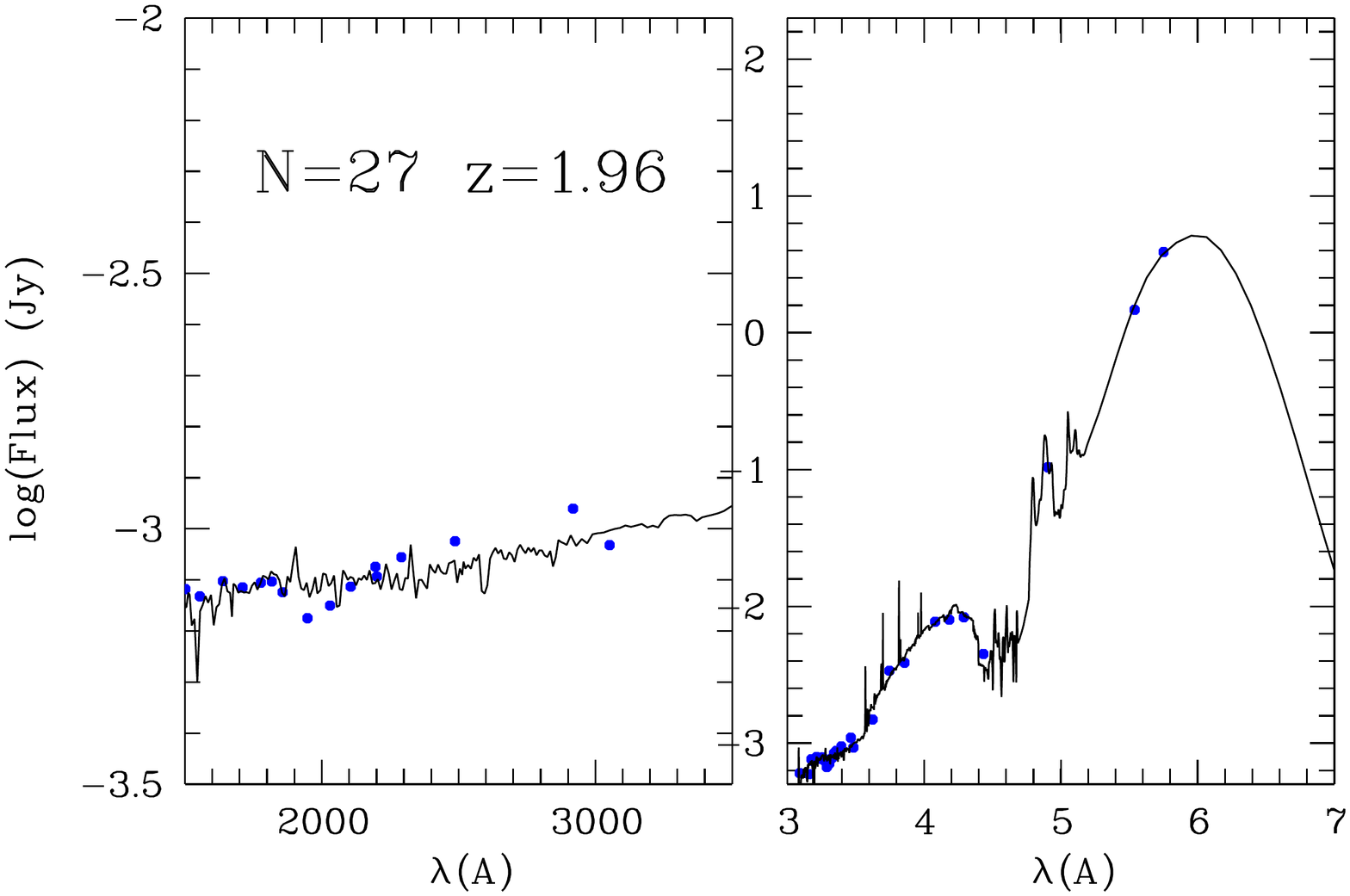}
\includegraphics[width=8cm]{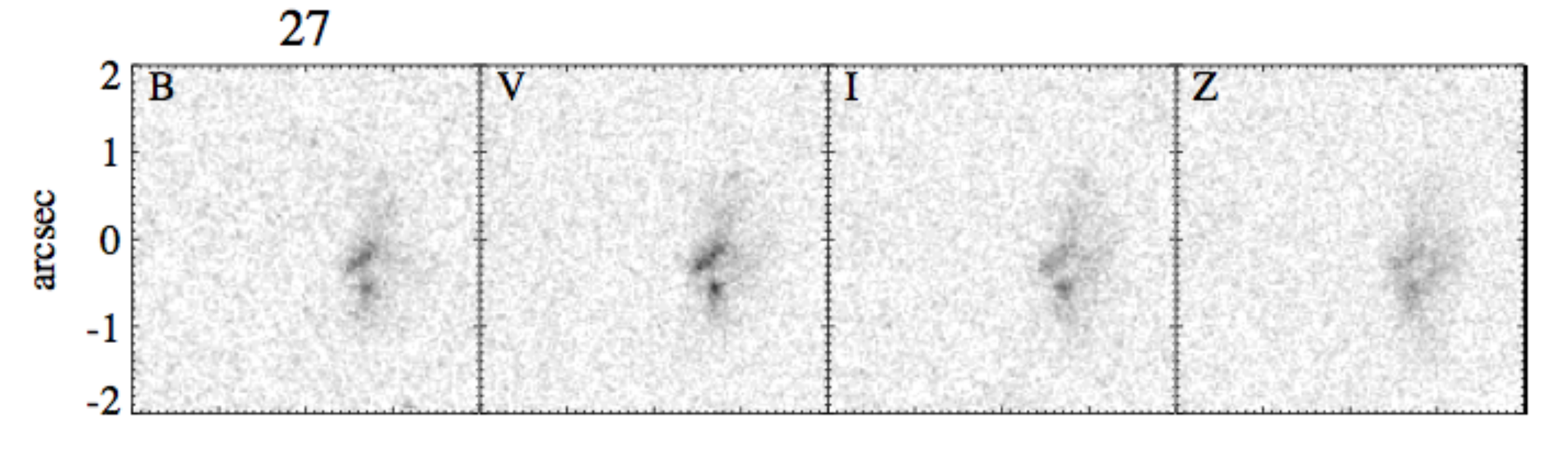}
\caption{ cont'd}
      \end{figure}

    \begin{figure}
    \ContinuedFloat
  \centering
  \includegraphics[width=5.8cm,angle=-90]{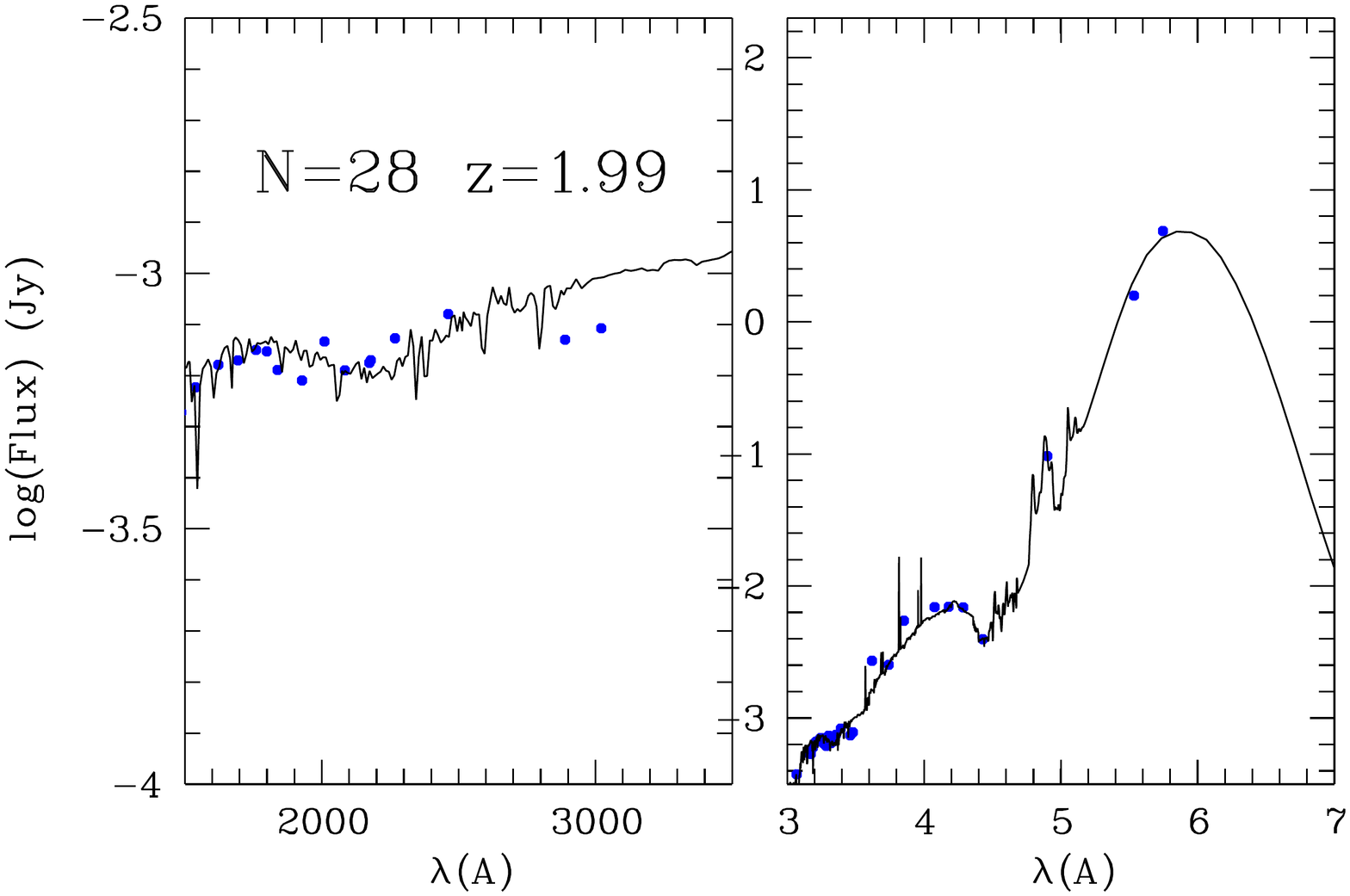}
  \includegraphics[width=8cm]{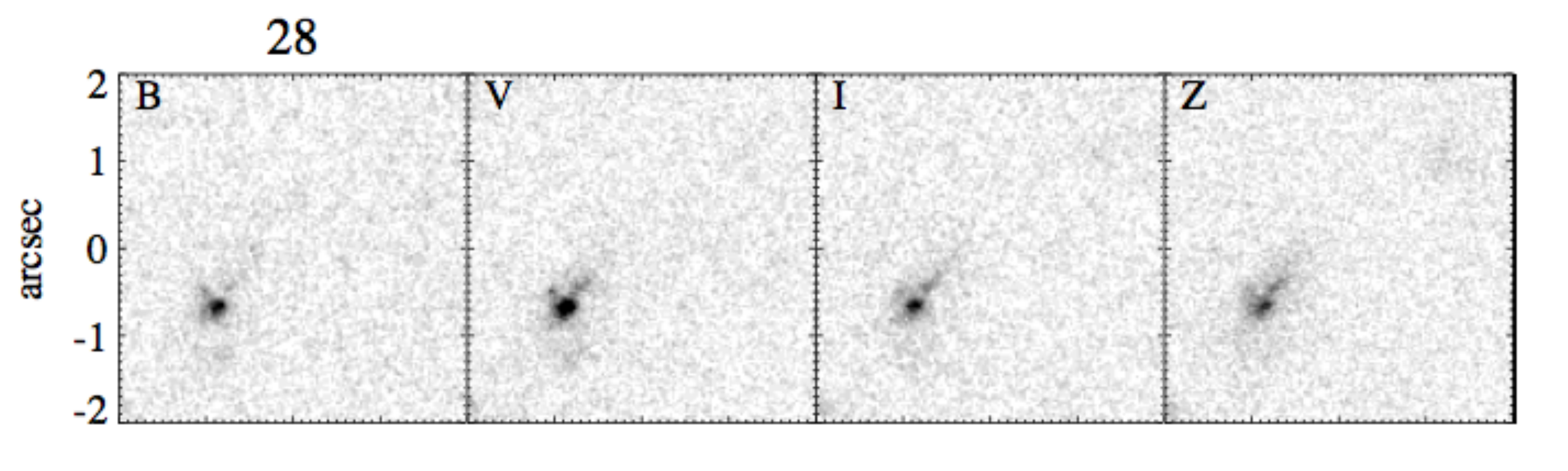}
\includegraphics[width=5.8cm,angle=-90]{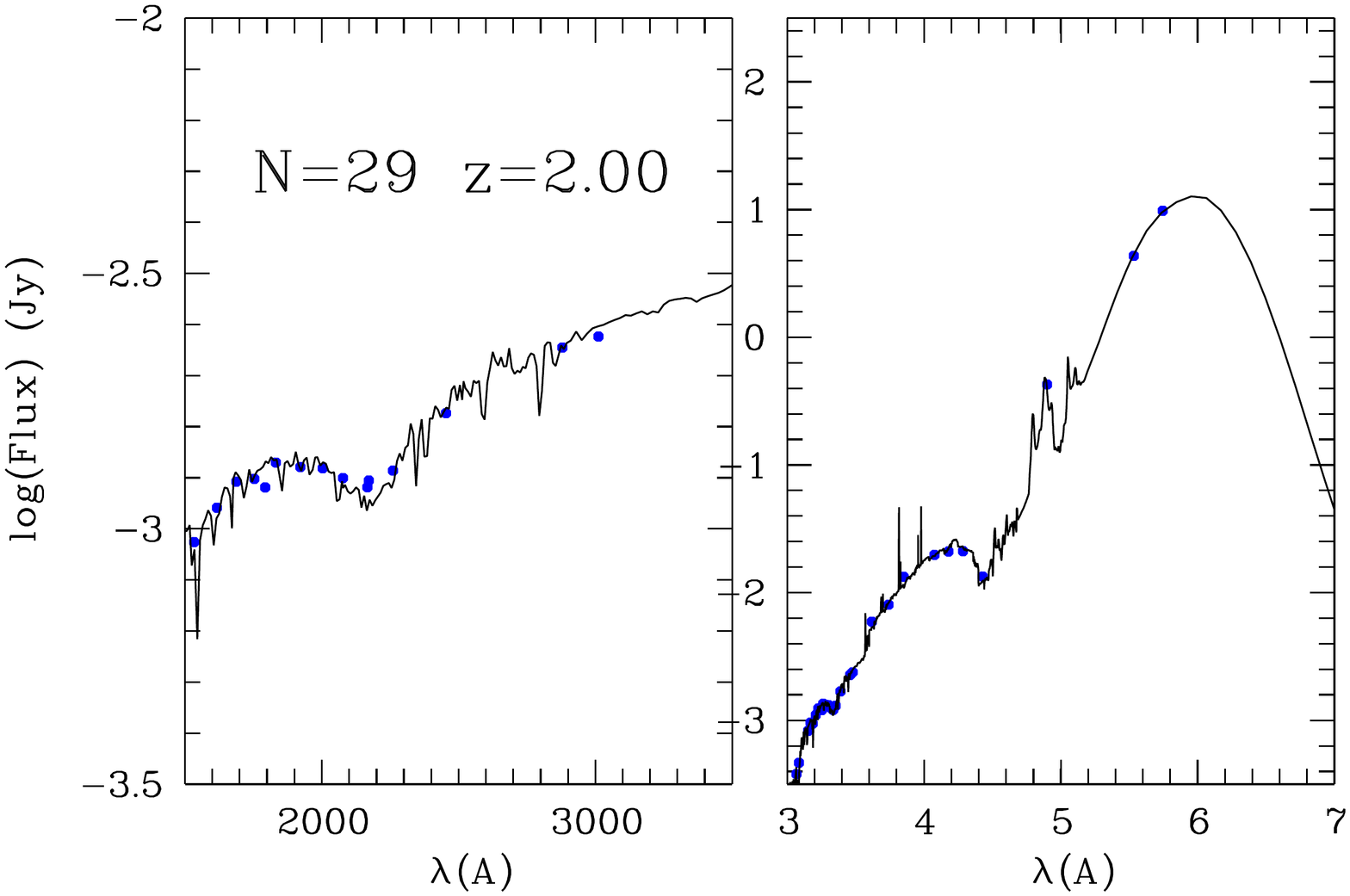}
\includegraphics[width=8cm]{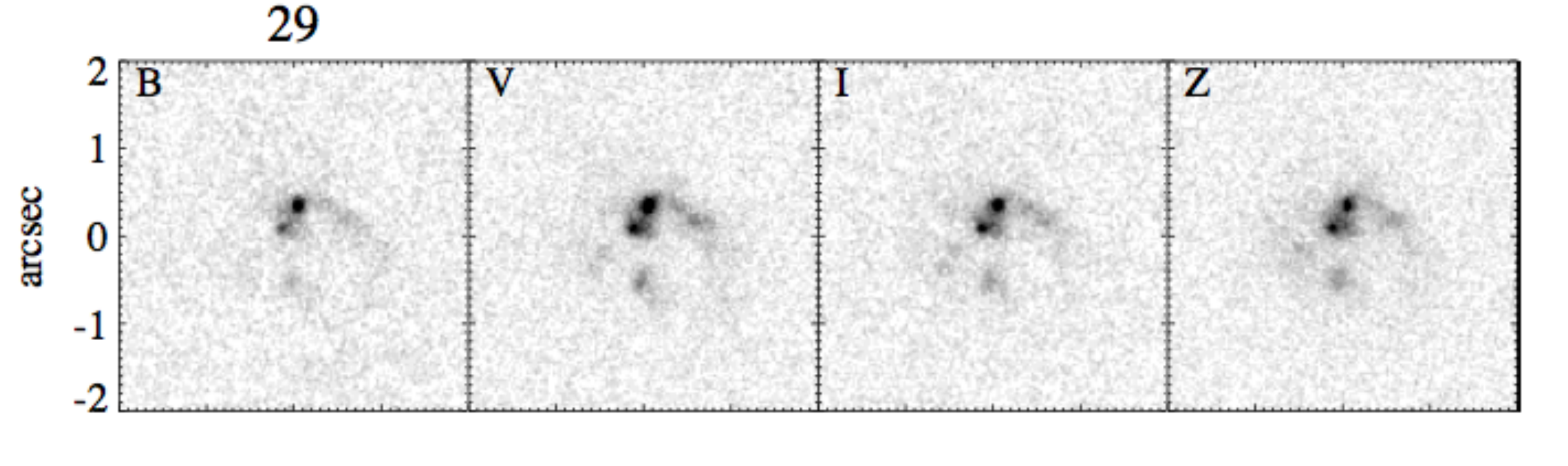}
\includegraphics[width=5.8cm,angle=-90]{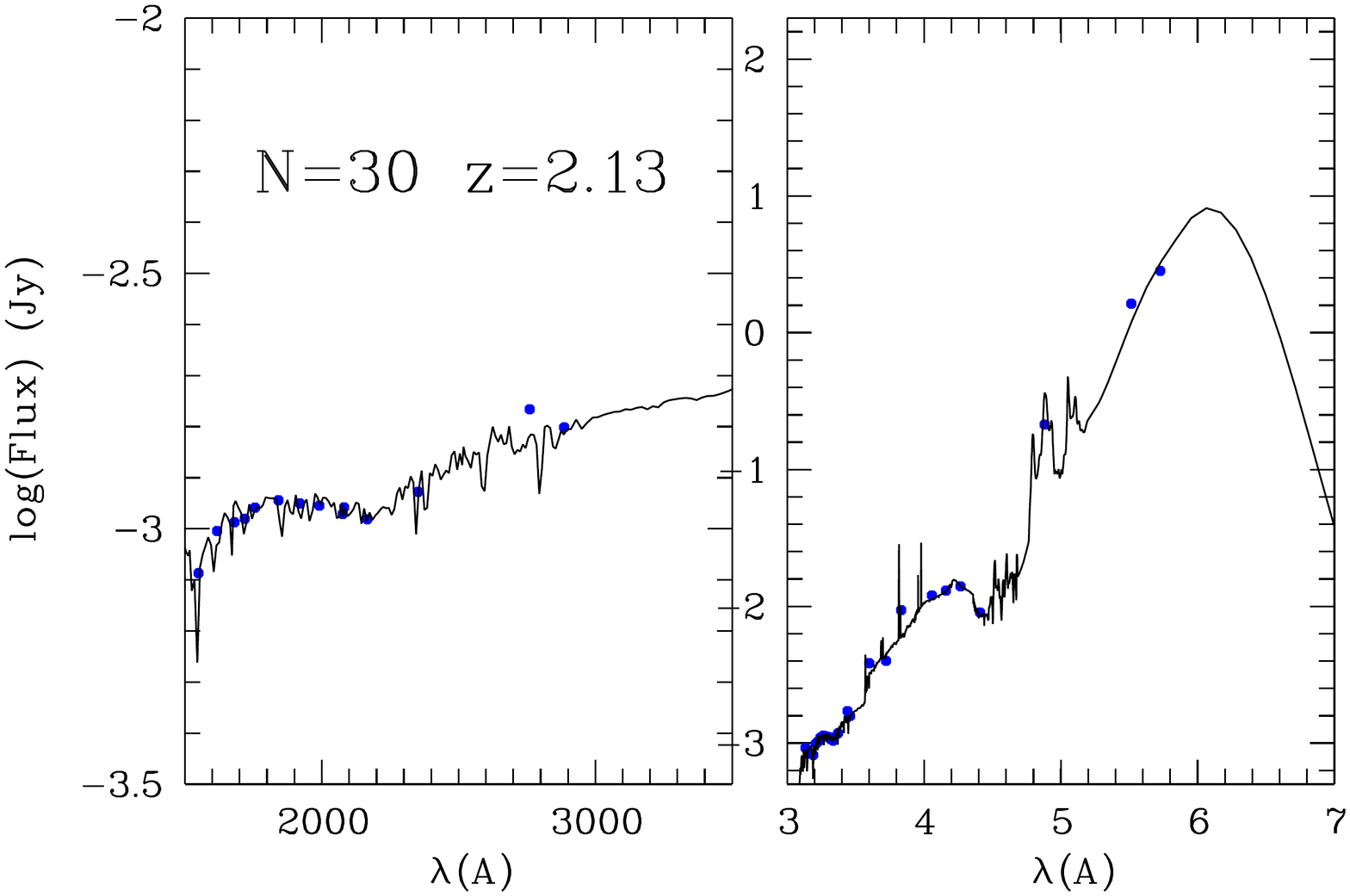}
\includegraphics[width=8cm]{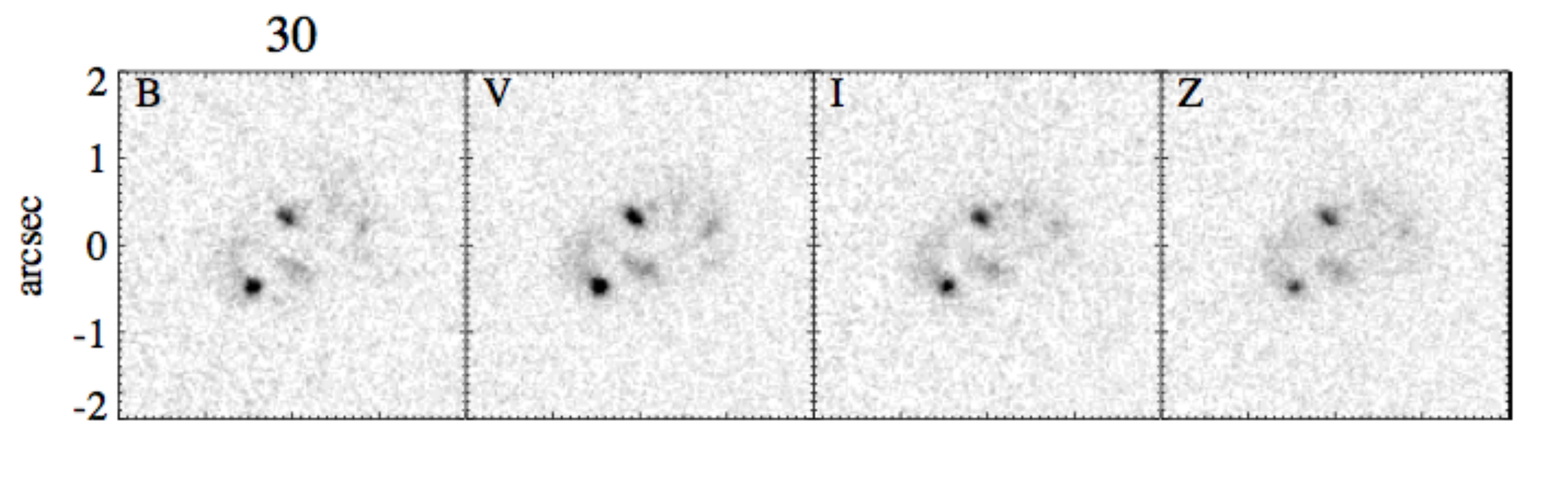}
\caption{ cont'd}
      \end{figure}

}

   \begin{acknowledgements}
     This work is partially supported by the French National Agency for research (ANR-09-BLAN-0224). 
 PACS has been developed by a consortium of institutes led by MPE (Germany)
and including UVIE (Austria); KU Leuven, CSL, IMEC (Belgium); CEA,
LAM (France); MPIA (Germany); INAFIFSI/ OAA/OAP/OAT, LENS, SISSA
(Italy); IAC (Spain). This development has been supported by the funding agencies
BMVIT (Austria), ESA-PRODEX (Belgium), CEA/CNES (France), DLR
(Germany), ASI/INAF (Italy), and CICYT/MCYT (Spain).
\end{acknowledgements}

\end{document}